\documentclass[useAMS,usenatbib,letterpaper]{mn2e}

\usepackage{times}
\usepackage{psfrag}
\usepackage[dvips]{graphicx}
\usepackage{amssymb}
\usepackage{bm}
\usepackage[french,english]{babel}
\usepackage{color}
\usepackage{amsmath}
\usepackage{algorithmic}
\usepackage{algorithm}
\usepackage{url}

\title[SARA for radio-interferometric imaging]
{Sparsity Averaging Reweighted Analysis (SARA): a novel algorithm for radio-interferometric imaging}

\author[Carrillo et al.]
{R. E. Carrillo$^{1}$\thanks{E-mail: rafael.carrillo@epfl.ch}, J. D. McEwen$^{2}$ and Y. Wiaux$^{1,3}$\\
$^{1}$Institute of Electrical Engineering, Ecole Polytechnique F{\'e}d{\'e}rale de Lausanne (EPFL),
      CH-1015 Lausanne, Switzerland\\
$^{2}$Department of Physics and Astronomy, University College London, London WC1E 6BT, UK\\
$^{3}$Department of Radiology and Medical Informatics, University of Geneva (UniGE), 
      CH-1211 Geneva, Switzerland\\}

\begin{document}

\date{\today}

\pagerange{\pageref{firstpage}--\pageref{lastpage}} \pubyear{2012}

\maketitle

\label{firstpage}

\begin{abstract}
We propose a novel algorithm for image reconstruction in radio interferometry. The ill-posed inverse problem associated with the incomplete Fourier sampling identified by the visibility measurements is regularized by the assumption of average signal sparsity over representations in multiple wavelet bases. The algorithm, defined in the versatile framework of convex optimization, is dubbed Sparsity Averaging Reweighted Analysis (SARA). We show through simulations that the proposed approach outperforms state-of-the-art imaging methods in the field, which are based on the assumption of signal sparsity in a single basis only.
\end{abstract}


\section{Introduction}
\label{sec:Introduction}

Aperture synthesis in radio interferometry is a powerful technique that dates back more than sixty years~\citep{ryle46,blythe57,ryle59,ryle60,thompson04}. It allows observation of the sky with otherwise inaccessible angular resolution and sensitivity (i.e.~dynamic range), providing a wealth of information for astrophysics and cosmology. The measurement equation for aperture synthesis provides incomplete linear information about the signal, thus defining an ill-posed inverse problem in the perspective of signal reconstruction. Under restrictive assumptions of narrow-band (i.e.~monochromatic) non-polarized imaging on small fields of view, the visibilities measured identify with Fourier measurements. Already powerful calibration and imaging techniques have been developed in the field. Standard imaging algorithms, such as CLEAN and its multi-scale variants~\citep{hogbom74,bhatnagar04,cornwell08b}, regularize the inverse problem through an implicit sparsity assumption of the signal in the spatial dimensions. 

The new science envisaged in astronomy in the coming decades requires that next-generation radio telescopes, such as the new LOw Frequency ARray (LOFAR\footnote{\url{http://www.lofar.org/}}), or the future Extended Very Large Array (EVLA\footnote{\url{http://www.aoc.nrao.edu/evla/}})  and Square Kilometer Array (SKA\footnote{\url{http://www.skatelescope.org/}}), achieve much higher dynamic range than current instruments, also at higher angular resolution. These telescopes will also have to consider wide-band (i.e.~hyper-spectral) polarized imaging on wide fields of view on the celestial sphere. Direction-dependent effects further complicate the measurement equation, and will have to be calibrated and accounted for in this high-dimensional imaging process, and calibrated. In this context, calibration and imaging techniques for radio interferometry literally need to be re-invented, thus triggering an intense research in the field.

The now famous theory of compressed sensing deals with the recovery of sparse signals from incomplete linear measurements~\citep{donoho06,candes06,baraniuk07a}. It acknowledges the fact that natural signals often exhibit a sparse representation in multi-scale bases. Compressed sensing proposes both optimization of the acquisition technique, and non-linear iterative algorithms for signal reconstruction regularizing the ill-posed inverse problem through a sparsity prior. These algorithms are defined either in the context of convex optimization, or greedy approaches. It is also important to note that, beyond the pure theory of compressed sensing, these frameworks are particularly versatile and can account for a large variety of priors.

The first application of compressed sensing to radio interferometry was performed by \citet{wiaux09}, where the problem of image reconstruction from incomplete visibility measurements was considered.  \citet{wiaux09} demonstrated the versatility of the approach and its superiority relative to standard interferometric imaging techniques.  The spread spectrum phenomenon, which arises by partially relaxing the small field-of-view (FOV) assumption and including a first order $w$ term, was introduced by \citet{wiaux09b} as a potential optimization of the acquisition, leading to enhanced image reconstruction quality for sparsity bases that are not maximally incoherent with the measurement basis.  Furthermore, a compressed sensing approach was developed and evaluated by \citet{wiaux10a} to recover the signal induced by cosmic strings in the cosmic microwave background. \citet{mcewen11a} generalise the compressed sensing imaging techniques developed by \citet{wiaux09} and \citet{wiaux09b} to a wide FOV, recovering interferometric images defined directly on the sphere, rather than a tangent plane.  All of these works consider uniformly random and discrete visibility coverage in order to remain as close to the theory of compressed sensing as possible.  First steps towards more realistic visibility coverages have been taken by \citet{suksmono09} and \citet{wenger10}, who consider coverages due to specific interferometer configurations but which remain discrete.  \cite{li11} studied a compressed sensing imaging approach based on the isotropic undecimated wavelet transform, reporting results from discrete simulated coverages of ASKAP. These preliminary works suggest that the performance of compressed sensing reconstructions is likely to hold for more realistic visibility coverages.

Convex optimization methods coupled with sparsity priors have proven to be a powerful framework for radio-interferometric imaging. Beyond the versatility that enables one to impose a wide range of sparsity priors, convex optimization provides significant improvements in the speed of the reconstruction process relative to state-of-the-art imaging methods in radio interferometry. This enhancement in speed is crucial for the scalability of the techniques to very high dimensions in the perspective of next-generation telescopes.  

In the present work, we propose a novel algorithm for radio interferometric imaging, defined in the framework of convex optimization, dubbed the sparsity averaging reweighted analysis (SARA) algorithm. The algorithm relies on the conjecture that astrophysical signals are simultaneously sparse in various bases, in particular the Dirac basis, wavelet bases, or in their gradient, so that promoting average signal sparsity over multiple wavelet bases represents an extremely powerful prior. For comparison, we also study a variety of fast image reconstruction algorithms designed in the frameworks of convex optimization and sparse signal modelling, some of which were identified as providing similar performance as CLEAN and its multi-scale versions. We show, through realistic simulations, the superiority of SARA compared to most radio-interferometric imaging techniques.

The organization of the remainder of the paper is the following. In Section~\ref{sec:Convex optimization for sparse reconstruction}, we review convex optimization methods for sparse inverse problems in the compressed sensing framework and discuss their versatility. In Section~\ref{sec:Radio interferometric imaging}, we recall the inverse problem for image reconstruction from radio-interferometric data and concisely describe the state-of-the-art image reconstruction techniques used in radio astronomy. Section~\ref{sec:SARA} presents the SARA algorithm. Numerical results of the SARA algorithm compared to state-of-the-art imaging techniques are presented in Section~\ref{sec:Simulations and results}. Finally we conclude in Section~\ref{sec:Conclusion} with closing thoughts.

\section{Compressed sensing and convex optimization}
\label{sec:Convex optimization for sparse reconstruction}

Convex optimization provides a powerful and versatile framework to solve sparse linear inverse problems such as those posed in radio interferometry. In this section, we concisely recall the inverse problem for sparse signals considered in the compressed sensing framework and proceed further with a discussion of the versatility offered by convex optimization approaches. Finally, we review the key ideas behind the methods to solve these convex problems.

\subsection{Compressed sensing}
In the framework of compressed sensing \citep{candes06,donoho06,baraniuk07a,donoho09,fornasier11} the signals probed are firstly assumed to be sparse or compressible in some basis. Consider a complex-valued signal denoted by the vector $\bm{x}\in\mathbb{C}^{N}$. An orthonormal basis $\mathsf{\Psi}\in\mathbb{C}^{N\times N}$ is also considered, in which the decomposition $\bm{\alpha}\in\mathbb{C}^{N}$ of the signal is defined by $\bm{x} \equiv \mathsf{\Psi}\bm{\alpha}$. The signal is said to be sparse if it contains only $K$ non-zero coefficients in its decomposition, where $K \ll N$, or compressible if its ordered set of coefficients decays rapidly and the signal can be well approximated by just the first $K$ coefficients.

Secondly, the signal is assumed to be probed by $M$ linear measurements denoted by a vector $\bm{y}\in\mathbb{C}^{M}$ in some sensing basis $\mathsf{\Phi}\in\mathbb{C}^{M\times N}$ and possibly affected by independent and identically distributed noise $\bm{n}\in\mathbb{C}^{M}$. This defines an inverse problem
\begin{equation}
\bm{y}\equiv\mathsf{\Theta}\bm{\alpha}+\bm{n}\textnormal{ with }\mathsf{\Theta}\equiv\mathsf{\Phi\Psi}\in\mathbb{C}^{M\times N},\label{cs2}
\end{equation}
where the matrix $\mathsf{\Theta}$ identifies the sensing basis as seen from the sparsity itself. Typically $M<N$ so that the inverse problem is ill-posed.

The ideal approach to recover $\bm{\alpha}$ from \eqref{cs2} is to find the sparsest representation $\bar{\bm{\alpha}}$ that is consistent with the measurements, posing the following problem:
\begin{equation}
\min_{\bar{\bm{\alpha}}\in\mathbb{C}^{N}}\|\bar{\bm{\alpha}}\|_{0}\textnormal{ subject to }\| \bm{y}-\mathsf{\Theta}\bar{\bm{\alpha}}\|_{2}\leq\epsilon,\label{cs3}
\end{equation}
where the $\ell_{0}$ norm, $\|\bar{\bm{\alpha}}\|_{0}$, counts the number of non-zero elements in $\bar{\bm{\alpha}}$ and $\epsilon$ is an upper bound on the $\ell_{2}$ norm $\|\bar{\bm{n}}\|_{2}$ of the residual noise, with $\bar{\bm{n}}\equiv \bm{y}-\mathsf{\Theta}\bar{\bm{\alpha}}$. Let us recall that the $\ell_{p}$ norm of a complex-valued vector $\bm{a}\in\mathbb{C}^{M}$ is defined as $\| \bm{a}\|_{p}\equiv(\sum_{i=1}^{M}|a_{i}|^{p})^{1/p}$, where $|\cdot|$ represents the modulus of a complex number.  

The problem in (\ref{cs3}) is combinatorial and NP-complete. The most common approach is to replace the $\ell_{0}$ norm by the $\ell_{1}$ norm and pose a convex problem to estimate a solution~\citep{chen01,candes06,donoho06}. In the presence of noise, the so-called Basis Pursuit denoise problem is the minimization of the $\ell_{1}$ norm $\|\bar{\bm{\alpha}}\|_{1}$ of the coefficients of the signal in the sparsity basis under a constraint on the $\ell_{2}$ norm $\|\bar{\bm{n}}\|_{2}$ of the residual noise:
\begin{equation}
\min_{\bar{\bm{\alpha}}\in\mathbb{C}^{N}}\|\bar{\bm{\alpha}}\|_{1}\textnormal{ subject to }\| \bm{y}-\mathsf{\Theta}\bar{\bm{\alpha}}\|_{2}\leq\epsilon.\label{cs4}
\end{equation}
The theory shows that the $\ell_0$ and Basis Pursuit denoise problems are equivalent under certain properties of the sensing matrix, $\mathsf{\Theta}$~\citep{candes06a,candes06,candes08}. The theory also offers various ways to design suitable sensing matrices, showing in particular that a small number of measurements is required relative to a naive Nyquist-Shannon sampling: $M\ll N$. Note that, in theory, an explicit $\ell_0$ minimization would require fewer measurements, $M\approx2K$~\citep{candes06a,candes06,candes08}. 

A family of iterative greedy algorithms are also proposed in the literature~\citep{mallat93,tropp07,needell08,blumensath09}. These algorithms are shown to enjoy similar approximate reconstruction properties, however, requiring more measurements for exact reconstruction than convex optimization approaches. 

\subsection{Convex optimization versatility}
While the theory of compressed sensing provides reconstruction guarantees for the $\ell_1$ minimization problem, convex optimization is extremely versatile and can account for many variations, which in practice can prove more effective for signal reconstruction. In the following we briefly describe these  variations.

\paragraph*{Positivity:} One of the main advantages of convex approaches is the flexibility that they provide to include prior information about the signal as convex constraints. In the case of image processing problems, where most images of interest are intensity images, the signals are real-valued and positive, i.e. $\bm{x}\in\mathbb{R}^{N}_{+}$. This constraint is convex and can be easily added to the optimization problems without much computational load increase and without affecting their convergence. This constraint has proven to be very effective in improving reconstruction quality in radio-interferometric imaging~\citep{wiaux09}.

\paragraph*{Constrained vs unconstrained problems:} The least squares $\ell_1$ regularized problem is an alternative formulation of the basis pursuit denoising problem that recovers a sparse signal as the solution of an unconstrained problem formulated as: \mbox{$\min_{\bar{\bm{\alpha}}\in\mathbb{C}^{N}}\frac{1}{2}\| \bm{y}-\mathsf{\Theta}\bar{\bm{\alpha}}\|_{2}^{2}+\lambda\|\bar{\bm{\alpha}}\|_{1}$}, where $\lambda$ is a regularization parameter that balances the weight between the fidelity term and the regularization term. It follows that determining the proper value of $\lambda$ is akin to determining the power limit of the noise~\citep{chen01}. However, there is no optimal strategy to fix the regularization parameter even if the noise level is known, therefore constrained problems, such as \eqref{cs4}, offer a stronger fidelity term when the noise power is known, or can be estimated \emph{a priori}. 

\paragraph*{Orthogonal vs overcomplete representations:} The techniques mentioned above hold for signals which are sparse in the standard coordinate basis or sparse with respect to some other orthonormal basis. However, there are numerous practical examples in which a signal of interest is not sparse in a single orthonormal basis but over several orthonormal bases or over an overcomplete dictionary~\citep{candes10}. In the generalized setting the signal $\bm{x}$ is now expressed in terms of a dictionary $\mathsf{\Psi}\in\mathbb{C}^{N\times D}$ ($N<D$) as $\bm{x} = \mathsf{\Psi}\bm{\alpha}$, $\bm{\alpha}\in\mathbb{C}^{D}$. Note that the problem is now severely underdetermined since $M\ll N <D$, therefore requiring greater sparsity or compressibility of $\bm{\alpha}$. \cite{rauhut08} find conditions on the compound matrix $\mathsf{\Phi \Psi}$ such that $\bm{\alpha}$ can be recovered accurately, which leads to a good estimate of $\bm{x}$. \cite{candes10} extend the compressed sensing theory to redundant dictionaries, providing theoretical stability guarantees based on general conditions on the sensing matrix $\mathsf{\Phi}$.

\paragraph*{Analysis vs synthesis problems:} The basis pursuit denoising problem defines the optimization in the sparse representation domain finding the optimal representation vector $\bar{\bm{\alpha}}$ and then recovering the true signal trough the synthesis relation $\bar{\bm{x}}=\mathsf{\Psi}\bar{\bm{\alpha}}$. These methods are known as synthesis based methods in the literature. Synthesis-based problems may also be substituted by analysis-based problems, where instead of estimating a sparse representation of the signal, the methods recover the signal itself~\citep{elad07}. In the case of orthonormal bases, $\mathsf{\Psi}$, the two approaches are equivalent. However, when $\mathsf{\Psi}$ is a frame or an overcomplete dictionary, the two problems are no longer equivalent. The geometry of the two problems are studied by \cite{elad07}, who show that because these geometrical structures exhibit substantially different properties, there is a large gap between the two formulations. One remark to make is that the analysis problem does not increase the dimensionality of the problem relative to solving for the signal itself, even in the case when overcomplete dictionaries are used. Empirical studies have shown very promising results for the analysis approach~\citep{elad07}. \cite{candes10} provide a theoretical analysis of the $\ell_1$ analysis problem coupled with redundant dictionaries. 

\paragraph*{Reweighted $\ell_1$ vs $\ell_1$ minimization:} 
As discussed above, the $\ell_1$ minimization problem is equivalent to $\ell_0$ minimization when the measurement matrix satisfies certain conditions defined in the context of compressed sensing. In general though, the key difference between the two problems, of course, is that $\ell_1$ depends on the magnitudes of the coefficients of a signal, whereas $\ell_0$ minimization does not. To reconcile this imbalance, a reweighted $\ell_1$ minimization algorithm was proposed by \cite{candes08a} to mimic the $\ell_0$ minimization behaviour. The algorithm replaces the $\ell_1$ norm in \eqref{cs4} by a weighted $\ell_1$ norm $\sum_{i=1}^Nw_i|\bar{\alpha}_i|$. The idea behind this formulation is that large weights will encourage small coordinates of the solution vector, and small weights will encourage larger coordinates. As a motivational example suppose that the sparse signal $\alpha$ is known \emph{a priori} and that we set the weights as $w_i=|\alpha_i|^{-1}$. In this case the weights are infinite at all locations where the signal is zero, forcing the coordinates of the solution vector at these locations to be zero. This set of weights makes the weighted norm independent of the precise value of the non-zero components and guarantees to recover the correct solution assuming only $K<M$. 

In practice, the original signal is not known in advance but we can compute the appropriate weights by solving sequentially weighted $\ell_1$ problems, each using as weights essentially the inverse of the values of the solution of the previous problem. Of course, it is not possible to have infinite weights where the estimated signal values are zero, so a stability parameter must also be added to the signal value in the selection of the weights. This procedure has been observed to be very effective in reducing the number of measurements needed for recovery, and to outperform standard $\ell_1$-minimization in many situations, see e.g. \citep{candes08a,needell09}.

\subsection{Convex optimization algorithms}
Unlike many generic optimization problems, convex optimization problems can be efficiently solved, both  in theory (i.e., via algorithms with worst-case  polynomial complexity) and in practice \citep{mattingley10}. There exists a broad range of methods to efficiently solve convex problems, e.g. interior point methods, primal-dual methods and proximal splitting methods. Among these, proximal splitting methods offer great flexibility and are shown to capture and extend several well-known algorithms in a unifying framework. Douglas-Rachford, iterative thresholding, projected Landweber, projected gradient, alternating projections, alternating direction method of multipliers, alternating split Bregman are special instances of proximal splitting algorithms \citep{combettes11}. Such methods offer a powerful framework for solving convex problems in terms of speed and scalability of the techniques to very high dimensions.

Proximal splitting methods solve optimization problems of the form
\begin{equation}\label{cvx1}
\min_{\bm{x}\in\mathbb{R}^{N}} f_1(\bm{x})+\ldots +f_n(\bm{x}),
\end{equation}
where $f_1(\bm{x}),\ldots,f_n(\bm{x})$ are convex functions from $\mathbb{R}^{N}$ to $(-\infty,\infty)$. Note that any convex constrained problem can be formulated as an unconstrained problem by using the indicator function of the convex constraint set as one of the functions in \eqref{cvx1}, i.e. $f_k(\bm{x})=i_{C}(\bm{x})$ where $C$ represents the constraint set. Also note that complex-valued vectors are treated as real-valued vectors with twice the dimension (accounting for real and imaginary parts). A major difficulty that arises in solving this problem stems from the fact that, typically, some of the functions are not differentiable, which rules out conventional smooth optimization techniques. The key concept in proximal splitting methods is the use of the proximity operator of a convex function, which is a natural extension of the notion of a projection operator onto a convex set. For example, the proximal operator of the $\ell_1$ norm is the soft-thresolding operator, and the proximal operator of the indicator function of a constraint is simply the projection operator onto the constraint set. Proximal splitting methods proceed by splitting the contribution of the functions $f_1,\ldots,f_n$ individually so as to yield an easily implementable algorithm. They are called proximal because each non-smooth function in \eqref{cvx1} is involved via its proximity operator. In essence, the solution to \eqref{cvx1} is reached iteratively by successive application of the proximity operator associated with each function. See \cite{combettes11} for a review of proximal splitting methods and their applications in signal and image processing. 

One remark to make is that there also exist proximal splitting algorithms that offer a parallel implementation structure where all the proximity operators can be computed in parallel rather than sequentially. Examples of these algorithms are the proximal parallel algorithm and the simultaneous-direction method of multipliers~\citep{combettes11}. Such a structure is useful when implementing the algorithms in multicore architectures, thus providing a significant gain in terms of speed.   

\section{Radio interferometric imaging}
\label{sec:Radio interferometric imaging}
In this section we recall the general form of the visibility measurements and also pose the corresponding interferometric inverse problem for image reconstruction under small FOV considerations. We also review the state-of-the-art imaging algorithms in radio interferometry.

\subsection{Visibilities}
\label{ssec:Visibilities}
In order to image a region of the sky, all radio telescopes of an interferometric array point in the same direction $\hat{\bm{s}}_0\in\mathbb{R}^{3}$ on the unit celestial sphere. We consider a Cartesian coordinate system centred on the earth aligned with the pointing direction. At each instant of observation, each telescope pair measures a complex visibility defined as the correlation between incoming electric fields at the positions of the two telescopes. This visibility only depends on the relative position between the two telescopes, defined as a baseline. We consider a monochromatic signal $x$, and made up of incoherent sources. Also, we consider non-polarized radiation and a small FOV such that the signal on the celestial sphere is well approximated by its projection on to plane orthogonal to $\hat{\bm{s}}_0$. In this context, each visibility  corresponds to the measurement of the Fourier transform of a planar signal at the spatial frequency $\bm{u}=(u,v)$ where $(u,v)$ identifies the baseline components in the image plane, and in units of the observation wavelength. This result is known as the van Cittert-Zernike theorem~\citep{thompson04}. The measured visibility reads as:
\begin{equation}
y\left(\bm{u}\right) = \int A\left(\bm{l}\right)x\left(\bm{l}\right) {\rm e}^{-2{\rm i}\pi \bm{u} \bm{\cdot} \bm{l} }\: {\rm d}^2\bm{l}, \label{ri1}
\end{equation}
where $\bm{l}=(l,m)$ denotes the coordinates on the image plane and $A\left(\bm{l}\right)$ is the so-called primary beam, which limits the observed FOV. The total number of points $\bm{u}$ probed by all telescope pairs of the array during the observation provides some incomplete coverage in the Fourier plane characterizing the interferometer. 

\subsection{Inverse problem in matrix form}
\label{ssec:Inverse problem in matrix form}
In a practical setting we want to represent the map $x$ by a discretized image. The band-limited functions considered are completely identified by their Nyquist-Shannon sampling on a discrete uniform grid of $N=N^{1/2}\times N^{1/2}$ points $\bm{l}_{i}\in\mathbb{R}^{2}$ in real space with $1\leq i\leq N$ and by their corresponding discrete spatial frequencies $\bm{u}_{i}$. The sampled intensity signal and primary beam are denoted by the vectors $\bm{x},\bm{A}\in\mathbb{R}^{N}$ respectively.

As in \citet{wiaux09}, we assume that the spatial frequencies $\bm{u}$ probed by all telescope pairs during the observation belong to the discrete uniform grid of points $\bm{u}_{i}$, thus bypassing gridding considerations for the sake of simplicity. The Fourier coverage provided by the $M$ spatial frequencies probed can simply be identified by a binary mask in the Fourier plane equal to $1$ for each spatial frequency probed and $0$ otherwise. The visibilities measured may be denoted by a vector of $M$ complex Fourier coefficients $\bm{y}\in\mathbb{C}^{M}\equiv\{y_{b}\equiv y(\bm{u}_b)\}_{1\leq b\leq M}$, possibly affected by complex noise of astrophysical or instrumental origin, identified by the vector $\bm{n}\in\mathbb{C}^{M}$. Since the signal $\bm{x}$ is real-valued, we could only take measurements in half of the plane and infer the measurements of the other half through conjugate relations.

In this discrete setting, the Fourier coverage is in general incomplete in the sense that the number of real constraints $2M$ is smaller than the number of unknowns $N$; complete coverage of the Fourier plane corresponds to $M=N/2$. An ill-posed inverse problem is thus defined for the reconstruction of the signal $\bm{x}$ from the measured visibilities $\bm{y}$:
\begin{equation}
\bm{y}\equiv\mathsf{\Phi}\bm{x}+\bm{n}\textnormal{~with~}\mathsf{\Phi}\equiv\mathsf{MF}\mathsf{A},\label{ri4}
\end{equation}
where the matrix $\mathsf{\Phi}\in\mathbb{C}^{M\times N}$ identifies the complete linear relation between the signal and the visibilities. The matrix $\mathsf{A}\in\mathbb{R}^{N\times N}$ is the diagonal matrix implementing the primary beam. The unitary matrix $\mathsf{F}\in\mathbb{C}^{N\times N}$ implements the discrete Fourier transform providing the Fourier coefficients. The matrix $\mathsf{M}\in\mathbb{R}^{M\times N}$ is the rectangular binary matrix implementing the mask characterizing the interferometer. The inverse transform of the binary mask, i.e. $\mathsf{F}^{T}\mathsf{M}^{T}\bm{1}_{M}$ with $\bm{1}_{M}\in\mathbb{R}^{M}$ defining the vector of ones, identifies the dirty beam and the inverse transform of the Fourier measurements with all non-observed visibilities set to zero, i.e. $\mathsf{F}^{T}\mathsf{M}^{T}\bm{y}$, is the dirty image.

For signal reconstruction, a regularization scheme that encompasses enough prior information on the original signal is needed in order to find a unique solution. All image reconstruction algorithms will differ through the kind of regularization considered.

\subsection{State-of-the-art imaging algorithms}
\label{ssec:State of the art imaging algorithms}

The most standard and otherwise already very effective image reconstruction algorithm from visibility measurements is called CLEAN, which is a non-linear deconvolution method based on local iterative beam removal \citep{hogbom74,schwarz78,thompson04}. A sparsity prior on the original signal in real space is implicitly introduced. Multi-scale versions of CLEAN (MS-CLEAN) have also been developed \citep{cornwell08b}, where the sparsity is improved by multi-scale decomposition, hence enabling better recovery of the signal. The MS-CLEAN method was shown to perform better than the standard CLEAN, but still suffers from an empirical choice of basis profiles and scales. An adaptive scale pixel decomposition method called ASP-CLEAN was also introduced to improve on multi-scale CLEAN by relying on an adaptive choice of scales \citep{bhatnagar04}. Note that these approaches are known to be slow, sometimes prohibitively. Another approach to the reconstruction of images from visibility measurements is the Maximum Entropy Method  (MEM). In contrast to CLEAN, MEM solves a global optimization problem in which the inverse problem is regularized by the introduction of an entropic prior on the signal, but sparsity is not explicitly required \citep{ables74,gull78,cornwell85}. In practice, CLEAN is often preferred to MEM.

Reconstruction techniques based on convex optimization and sparse models have also been proposed. Approaches based on $\ell_1$ minimization coupled with the Dirac basis have been previously studied in the field~\citep{marsh87,wiaux09,wiaux09b,mcewen11a,li11}. The equivalence between CLEAN and $\ell_1$ minimization has been studied in \citep{marsh87}. \cite{wiaux09} and \cite{li11} report that $\ell_1$ minimization yields similar reconstruction quality to CLEAN, while including a positivity constraint in a convex formulation significantly enhances the reconstruction quality relative to CLEAN. 

Extended structures do not have an optimal sparse representation in Dirac basis. Thus, wavelet bases have also been considered in order to provide a sparser representation. Synthesis-based approaches with redundant representations have been proposed by \cite{wiaux09} and \cite{li11}. \cite{wiaux09} use a reweighted $\ell_1$ approach coupled with a steerable wavelet frame as sparsity dictionary. \cite{li11} use a least squares $\ell_1$ regularized problem with the isotropic undecimated wavelet transform (IUWT) as sparsity dictionary. The reconstruction quality of the IUWT method was reported to be superior than those of CLEAN and MS-CLEAN. 

Many signals in nature are also sparse or compressible in the magnitude of their gradient space, in which case the TV minimization problem has been shown to yield superior reconstruction results~\citep{rudin92}. The TV norm is defined by the $\ell_1$ norm of the magnitude of the gradient of the signal $\|\bar{\bm{x}}\|_{\rm{TV}} = \| \nabla \bar{\bm{x}} \|_1$, where $\nabla \bar{\bm{x}}$ denotes the gradient magnitude \citep{rudin92}. From this formulation, it can be seen that the TV problem might be modelled as an analysis $\ell_1$ minimization problem with the discrete gradient operator as the sparsity inducing transform. TV minimization was already proposed for radio-interferometric imaging by \cite{wiaux10a} and \cite{mcewen11a} showing promising results. Moreover, \cite{wiaux10a} used a reweighted TV minimization approach to recover the signal induced by cosmic strings in the cosmic microwave background.

\section{Sparsity Averaging Reweighted Analysis}
\label{sec:SARA}
In this section we propose a novel algorithm for radio-interferometric imaging based on the prior of average signal sparsity over multiple wavelet bases. We start by discussing our conjecture of average signal sparsity. Then, we propose the reweighted $\ell_1$ analysis method as a means to promote average sparsity. Finally, we present the resulting algorithm.

\subsection{Sparsity average conjecture}
As already discussed in the previous sections, while point and compact sources have a sparse representation in the Dirac basis, piecewise smooth structures exhibit gradient sparsity, and continuous extended structures are better encapsulated in wavelet bases. Astronomical images are often complex and all these types of structures can be present at once. Therefore, we here conjecture that promoting average sparsity or compressibility over multiple bases rather single bases represents an extremely powerful prior. Note on a theoretical level that a single signal cannot be arbitrarily sparse simultaneously in any pair of bases, due to the incoherence between these bases (see \cite{wiaux09} and references therein for a definition of incoherence). For illustration, a signal extremely sparse in the Dirac basis is completely spread in the Fourier basis. We hypothesize that,  for any pair of bases,  there might exist a lower bound on the average sparsity of a signal, which identifies a generalized ``uncertainty relation''. In essence, our prior consists of assuming that the signals of interest are those that saturate this uncertainty relation between multiple pairs of bases.

We propose using a dictionary composed of a concatenation of orthonormal bases, i.e.
\begin{equation}
\mathsf{\Psi}=\frac{1}{\sqrt{q}}[\mathsf{\Psi}_1, \mathsf{\Psi}_2, \ldots, \mathsf{\Psi}_q],
\end{equation}
thus $\mathsf{\Psi}\in\mathbb{R}^{N\times D}$ with $D=qN$. Given the previous considerations on astronomical images, the dictionary should be composed of the Dirac basis and wavelet bases. In particular, the Haar wavelet basis can be used as an alternative to gradient sparsity (usually imposed by a TV prior) to promote piecewise smooth signals\footnote{In fact, the compressibility of Haar wavelet coefficients is controlled by the image TV norm~\citep{devore92}.}. See Section~\ref{ssec:saraimp} for further details on a practical selection of these bases.

\subsection{Reweighted $\ell_1$ analysis problem}
\label{ssec:rwl1}
In the light of our previous discussions on the versatility of convex optimization, we promote this average sparsity through a reweighted $\ell_1$ analysis method. Let us define the weighted $\ell_1$ problem 
\begin{align}\label{delta}
\min_{\bar{\bm{x}}\in\mathbb{R}^{N}}&\|\mathsf{W\Psi^T}\bar{\bm{x}}\|_{1}\\ \nonumber
\textnormal{ subject to }&\| \bm{y}-\mathsf{\Phi}\bar{\bm{x}}\|_{2}\leq\epsilon\\ \nonumber
\textnormal{and }&\bar{\bm{x}}\geq 0,
\end{align}
where $\mathsf{W}\in\mathbb{R}^{D\times D}$ is a diagonal matrix with positive weights and the constraint $\bar{\bm{x}}\geq 0$ represents the positivity prior on $\bm{x}$. Assuming i.i.d.~complex Gaussian noise with variance $\sigma_n$, the $\ell_{2}$ norm term in \eqref{delta} is identical to a bound on the $\chi^{2}$ distribution with $2M$ degrees of freedom governing the noise level estimator. Therefore, we set this bound as
 $\epsilon^2=(2M+4\sqrt{M})\sigma_n^2/2$, where $\sigma^2_n/2$ is the variance of both the real and imaginary part of the noise. This choice provides a likely bound for $\|\bm{n}\|_2$, since the probability that $\|\bm{n}\|_2^2$ exceeds $\epsilon^2$ is the probability that a $\chi^{2}$ with $2M$ degrees of freedom exceeds its mean, $2M$, by at least two times the standard deviation $2\sqrt{M}$, which is very small. The solution to \eqref{delta} is denoted as $\Delta(\bm{y}, \mathsf{\Phi},\mathsf{W},\epsilon)$, which is a function of the data vector $\bm{y}$, the measurement and weight matrices $\mathsf{\Phi}$ and $\mathsf{W}$, and the bound $\epsilon$ on the noise level estimator.

Recall that in the reweighting approach a sequence of weighted $\ell_1$ problems is solved, each using as weights essentially the inverse of the values of the solution of the previous problem. In practice, we update the weights at each iteration, i.e. after solving a complete weighted $\ell_1$ problem, by the function
\begin{equation}
f(\gamma,x)\equiv \frac{\gamma}{\gamma+|x|},
\end{equation}
where $\gamma$ plays the role of a stabilization parameter (ideally zero). Note that as $\gamma\rightarrow 0$ the weighted $\ell_1$ norm approaches the $\ell_0$ norm. To approximate the $\ell_0$ norm by the reweighted $\ell_1$ algorithm, we use a homotopy strategy~\citep{nocedal06} and solve a sequence of weighted $\ell_1$ problems using a decreasing sequence $\{\gamma^{(t)}\}$, with $t$ denoting the iteration time variable. Under this scheme, a weighted $\ell_1$ problem is first solved and its solution is used as the warm start initialization for the next problem that is geometrically closer to the $\ell_0$  problem. This process is then repeated until the solution becomes stationary~\citep{nocedal06}. 

\subsection{The SARA algorithm}
The resulting algorithm, dubbed sparsity averaging reweighted analysis (SARA), is defined in Algorithm~\ref{alg3}. A rate parameter $\beta$, with $0<\beta<1$, controls the decrease of the sequence $\gamma^{(t)}=\beta\gamma^{(t-1)}=\beta^{t}\gamma_0$ such that $\gamma^{(t)}\rightarrow 0$ as $t\rightarrow\infty$. Ideally, $\gamma^{(t)}$ should decrease to zero, but since we have noise, we set a lower bound as $\gamma^{(t)}\geq\sigma_c$. The standard deviation of the noise in the representation domain is computed as $\sigma_c=\sqrt{M/qN}\sigma_n$, which gives a rough estimate for a baseline above which significant signal components could be identified. As a starting point we set $\hat{\bm{x}}^{(0)}$ as the solution of the $\ell_1$ problem and $\gamma^{(0)}=\sigma_s\left(\mathsf{\Psi}^T\hat{\bm{x}}^{(0)}\right)$, where $\sigma_s(\cdot)$ stands for the empirical standard deviation of the signal, fixing the signal scale. The reweighting process ideally stops  when the relative variation between successive solutions $\|\hat{\bm{x}}^{(t)}-\hat{\bm{x}}^{(t-1)}\|_2/\|\hat{\bm{x}}^{(t-1)}\|_2$ is smaller than some bound $\eta$, with $0<\eta<1$, or after the maximum number of iterations allowed, $N_{\rm{max}}$, is reached. In our implementation, which will be detailed in Section~\ref{ssec:saraimp}, we fix $\eta=10^{-3}$ and $\beta=10^{-1}$. 

\begin{algorithm}[h!]
\caption{SARA algorithm for RI imaging}\label{alg3}
\begin{algorithmic}[1]
\REQUIRE $\bm{y}$, $\mathsf{\Phi}$, $\epsilon$, $\sigma_c$, $\beta$, $\eta$ and $N_{\rm{max}}$.
\ENSURE Reconstructed image $\hat{\bm{x}}$.
\STATE Initialize $t=1$, $\mathsf{W}^{(0)}=\mathsf{I}$ and $\rho=1$.
\STATE Compute\\
$\hat{\bm{x}}^{(0)}=\Delta(\bm{y}, \mathsf{\Phi},\mathsf{W}^{(0)},\epsilon)$,\\
$\gamma^{(0)}=\sigma_s\left(\mathsf{\Psi}^T\hat{\bm{x}}^{(0)}\right)$.
\WHILE{$\rho>\eta$ and $t<N_{\rm{max}}$}
\STATE Update the weight matrix\\
$\mathsf{W}_{ij}^{(t)}=f\left (\gamma^{(t-1)},\hat{\alpha}_{i}^{(t-1)}\right)\delta_{ij}$, for $i,j=1,\ldots,D$\\
with $\hat{\bm{\alpha}}^{(t-1)}=\mathsf{\Psi}^T\hat{\bm{x}}^{(t-1)}$
\STATE Compute a solution\\
$\hat{\bm{x}}^{(t)}=\Delta(\bm{y}, \mathsf{\Phi},\mathsf{W}^{(t)},\epsilon)$.\\
\STATE Update $\gamma^{(t)}=\max(\beta\gamma^{(t-1)},\sigma_c)$.
\STATE Update $\rho=\| \hat{\bm{x}}^{(t)}-\hat{\bm{x}}^{(t-1)}\|_2/\|\hat{\bm{x}}^{(t-1)}\|_2$
\STATE $t\leftarrow t+1$
\ENDWHILE
\RETURN $\hat{\bm{x}}$
\end{algorithmic}
\end{algorithm} 
\vspace{-0.5cm}
\section{Simulations and results}
\label{sec:Simulations and results}
In this section we evaluate the performance of the SARA algorithm through numerical simulations. Firstly, we describe the practical implementation of SARA and state-of-the-art algorithms used as benchmarks. Secondly, we describe the simulation set up in the context of the inverse problem associated with \eqref{ri4}. Thirdly, we report the results of the comparison of SARA to the state-of-the-art. Finally, we present an illustrative example of the performance of SARA in the presence of the spread spectrum phenomenon.

\subsection{SARA implementation and benchmarck algorithms}
\label{ssec:saraimp}
For all the experiments we consider a concatenation of nine bases ($q=9$), thus $\mathsf{\Psi}\in\mathbb{R}^{N\times D}$ with $D=9N$, as the dictionary for SARA. The first basis is the Dirac basis. The eight remaining bases are the first eight Daubechies wavelets, Db1-Db8~\citep{daubechies92}. The first Daubechies wavelet basis, Db1, is the Haar wavelet basis. We use a fourth order decomposition depth for all wavelet bases\footnote{Experimental results have shown that the performance of SARA degrades if one of the bases is withdrawn from the dictionary.~We do not present these results in detail here for the sake of space.}. 

We compare SARA to state-of-the-art algorithms for $\ell_1$ and TV minimization problems, as well as their reweighted versions, in terms of reconstruction quality and computation time. Firstly, the reweighted $\ell_1$ problems considered are defined through the reweighting procedure described in Section \ref{ssec:rwl1} based on \eqref{delta}, with specific choices of the sparsity dictionary $\mathsf{\Psi}$. We consider three different options for $\mathsf{\Psi}$: the Dirac basis, the Daubechies 8 wavelet basis and the isotropic undecimated wavelet redundant dictionary. The associated methods are respectively denoted R-BP, R-BPDb8 and R-BPIU. The (non-reweighted) $\ell_1$ problems are defined  through \eqref{delta} with $\mathsf{W}=\mathsf{I}$ and again different choices of the sparsity dictionary $\mathsf{\Psi}$. We here consider four different options for $\mathsf{\Psi}$: the Dirac basis, the Daubechies 8 wavelet basis, the isotropic undecimated wavelet dictionary and the concatenation of 9 bases described above for SARA. The associated methods are respectively denoted BP, BPDb8, BPIU and BPSA.

Secondly, the TV minimization problem is formulated as:
\begin{align}\label{alg2}
\min_{\bar{\bm{x}}\in\mathbb{R}^{N}}&\|\bar{\bm{x}}\|_{\rm{TV}}\\ \nonumber
\textnormal{ subject to }&\| \bm{y}-\mathsf{\Phi}\bar{\bm{x}}\|_{2}\leq\epsilon\\ \nonumber
\textnormal{and }&\bar{\bm{x}}\geq 0.
\end{align}
We have also implemented a reweighted version of TV (still through the procedure defined in Section \ref{ssec:rwl1}), denoted as R-TV. Finally, we also use as benchmark the synthesis-based IUWT method of \cite{li11} and we denote it as IUWT. In the light of the discussion in Section \ref{ssec:State of the art imaging algorithms} we assume that the reconstruction quality provided by BP is essentially equivalent to that of the standard CLEAN algorithm, and that the reconstruction quality provided by IUWT is an upper bound on the reconstruction quality of any multi-scale implementation of CLEAN.

To solve \eqref{delta} and \eqref{alg2}, we use the Douglas-Rachford splitting algorithm, which is tailored to solve problems of the form \eqref{cvx1} for $n=2$ and with the additional property of not requiring differentiability of any of the functions~\citep{combettes07}. 

\subsection{Simulations}
\begin{figure*}

\centering
    \begin{tabular}{ccc}
   
    \includegraphics[trim = 2cm 1cm 1cm 1cm, clip, keepaspectratio, width = 5.8cm]{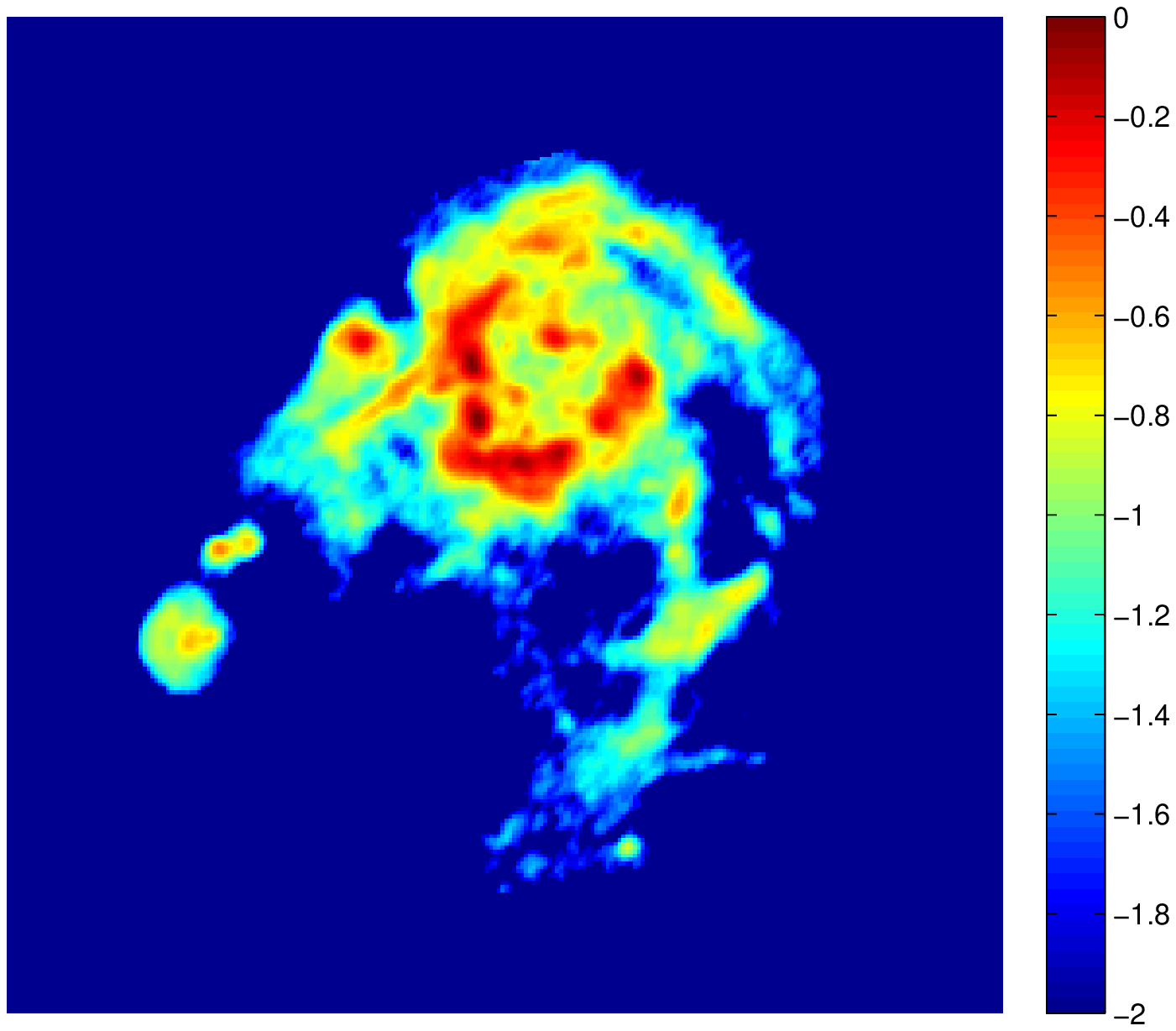}&
    \includegraphics[trim = 2cm 1cm 1cm 1cm, clip, keepaspectratio, width = 5.8cm]{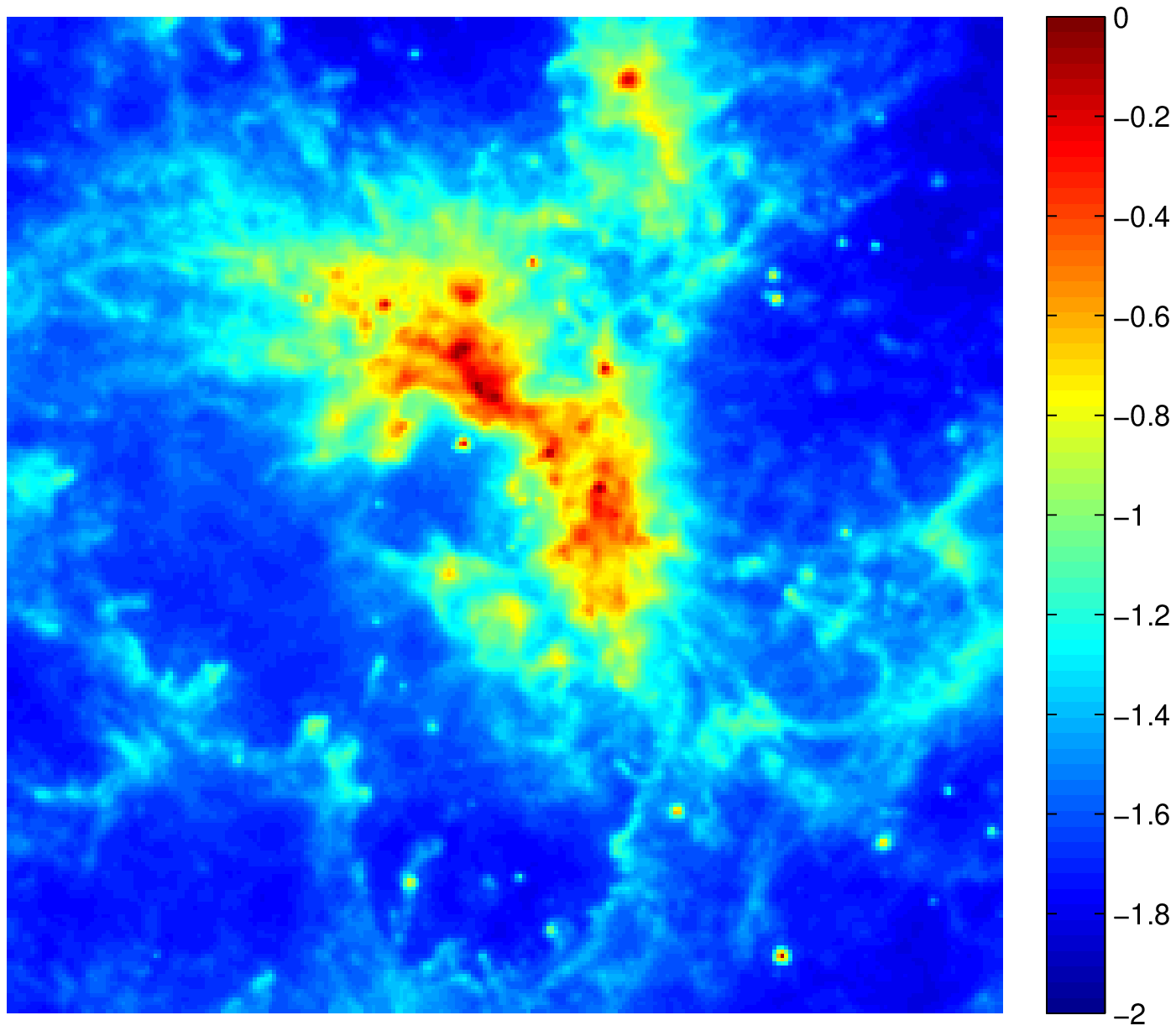}\\
    
    \end{tabular}

\caption{(color online). Test images with brightness values in the interval $[0.01, 1]$ shown in a $\log_{10}$ scale. Left: M31. Right: 30Dor.}
\label{fig:1}
\end{figure*}

\begin{figure*}

\centering
    \begin{tabular}{ccc}
    \includegraphics[trim = 2cm 1cm 1cm 1cm, clip, keepaspectratio, width = 5.8cm]{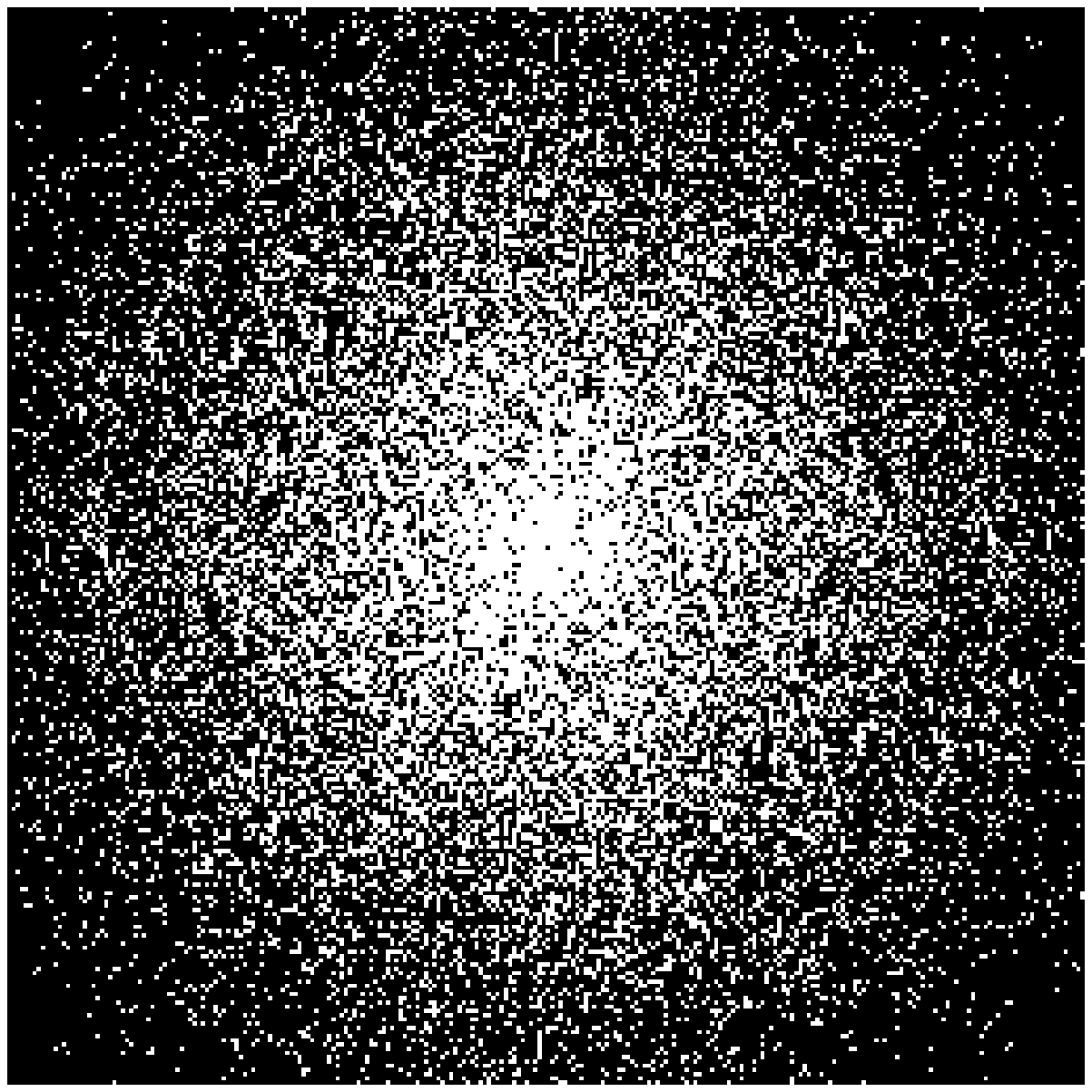}&
     \includegraphics[trim = 2cm 1cm 1cm 1cm, clip, keepaspectratio, width = 5.8cm]{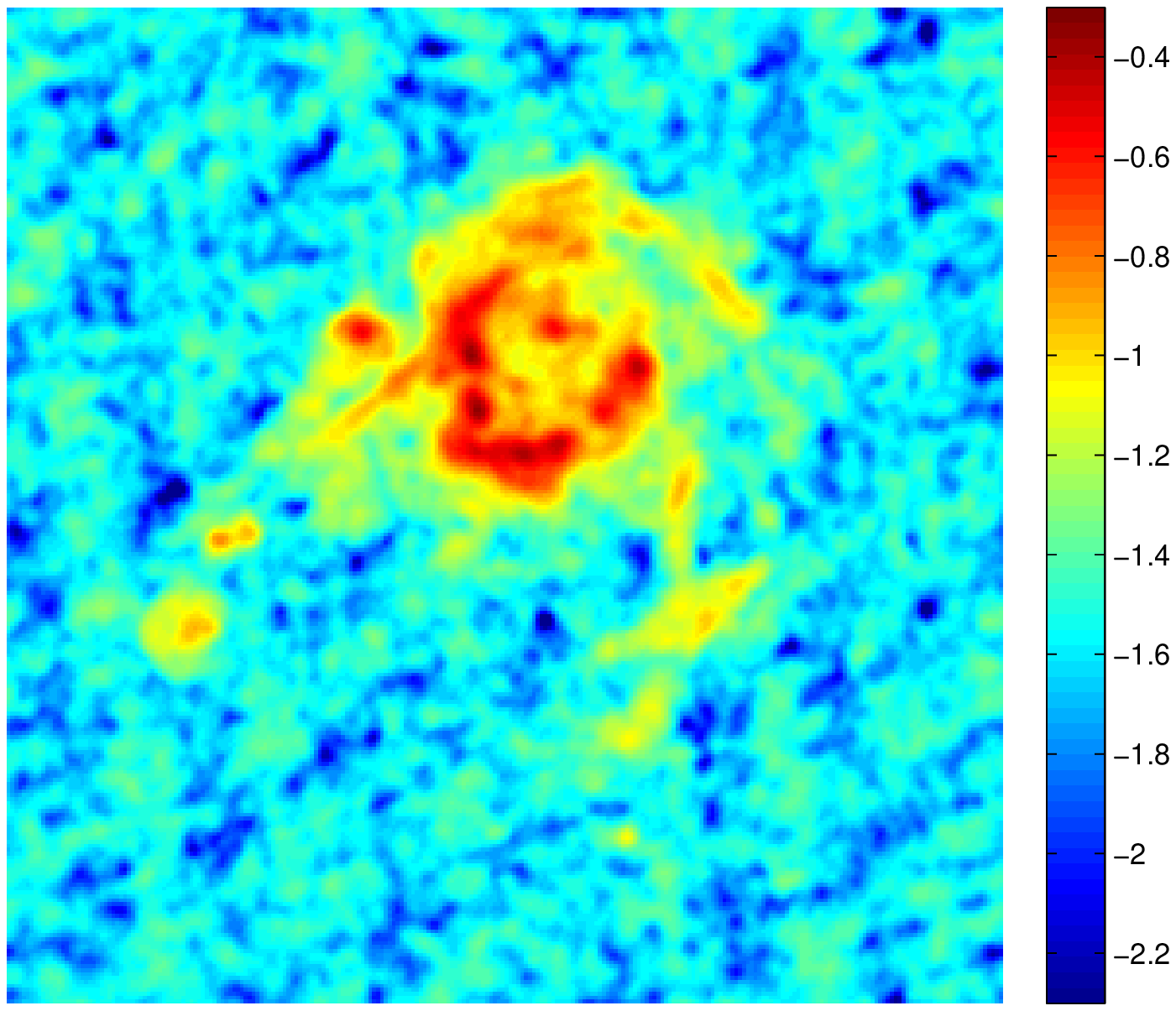}&
    \includegraphics[trim = 2cm 1cm 1cm 1cm, clip, keepaspectratio, width = 5.8cm]{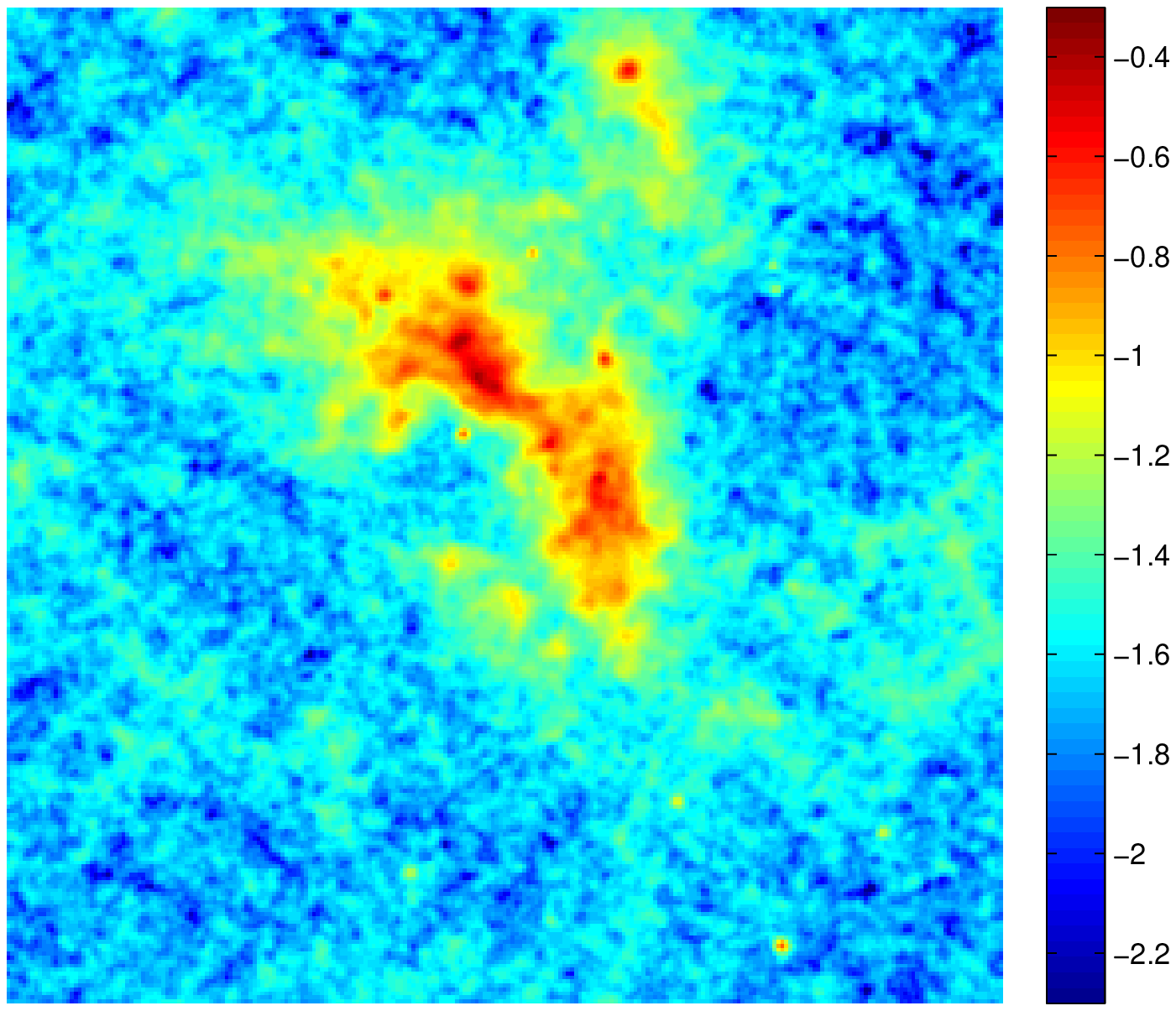}\\
    \end{tabular}

\caption{(color online). Example of a simulated visibility coverage. Left panel: Variable density sampling pattern for 30\% of coverage of the Fourier plane. The mask is symmetrized for visualization purposes: each pair of symmetric points represents one measured visibility. The real part of the corresponding dirty images for M31 and 30Dor are shown in the middle and right panels, respectively, in a $\log_{10}$ scale.}
\label{fig:2}
\end{figure*}

We evaluate the reconstruction performance of SARA by recovering well-known test images from simulated incomplete visibilities following the model in \eqref{ri4} with $\mathsf{A}=\mathsf{I}$. The test images used in all simulations are based on a HII region in M31 and the 30 Doradus (30Dor) in the Large Magellanic Cloud. We use discrete models of size 256$\times$256 as ground truth images\footnote{Available at \url{http://casaguides.nrao.edu/index.php}.}. The test images with brightness values in the interval $[0.01, 1]$ are shown in Figure~\ref{fig:1} in a $\log_{10}$ scale.  We consider incomplete visibility coverages generated by random variable density sampling profiles. Such profiles are characterized by denser sampling at low spatial frequencies than at high frequencies. This choice allows one to take into account the fact that most of the signal energy is usually concentrated around low frequencies, also mimicking common generic sampling patterns in radio interferometry (see \cite{puy11} for the exact shape of the density profile). In order to make the simulated coverages more realistic we suppress the $(0,0)$ component of the Fourier plane from the measured visibilities. This generic profile approach allows us to evaluate the reconstruction quality for arbitrary percentages of visibility coverage and without concern for various telescope configurations. Let us recall that, accounting for image reality, we only take measurements in the half Fourier plane. A complete coverage of the half plane is referred to as a 100\% coverage. The left panel in Figure~\ref{fig:2} shows an example of a sampling pattern for 30\% of coverage of the Fourier plane (symmetrized mask for visualization purposes). The middle and right panels in Figure~\ref{fig:2} show the real part of the dirty images generated by this sampling pattern for M31 and 30Dor respectively. We evaluate numerically the reconstruction quality and computational speed of the algorithms considered for coverages between 10\% and 90\%.

We use as reconstruction quality metric the signal to noise ratio (SNR), which is defined as: 
\begin{equation}\label{snrdef}
\rm{SNR}=20\log_{10}\left( \frac{\sigma_{\bm{x}}}{\sigma_{\bm{x}-\hat{\bm{x}}}}\right), 
\end{equation}
where $\sigma_{\bm{x}}$ and $\sigma_{\bm{x}-\hat{\bm{x}}}$ denote the standard deviation of the original image and the standard deviation of the error image respectively. The visibilities are corrupted by complex Gaussian noise with a fixed input SNR of 30~dB, with the input SNR defined as in \eqref{snrdef} with $\sigma_{\bm{x}-\hat{\bm{x}}}$ substituted by the standard deviation of the noise on each visibility.

\subsection{Results}
\label{ssec:results}

\begin{figure*}

\centering
    \begin{tabular}{cc}
   
    \includegraphics[trim = 1.3cm 0.6cm 0.7cm 0.8cm, clip, keepaspectratio, width = 8.7cm]{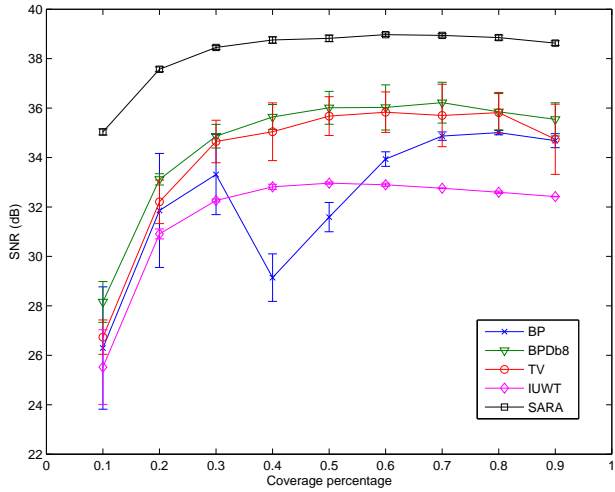}&
    \includegraphics[trim = 1.3cm 0.6cm 0.7cm 0.8cm, clip, keepaspectratio, width = 8.7cm]{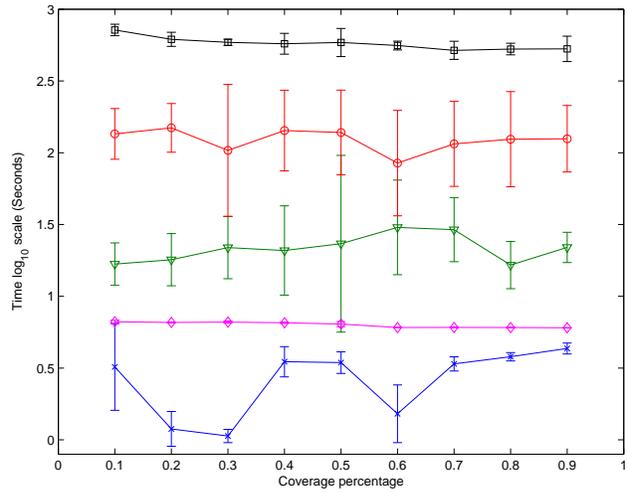}\\
    
    \end{tabular}

\caption{(color online). Reconstruction results for M31. Left: Average reconstruction SNR against percentage of coverage. Right: Average computation time. Vertical bars identify one standard deviation errors around the mean over 100 simulations. The input SNR is set to 30~dB.}
\label{fig:3}
\end{figure*}

\begin{figure*}

\centering
    \begin{tabular}{cc}
   
    \includegraphics[trim = 1.3cm 0.6cm 0.7cm 0.8cm, clip, keepaspectratio, width = 8.7cm]{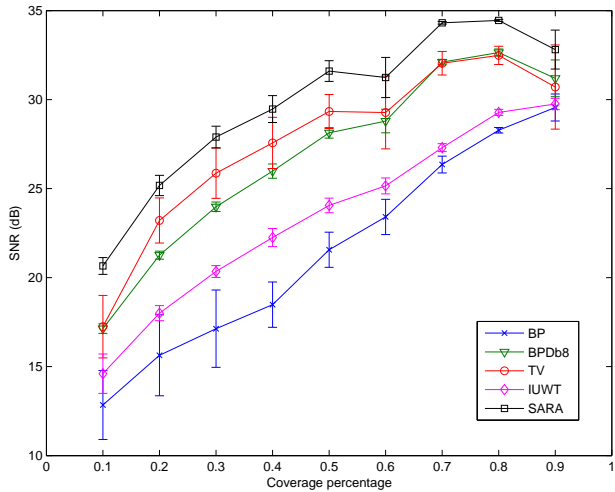}&
    \includegraphics[trim = 1.3cm 0.6cm 0.7cm 0.8cm, clip, keepaspectratio, width = 8.7cm]{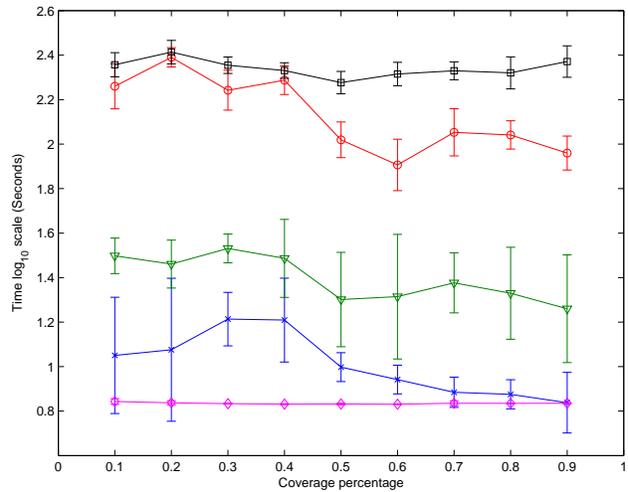}\\
    
    \end{tabular}

\caption{(color online). Reconstruction results for 30Dor. Left: Average reconstruction SNR against percentage of coverage. Right: Average computation time. Vertical bars identify one standard deviation errors around the mean over 100 simulations. The input SNR is set to 30~dB.}
\label{fig:4}
\end{figure*}

The left panels in Figure~\ref{fig:3} and Figure~\ref{fig:4} show, for M31 and 30Dor respectively, the SNR results against percentage of coverage for BP, BPDb8, TV, IUWT and SARA. Average values over 100 simulations and associated one standard deviation error bars are reported. The results demonstrate that SARA outperforms state-of-the-art methods for all coverages. Moreover, the results for M31 show considerable enhancement provided by SARA, with a gain of more than 6~dB for 10\% of coverage and at least 3~dB for the rest of the coverages relative to other methods. The results for 30Dor, which is a more complicated image with both extended structures and compact structures, show a SNR improvement of SARA of at least 2~dB over all other methods. These results confirm the conjecture that average sparsity over multiple orthonormal bases represents a stronger prior than sparsity over a single representation.

It was found that the reweighting process never enhances the results for the benchmark algorithms significantly. R-BP provides worse results than BP for both test images achieving a SNR at least 3 dB below BP. R-TV does not show any improvement over TV for 30Dor. For M31, R-TV reconstructions exhibit a gain of at most 1 dB for coverages above 70\%, and lower SNRs than TV for coverages below 70\%. The reconstruction quality of R-BPDb8 is worse than that obtained by BPDb8 for 30Dor. For M31, R-BPDb8 provides a SNR improvement of at most 1 dB over BPDb8 for coverages above 50\% and lower SNRs than BPDb8 for coverages below 50\%. The analysis-based BPIU and R-BPIU did not show any improvement with respect to IUWT for 30Dor. For M31, BPIU and R-BPIU provide a gain of at most 1 dB and 3 dB, respectively, compared to IUWT. Also, the reconstruction quality of BPSA is always worse than that achieved by SARA, being at least 3 dB below. Therefore, results for R-BP, R-BPDb8, R-TV, R-BPIU, BPIU and BPSA are not shown in Figure~\ref{fig:3} and Figure~\ref{fig:4}.

Computation times (on a 2.4 GHz Xeon quad core) are reported in the right panels of Figure~\ref{fig:3} and Figure~\ref{fig:4}, for M31 and 30Dor respectively, in a $\log_{10}$ scale, for the same algorithms as those considered in the left panels. Again, average values over 100 simulations and associated one standard deviation error bars are reported. Even though the concatenation of multiple bases and the reweighting process render the algorithm structure more costly, we see that the computation times are of the order of minutes for SARA, and very similar to those required for TV minimization and those reported for MS-CLEAN in the literature~\citep{cornwell08b,li11}. Note that all the preliminary implementations for these experiments are made in MATLAB. Therefore, significantly faster implementations can be achieved using a lower level programming language with custom optimized code. Also, the versatile framework of convex optimization offers a lot of room for improvement in terms of computational speed and efficiency. The Douglas-Rachford algorithm provides nice properties but, as emphasised in Section~\ref{sec:Convex optimization for sparse reconstruction}, other proximal splitting methods exist that offer a parallel implementation structure, such as the proximal parallel algorithm and the simultaneous-direction method of multipliers, where all the proximity operators can be computed in parallel rather than sequentially \citep{combettes11}. Furthermore, the simultaneous-direction method of multipliers offers a distributed implementation structure. 

\begin{figure*}

\centering
    \begin{tabular}{ccc}
   
    \includegraphics[trim = 2cm 1.3cm 1cm 0.5cm, clip, keepaspectratio, width = 5.8cm]{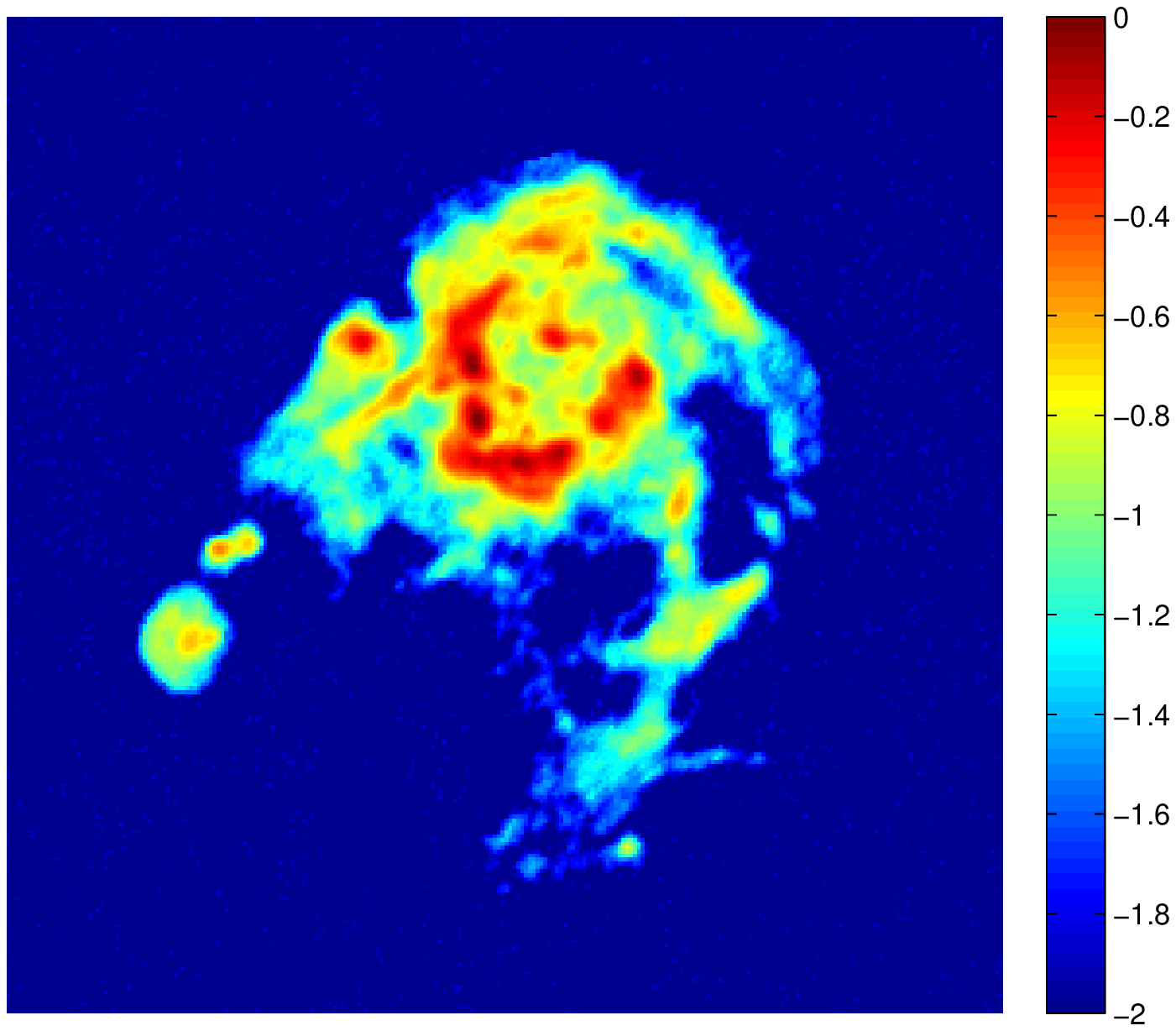}&
    \includegraphics[trim = 2cm 1.3cm 1cm 0.5cm, clip, keepaspectratio, width = 5.8cm]{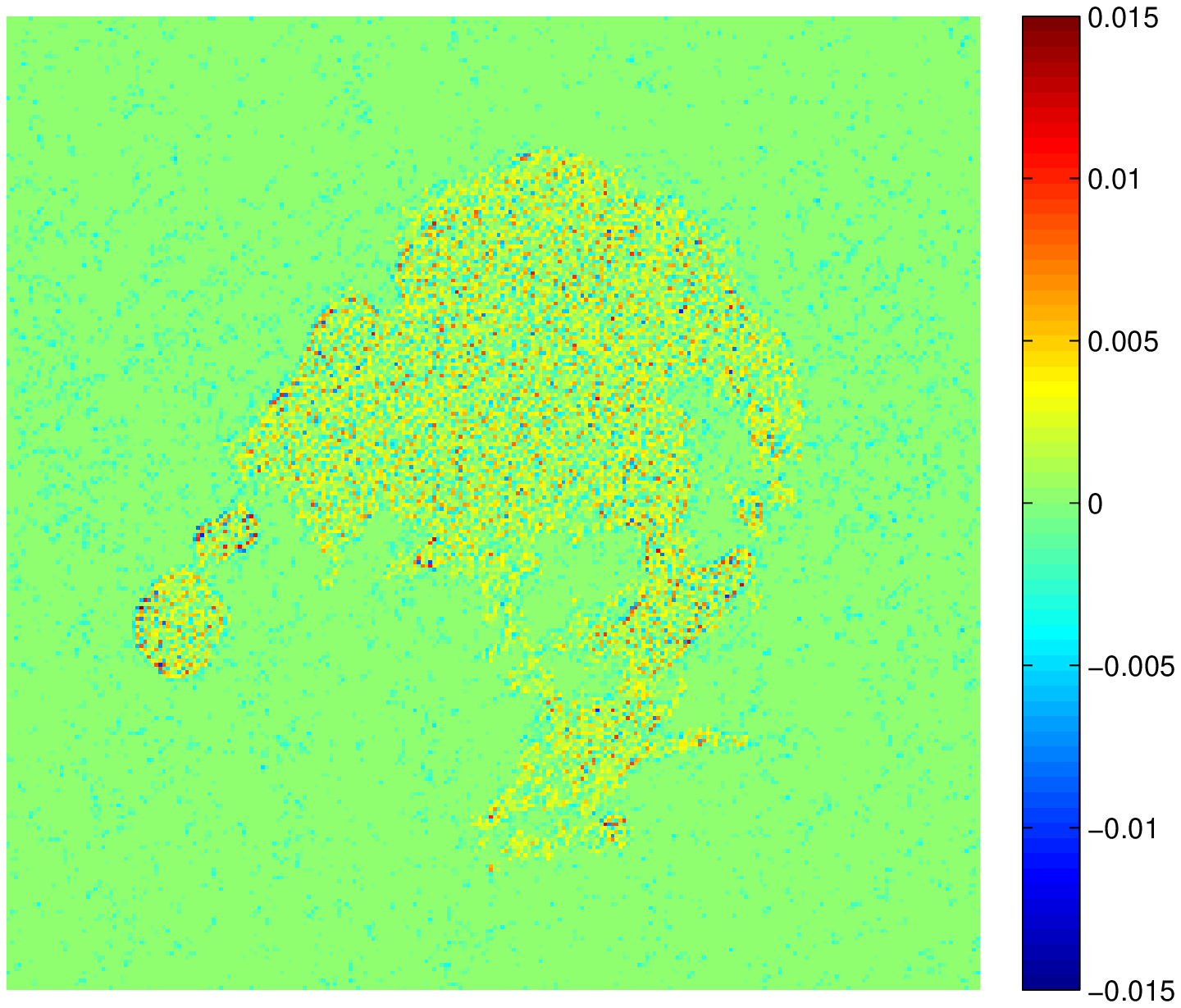}&
    \includegraphics[trim = 2cm 1.3cm 1cm 0.5cm, clip, keepaspectratio, width = 5.8cm]{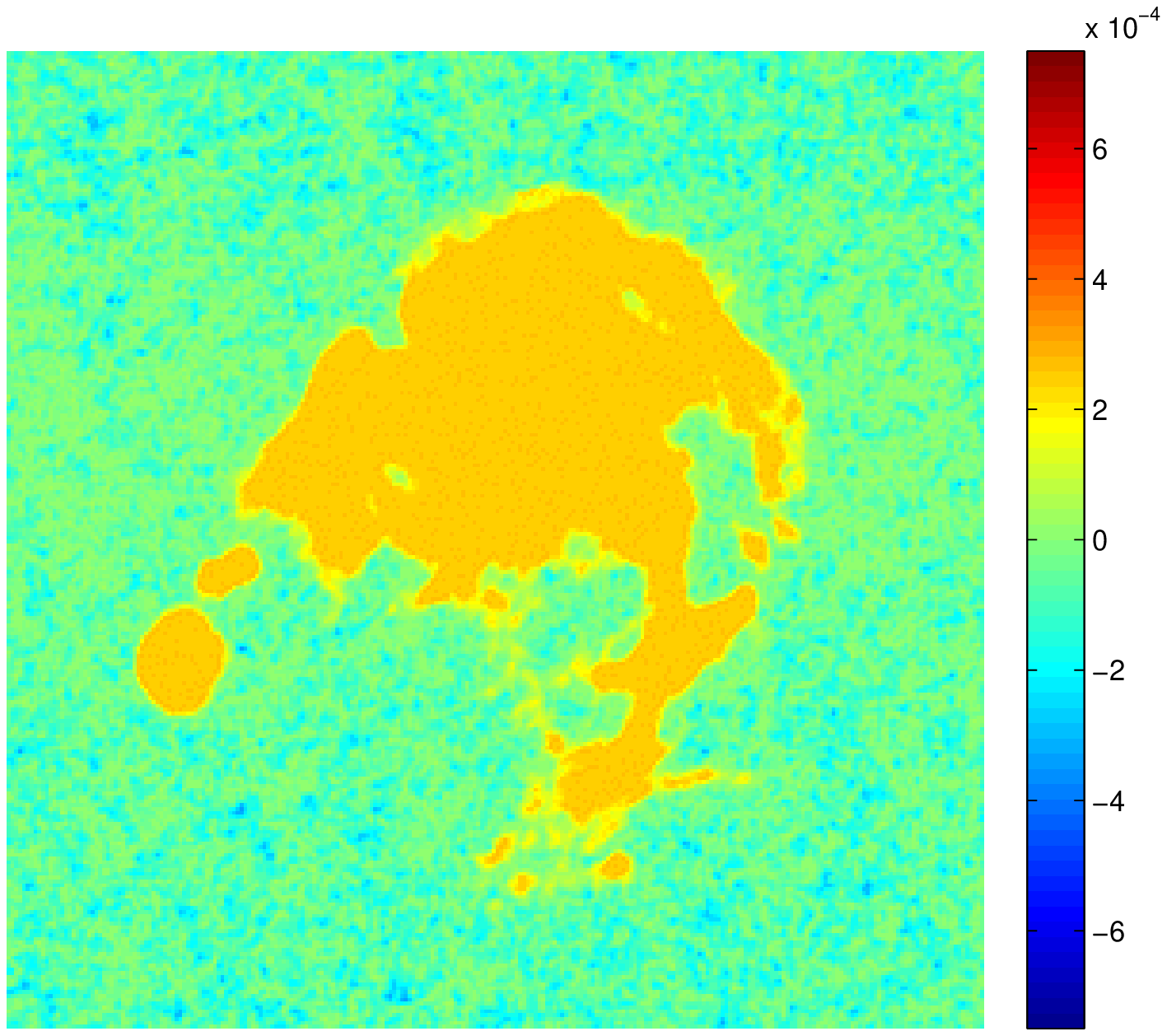}\\
    
    \includegraphics[trim = 2cm 1.3cm 1cm 0.5cm, clip, keepaspectratio, width = 5.8cm]{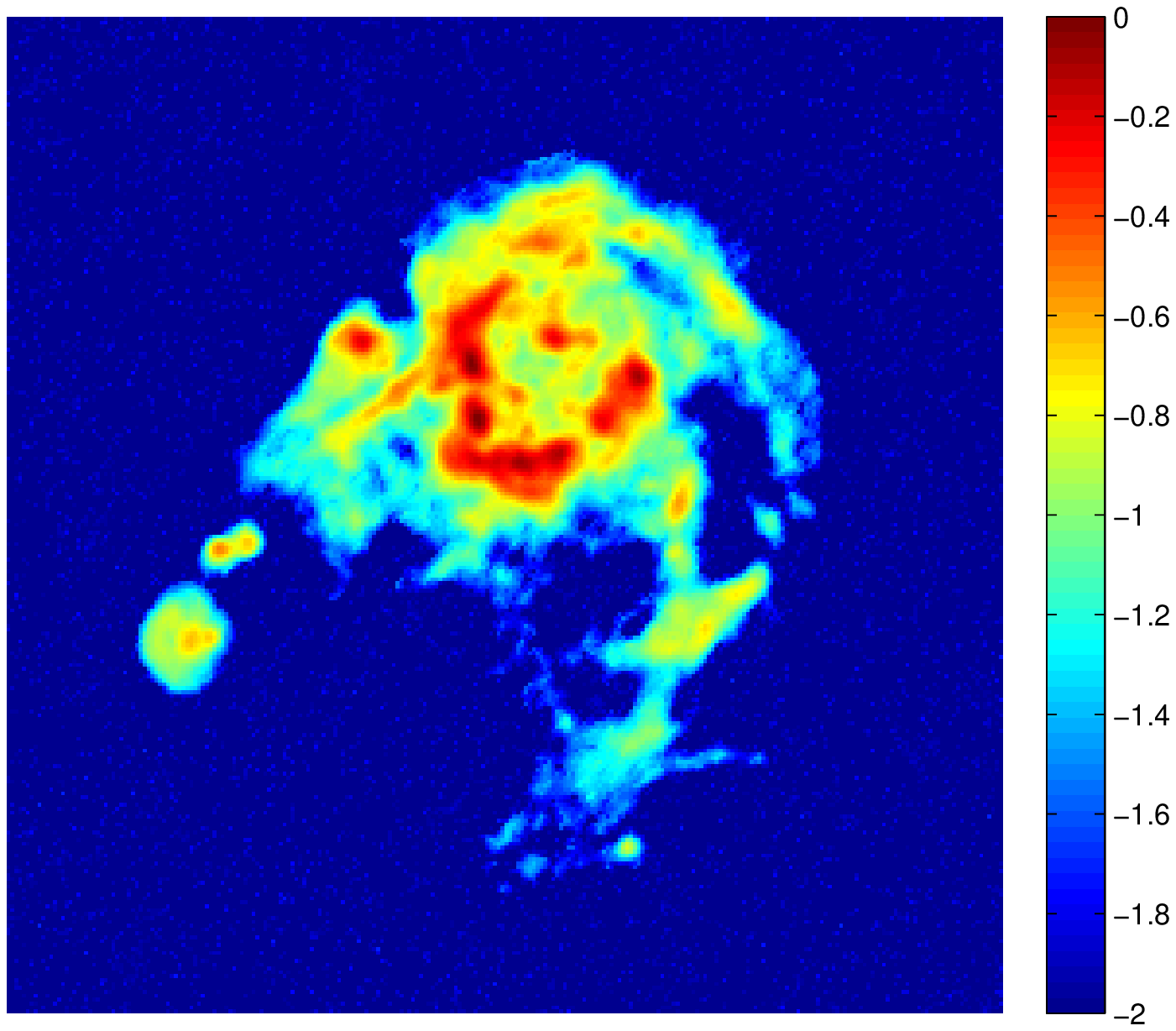}&
    \includegraphics[trim = 2cm 1.3cm 1cm 0.5cm, clip, keepaspectratio, width = 5.8cm]{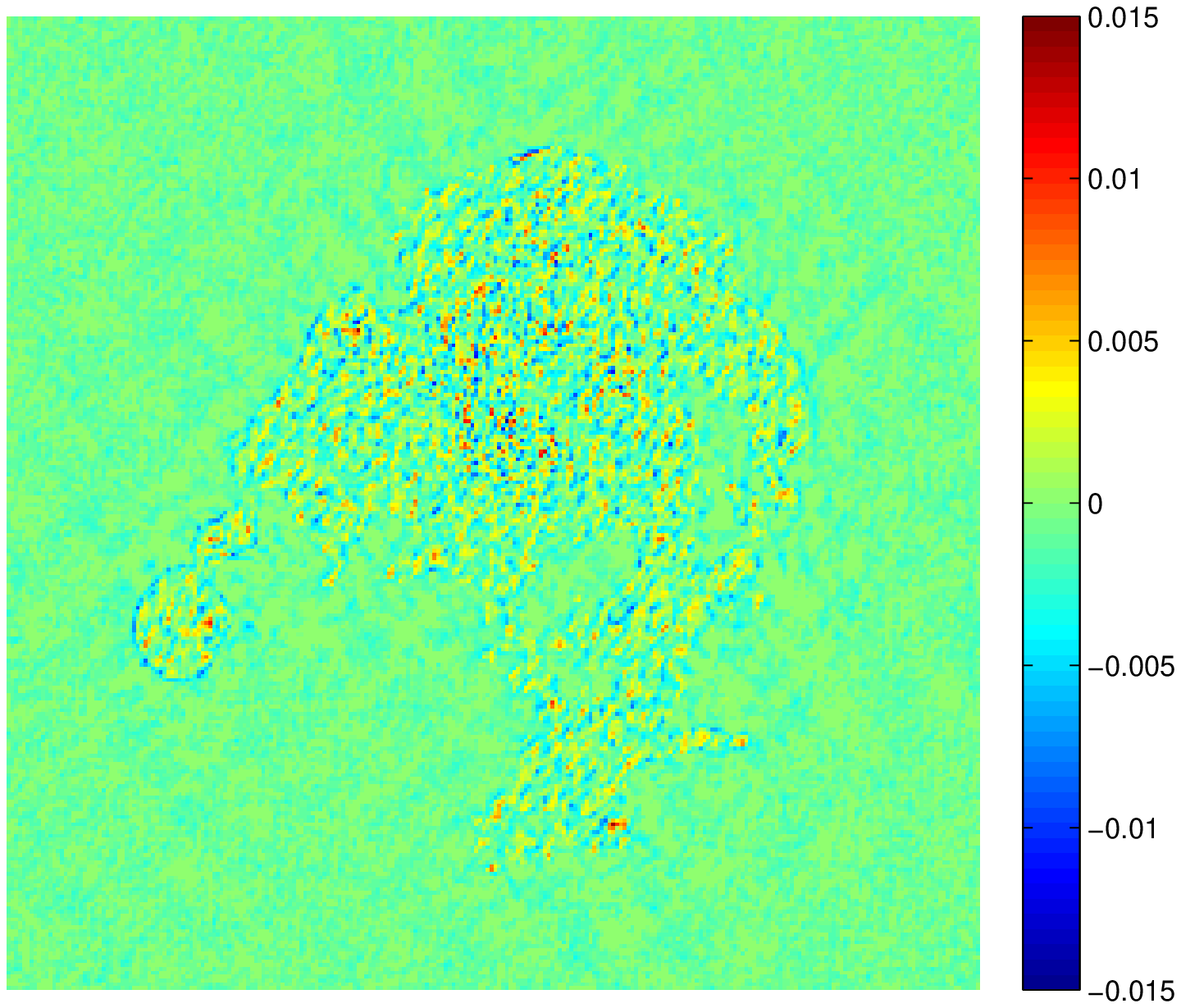}&
    \includegraphics[trim = 2cm 1.3cm 1cm 0.5cm, clip, keepaspectratio, width = 5.8cm]{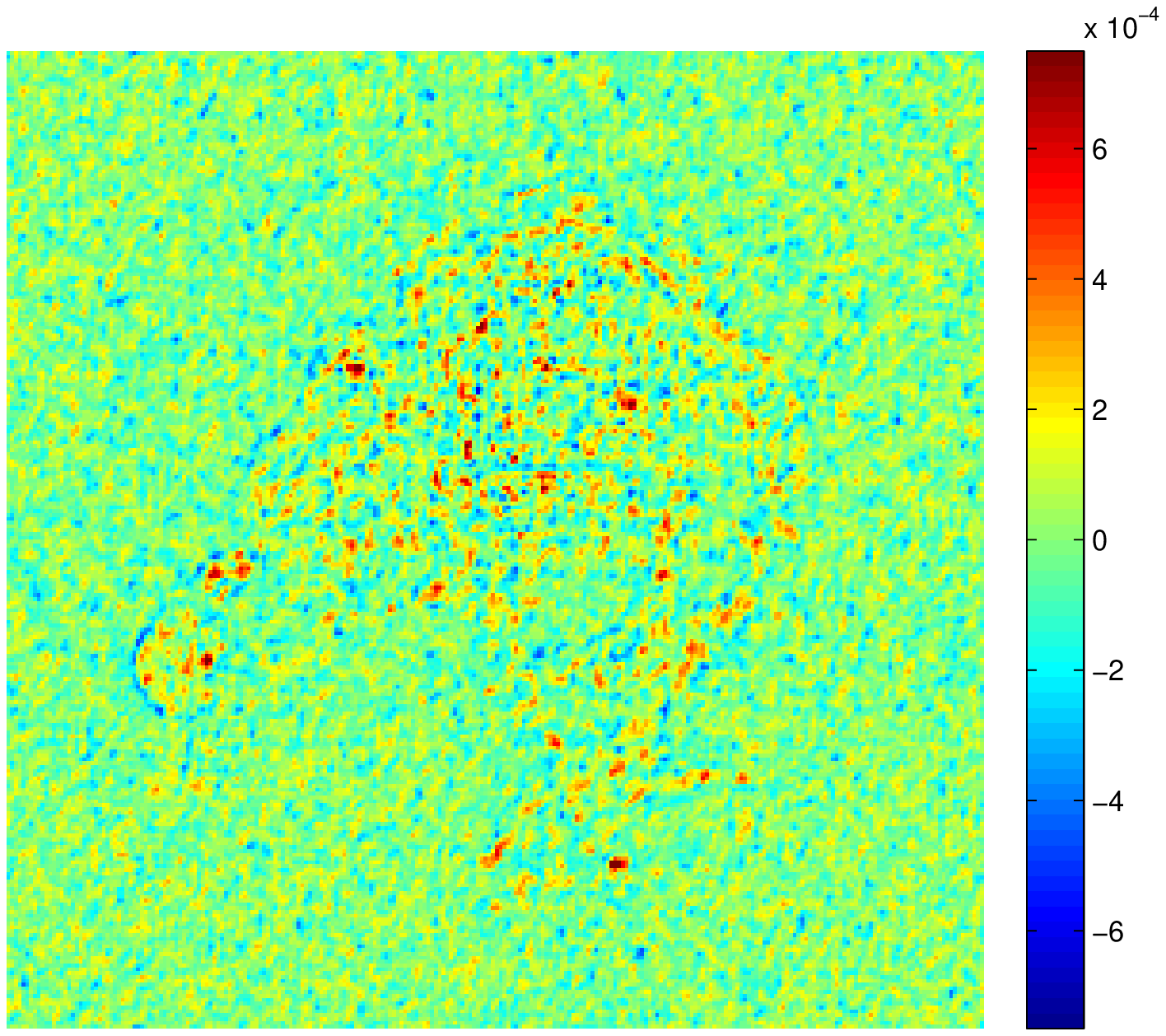}\\
    
    \includegraphics[trim = 2cm 1.3cm 1cm 0.5cm, clip, keepaspectratio, width = 5.8cm]{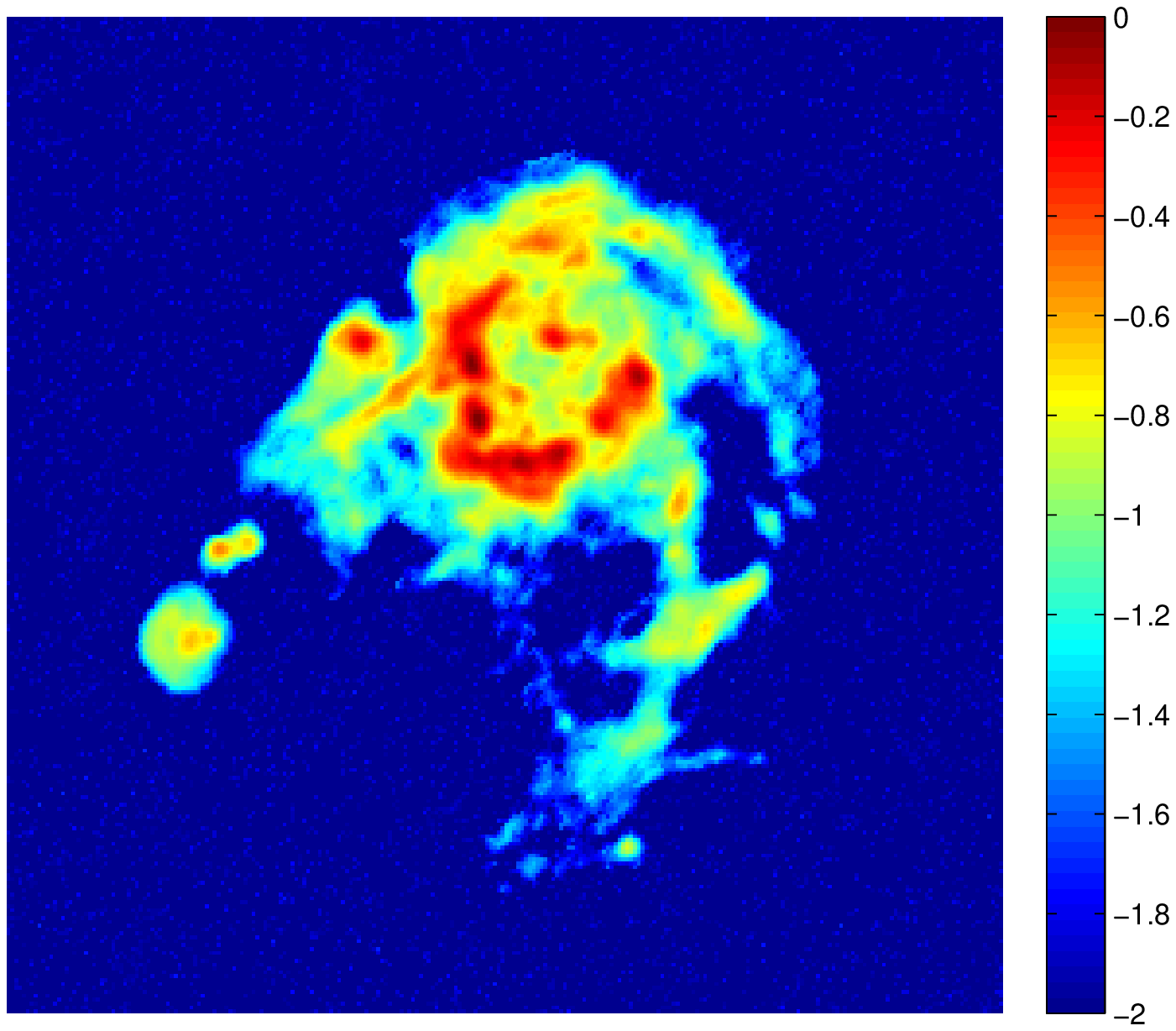}&
    \includegraphics[trim = 2cm 1.3cm 1cm 0.5cm, clip, keepaspectratio, width = 5.8cm]{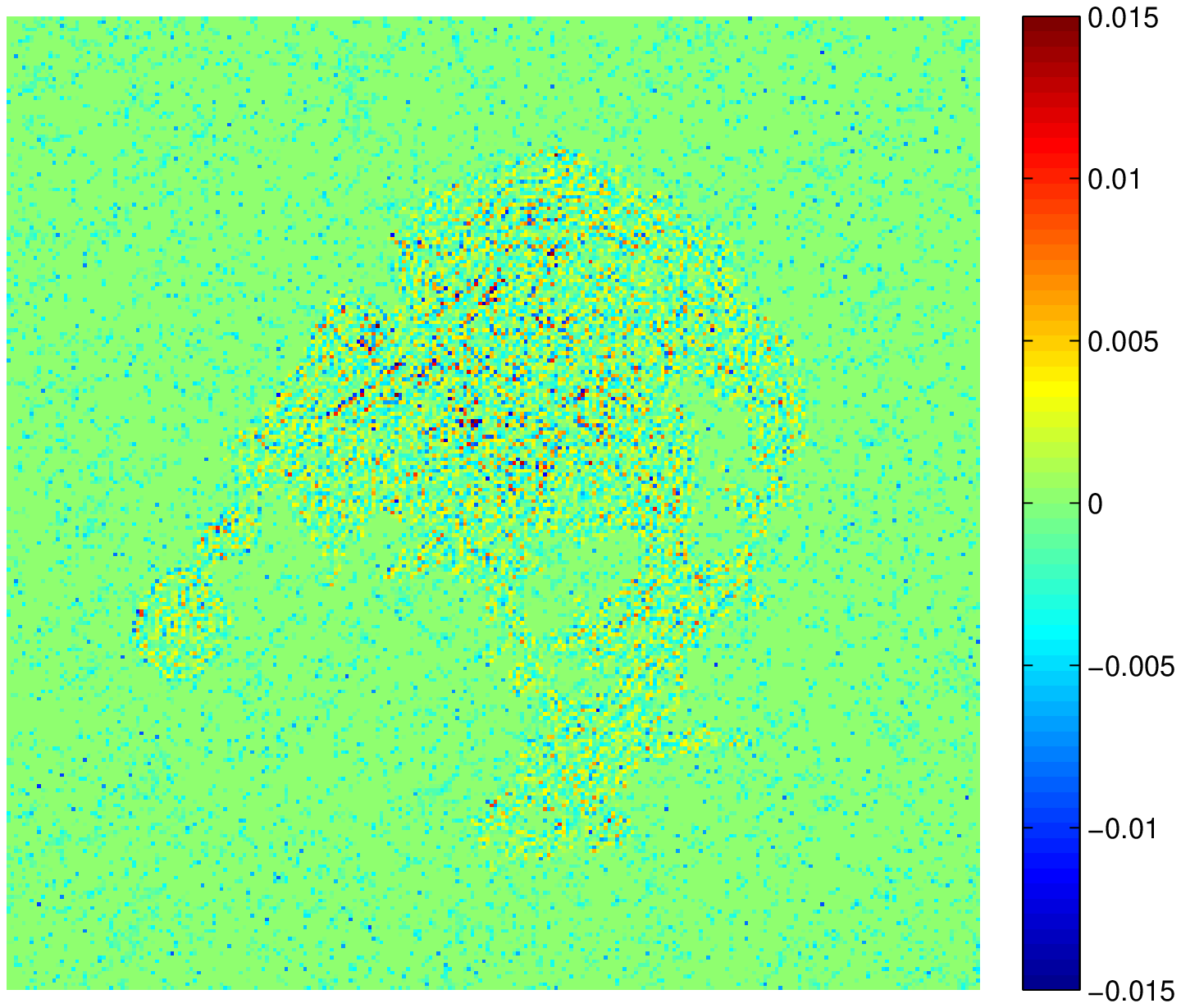}&
    \includegraphics[trim = 2cm 1.3cm 1cm 0.5cm, clip, keepaspectratio, width = 5.8cm]{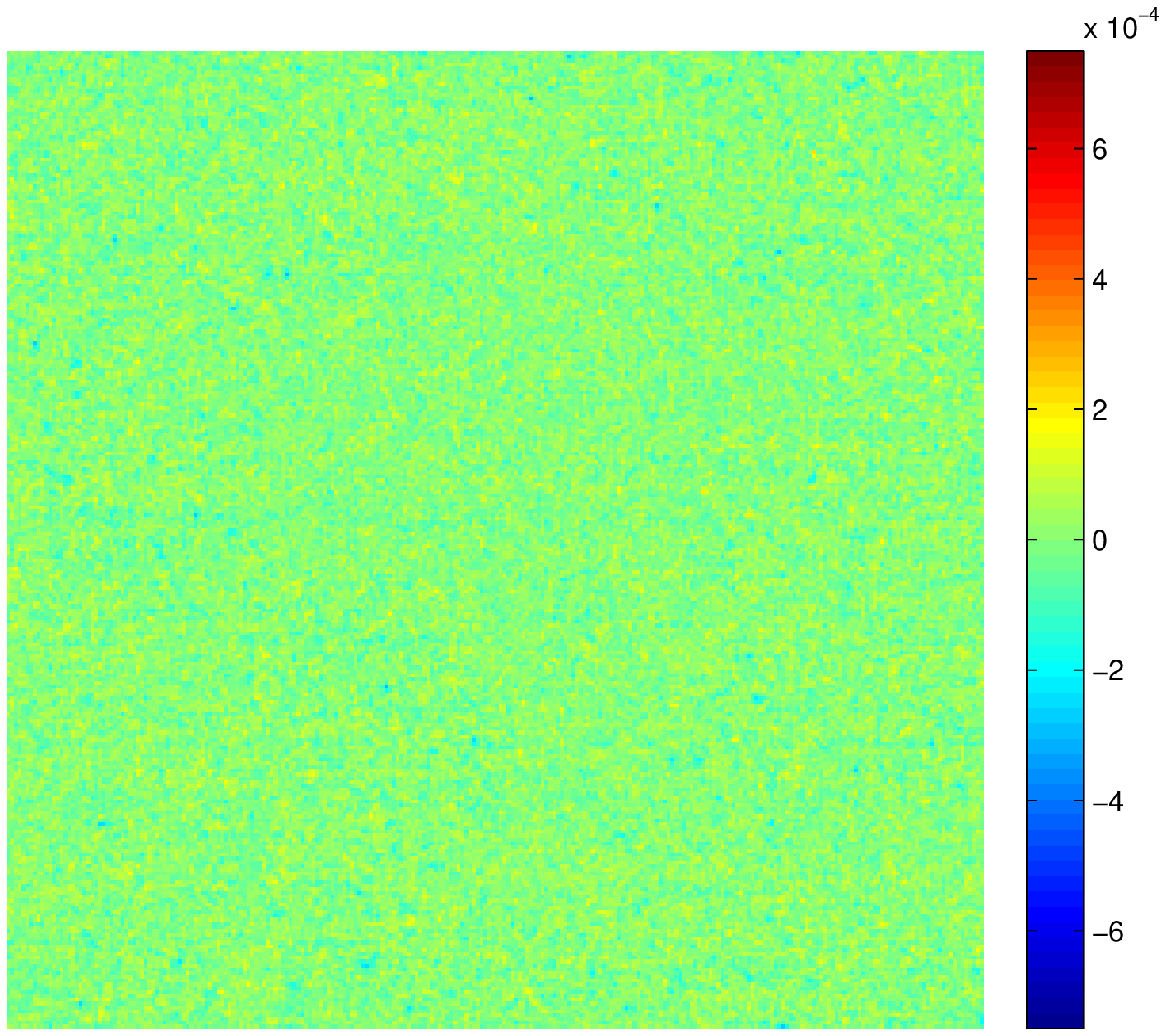}\\
    
    \includegraphics[trim = 2cm 1.3cm 1cm 0.5cm, clip, keepaspectratio, width = 5.8cm]{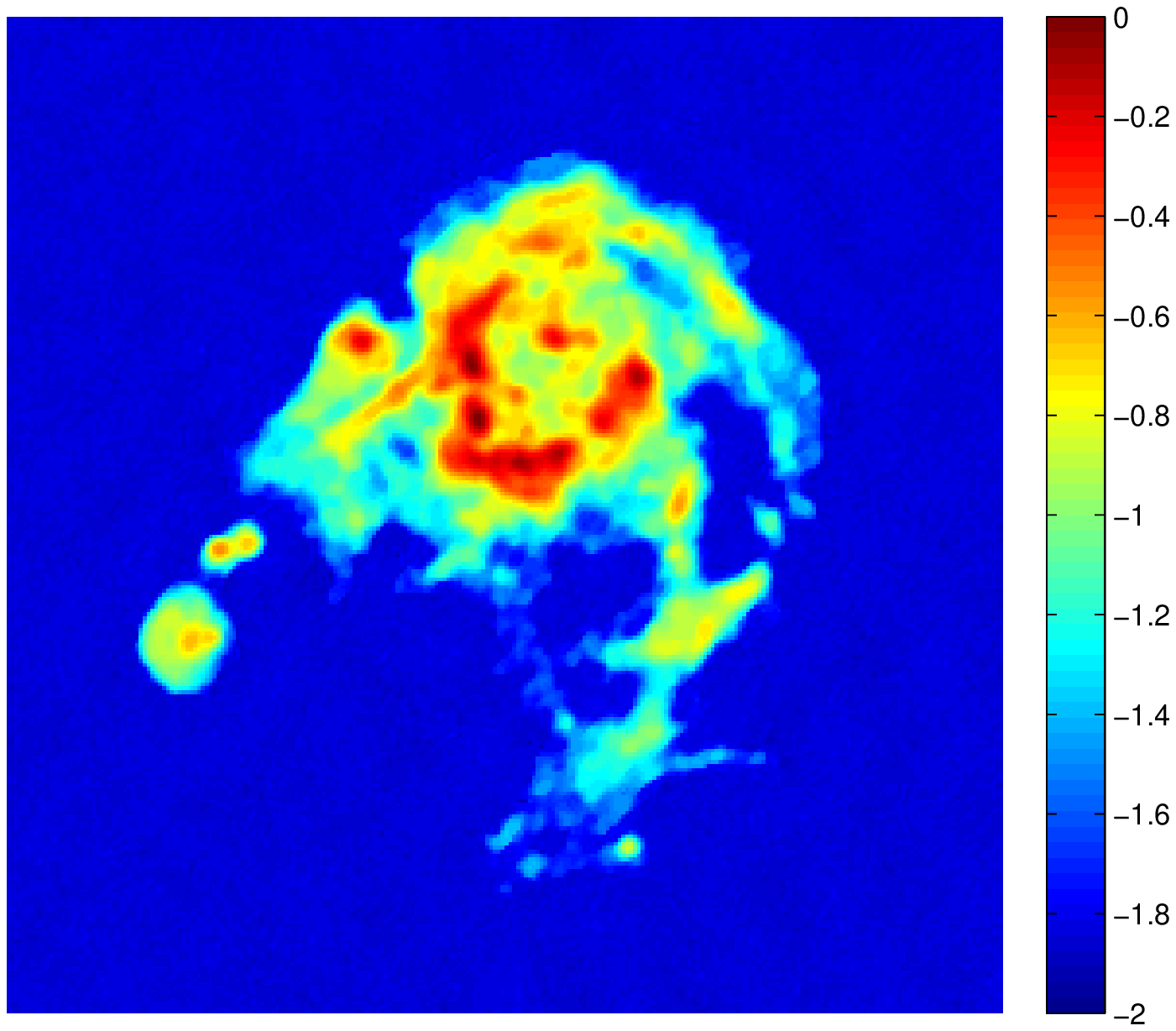}&
    \includegraphics[trim = 2cm 1.3cm 1cm 0.5cm, clip, keepaspectratio, width = 5.8cm]{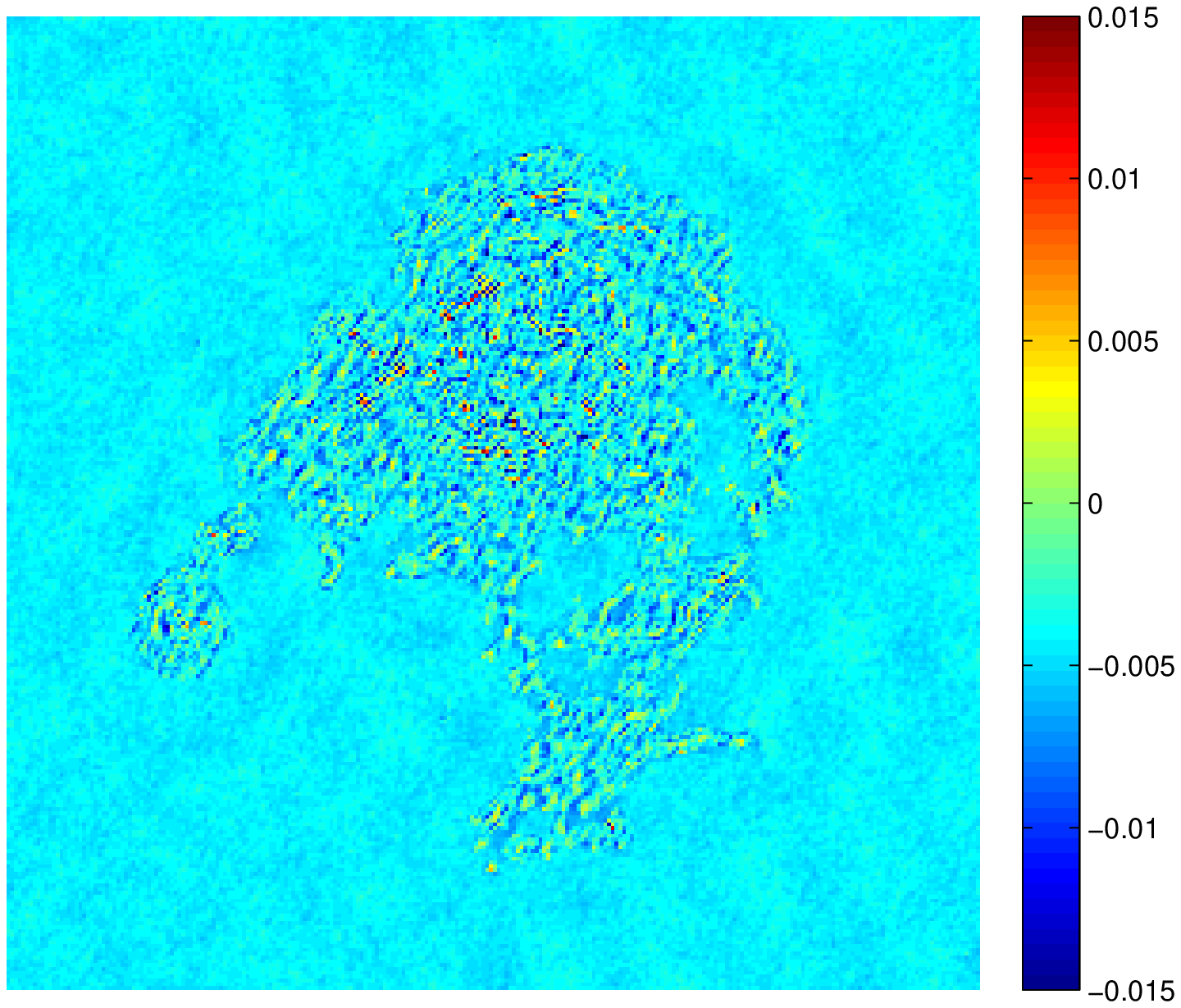}&
    \includegraphics[trim = 2cm 1.3cm 1cm 0.5cm, clip, keepaspectratio, width = 5.8cm]{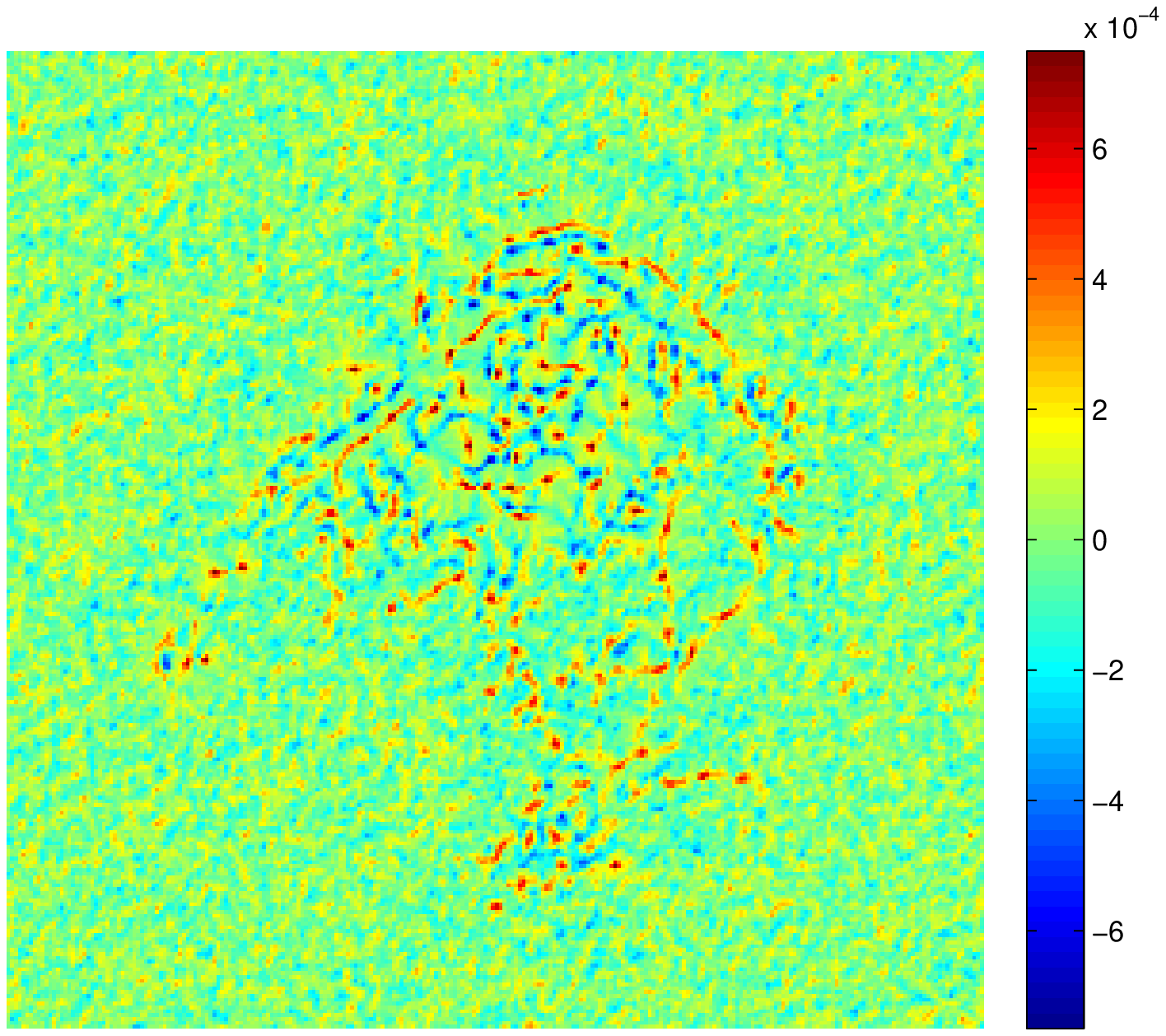}\\
    
    \includegraphics[trim = 2cm 1.3cm 1cm 0.5cm, clip, keepaspectratio, width = 5.8cm]{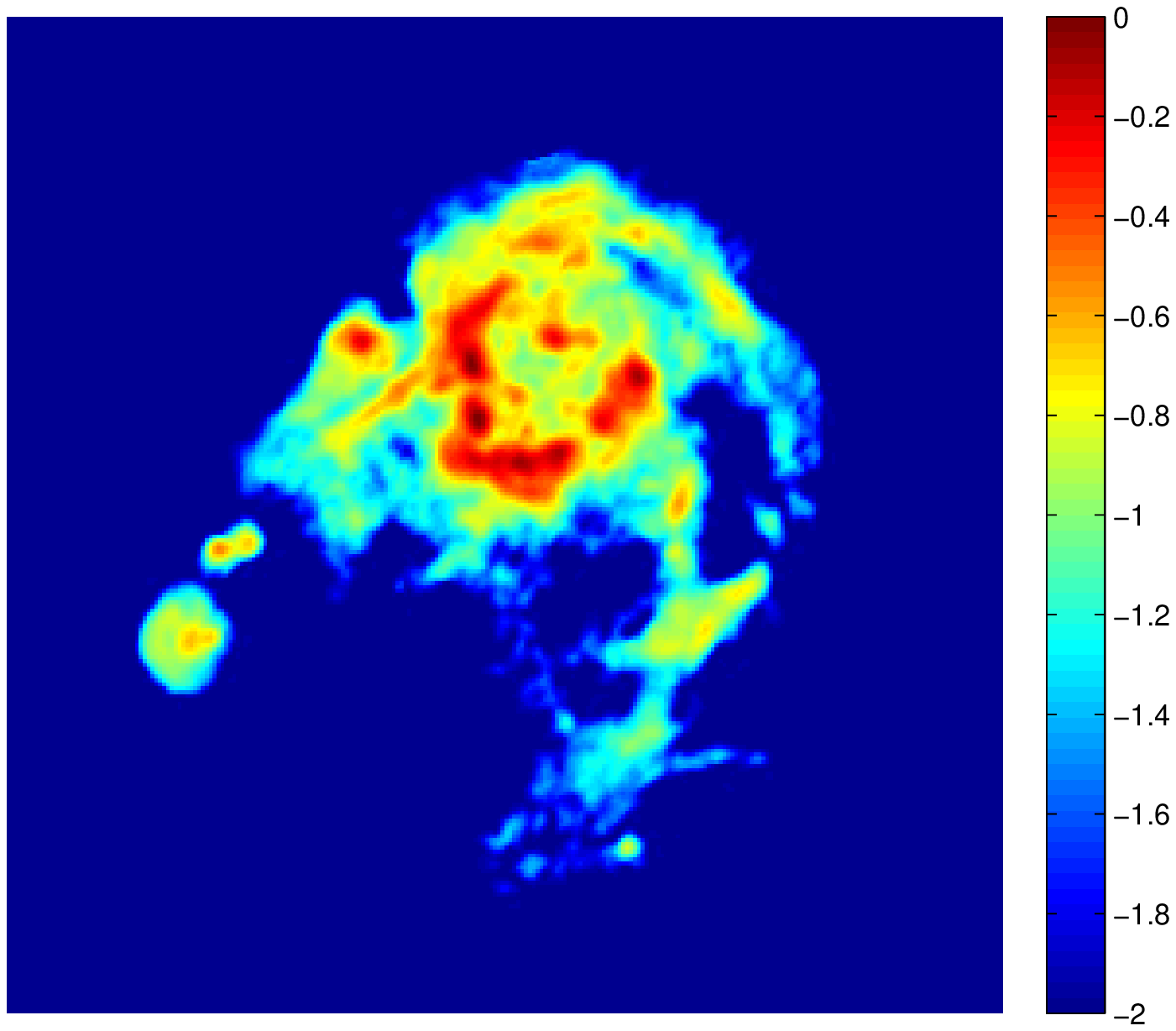}&
    \includegraphics[trim = 2cm 1.3cm 1cm 0.5cm, clip, keepaspectratio, width = 5.8cm]{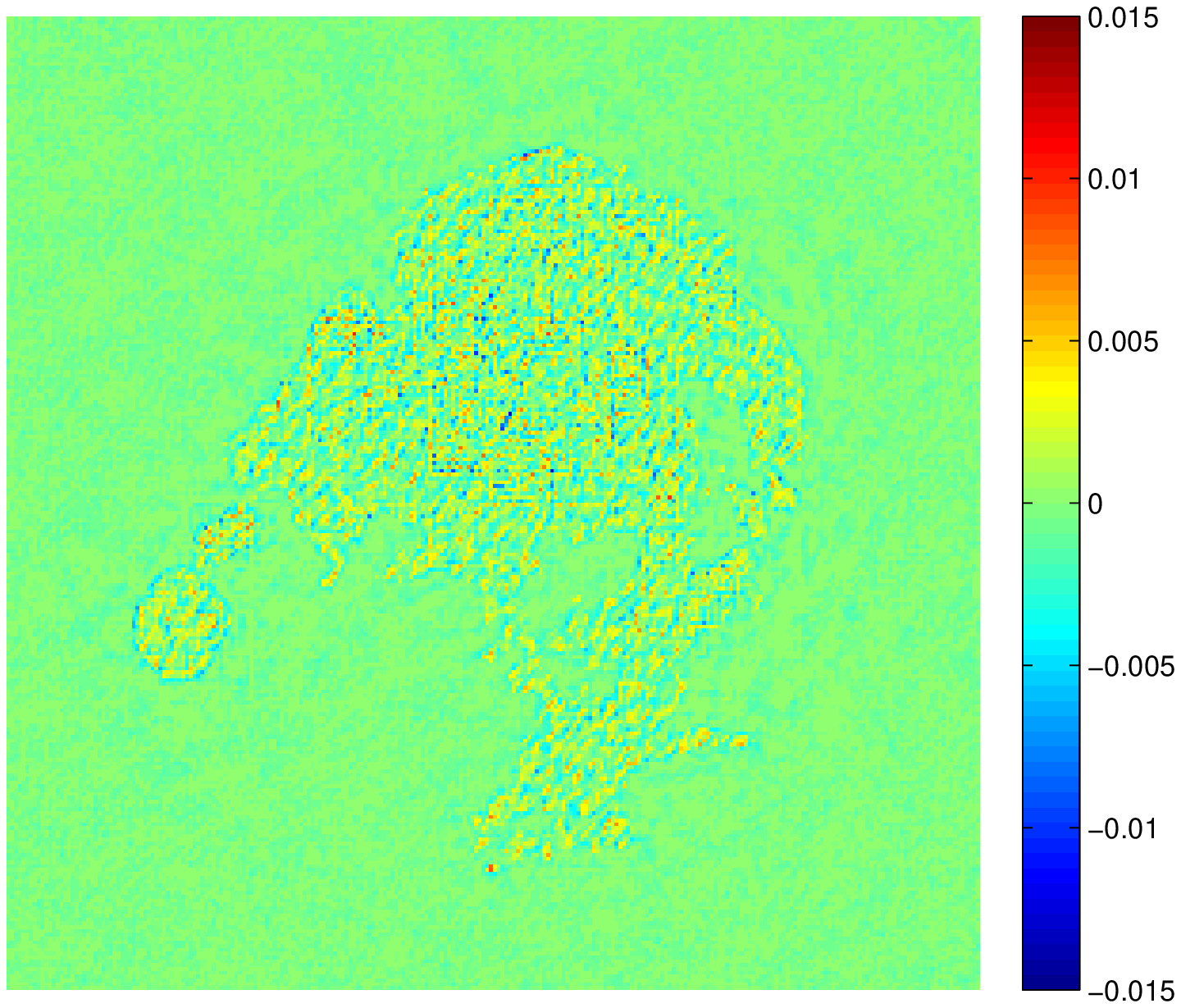}&
    \includegraphics[trim = 2cm 1.3cm 1cm 0.5cm, clip, keepaspectratio, width = 5.8cm]{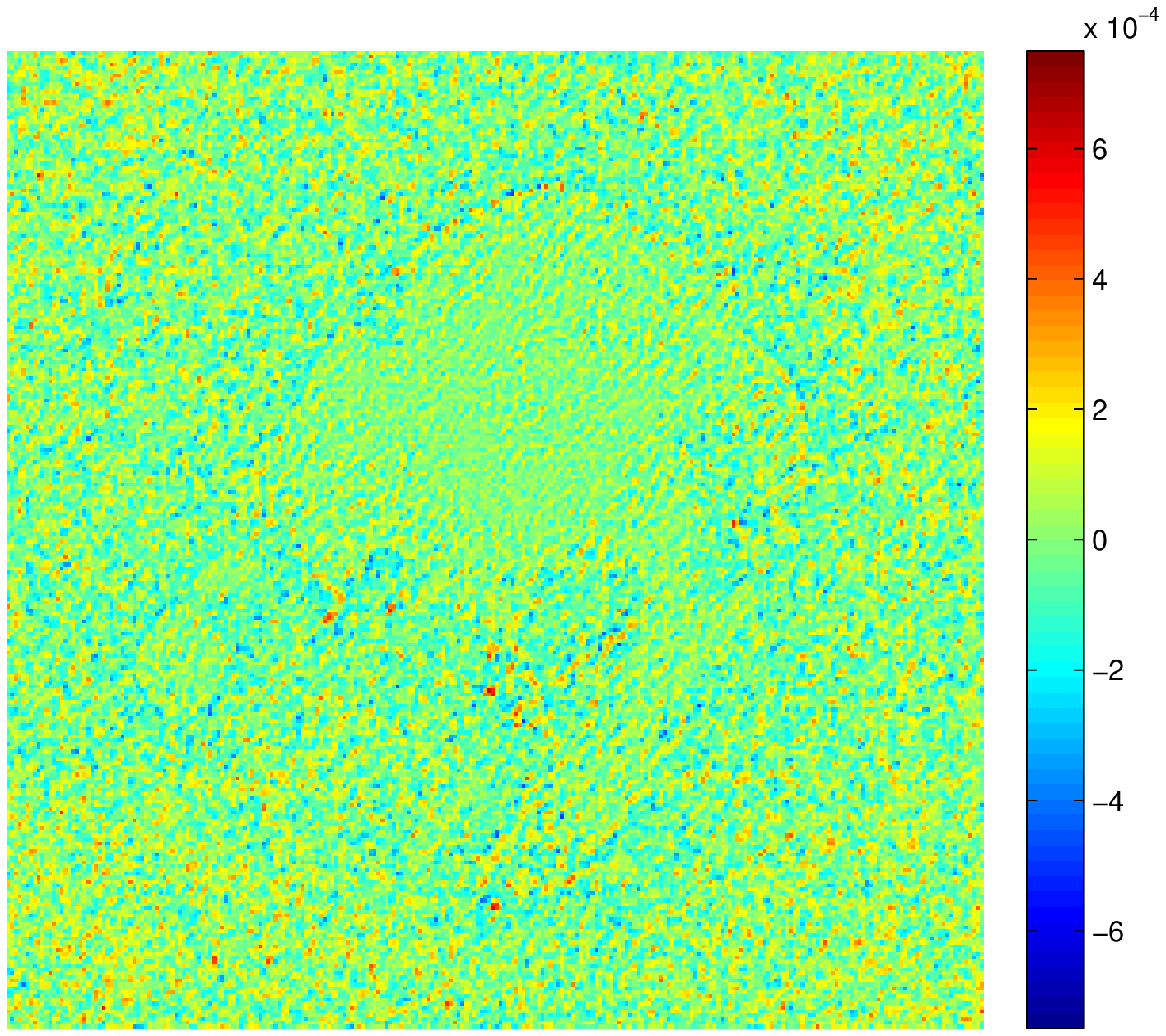}\\

    \end{tabular}

\caption{(color online). Reconstruction example of M31. The results are shown from top to bottom for BP (SNR=32.82~dB), BPDb8 (SNR=33.70~dB), IUWT (SNR=32.12~dB), TV (SNR=33.89~dB) and SARA (SNR=38.43~dB) respectively. The first column shows the reconstructed images in a $\log_{10}$ scale, the second column shows the error images in linear scale, and the third column shows the residual images also in linear scale. }
\label{fig:5}
\end{figure*}

\begin{figure*}

\centering
    \begin{tabular}{ccc}
   
    \includegraphics[trim = 2cm 1.3cm 1cm 0.5cm, clip, keepaspectratio, width = 5.8cm]{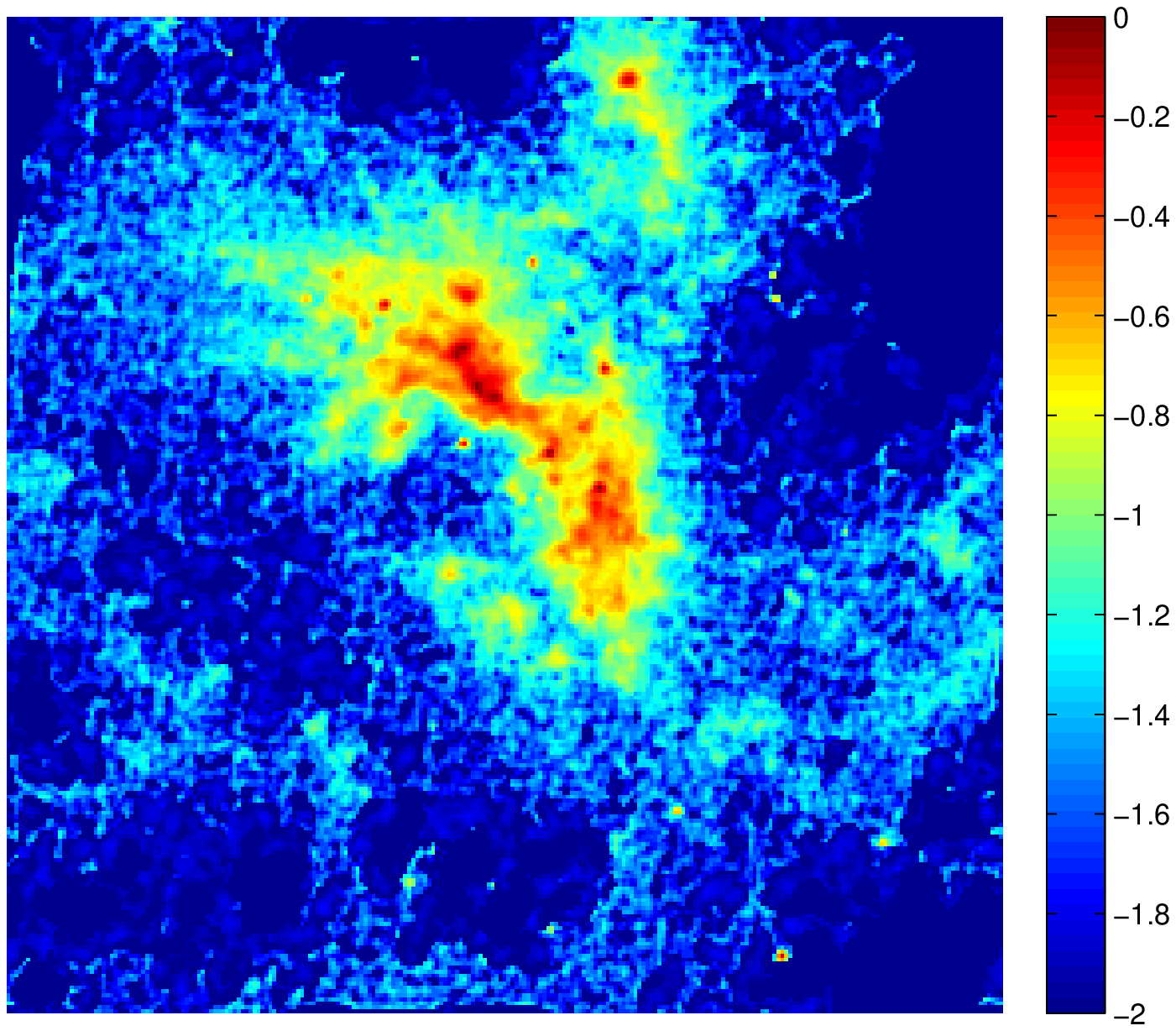}&
    \includegraphics[trim = 2cm 1.3cm 1cm 0.5cm, clip, keepaspectratio, width = 5.8cm]{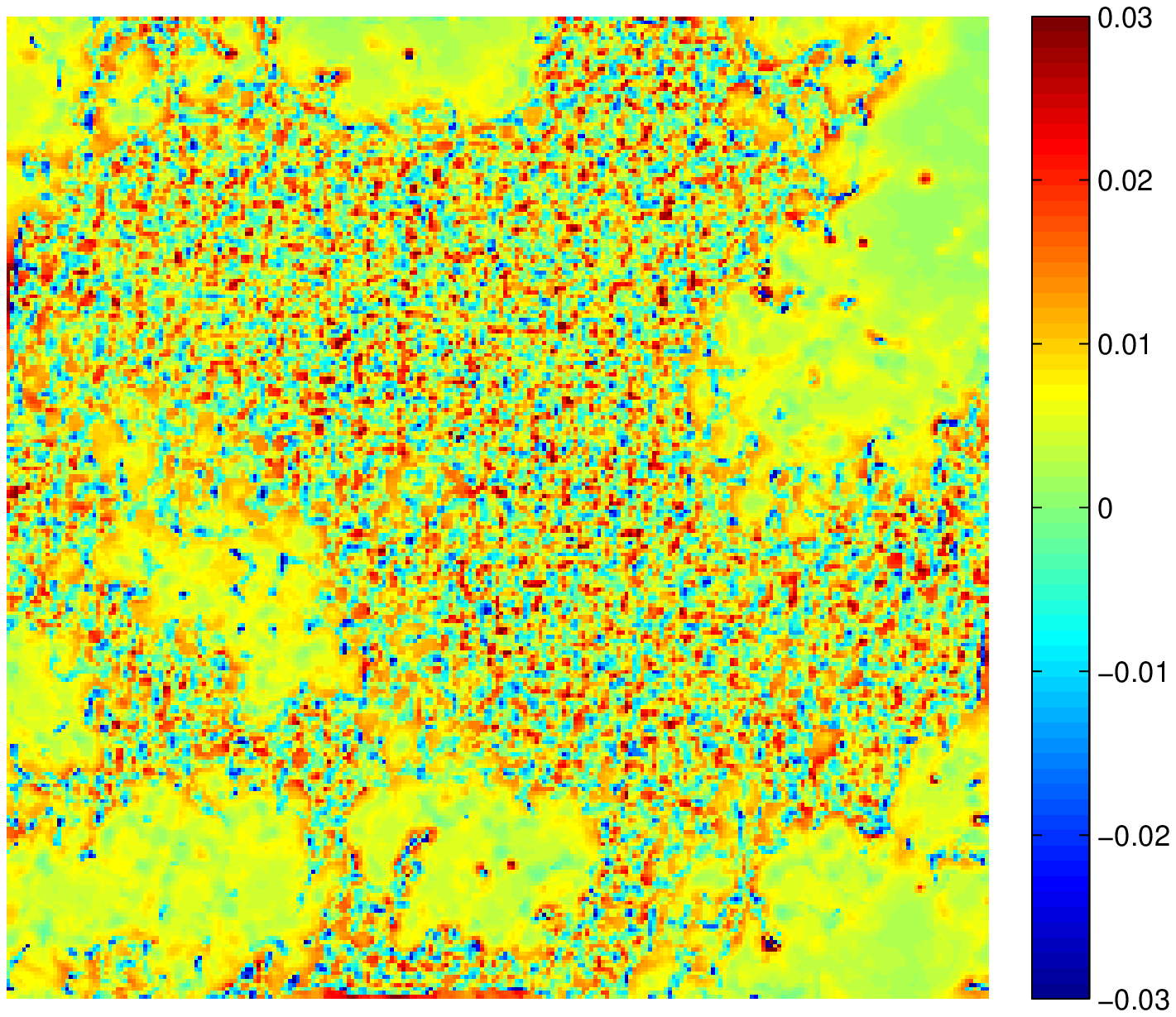}&
    \includegraphics[trim = 2cm 1.3cm 1cm 0.5cm, clip, keepaspectratio, width = 5.8cm]{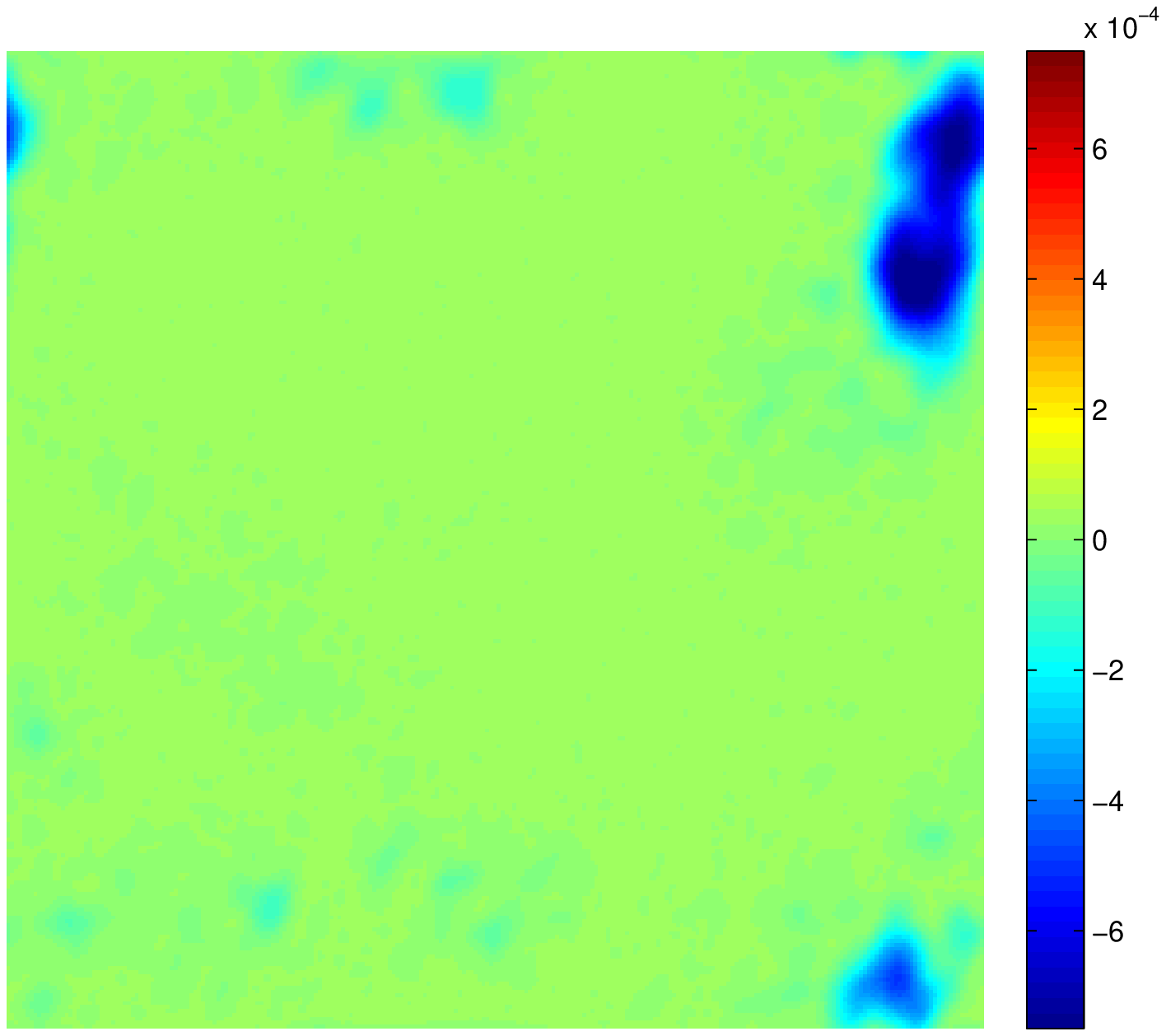}\\
    
    \includegraphics[trim = 2cm 1.3cm 1cm 0.5cm, clip, keepaspectratio, width = 5.8cm]{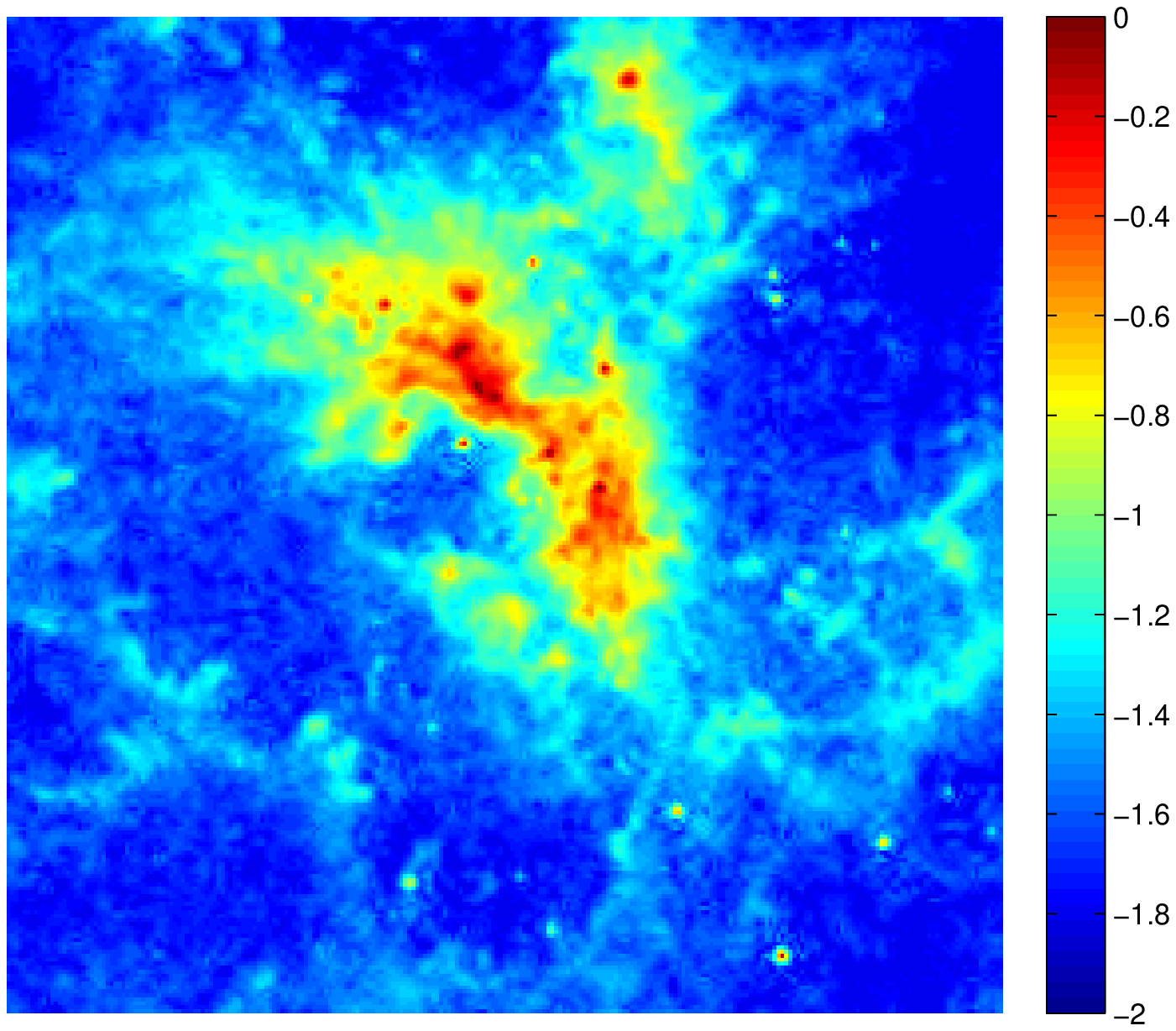}&
    \includegraphics[trim = 2cm 1.3cm 1cm 0.5cm, clip, keepaspectratio, width = 5.8cm]{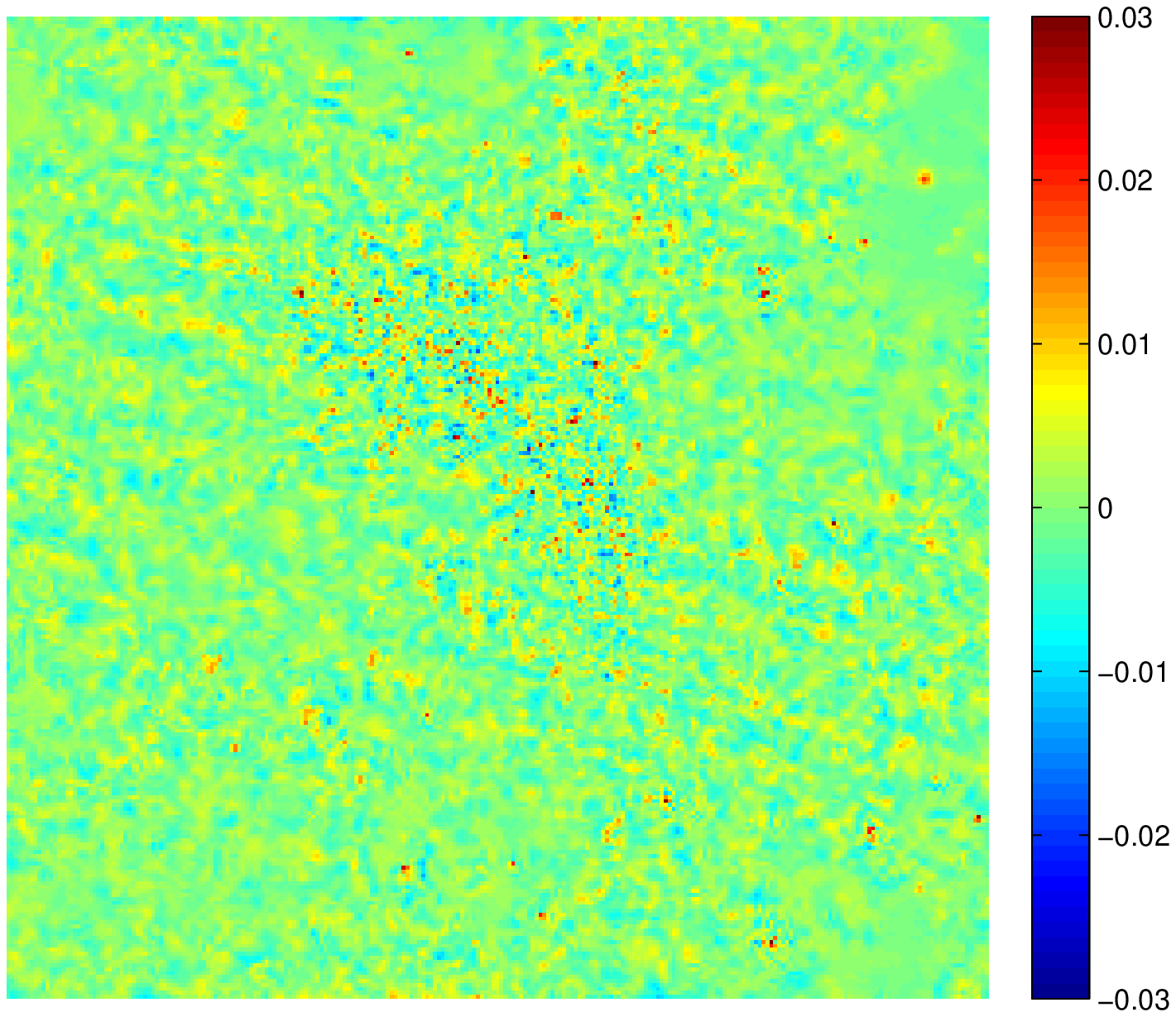}&
    \includegraphics[trim = 2cm 1.3cm 1cm 0.5cm, clip, keepaspectratio, width = 5.8cm]{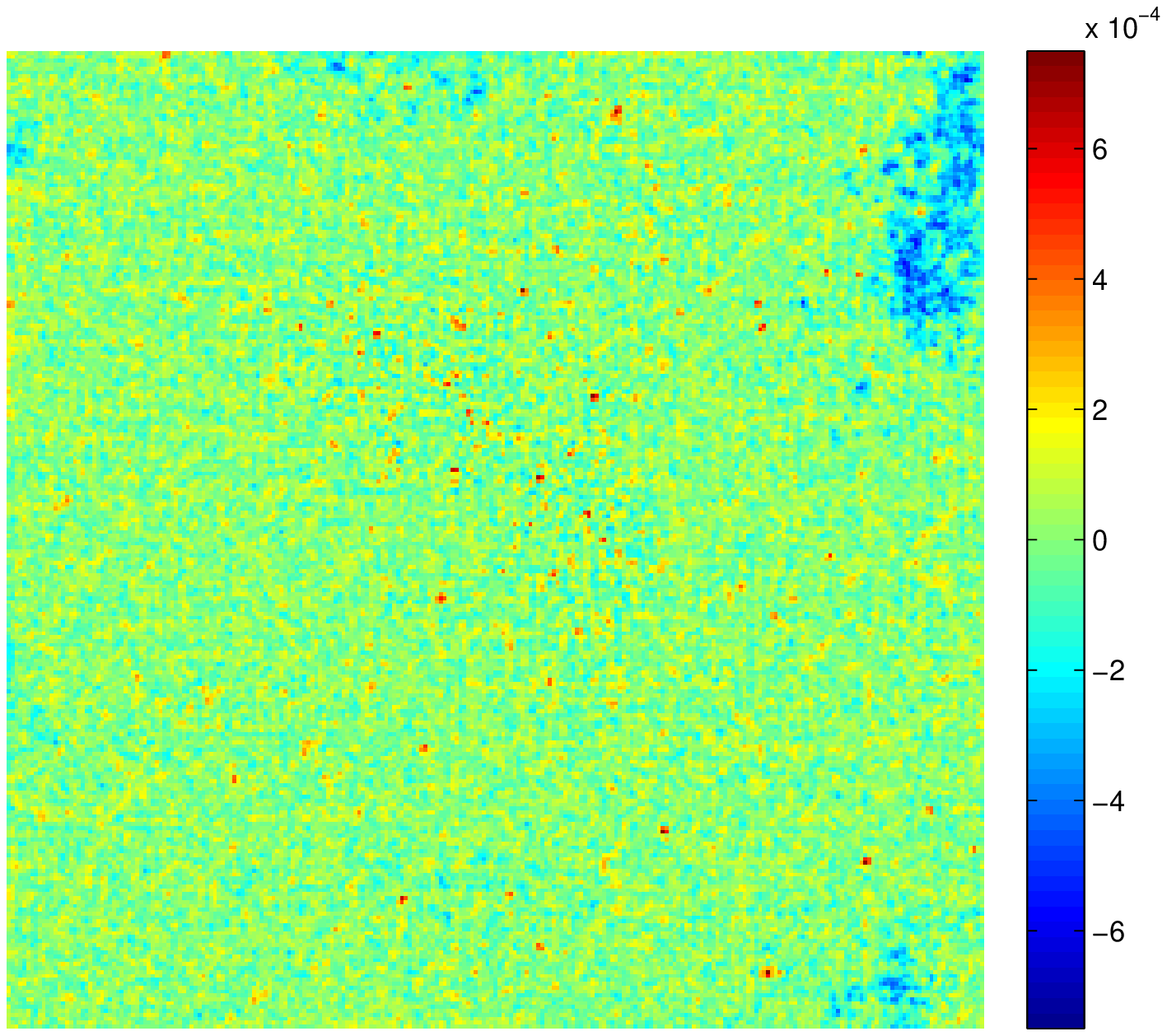}\\
    
    \includegraphics[trim = 2cm 1.3cm 1cm 0.5cm, clip, keepaspectratio, width = 5.8cm]{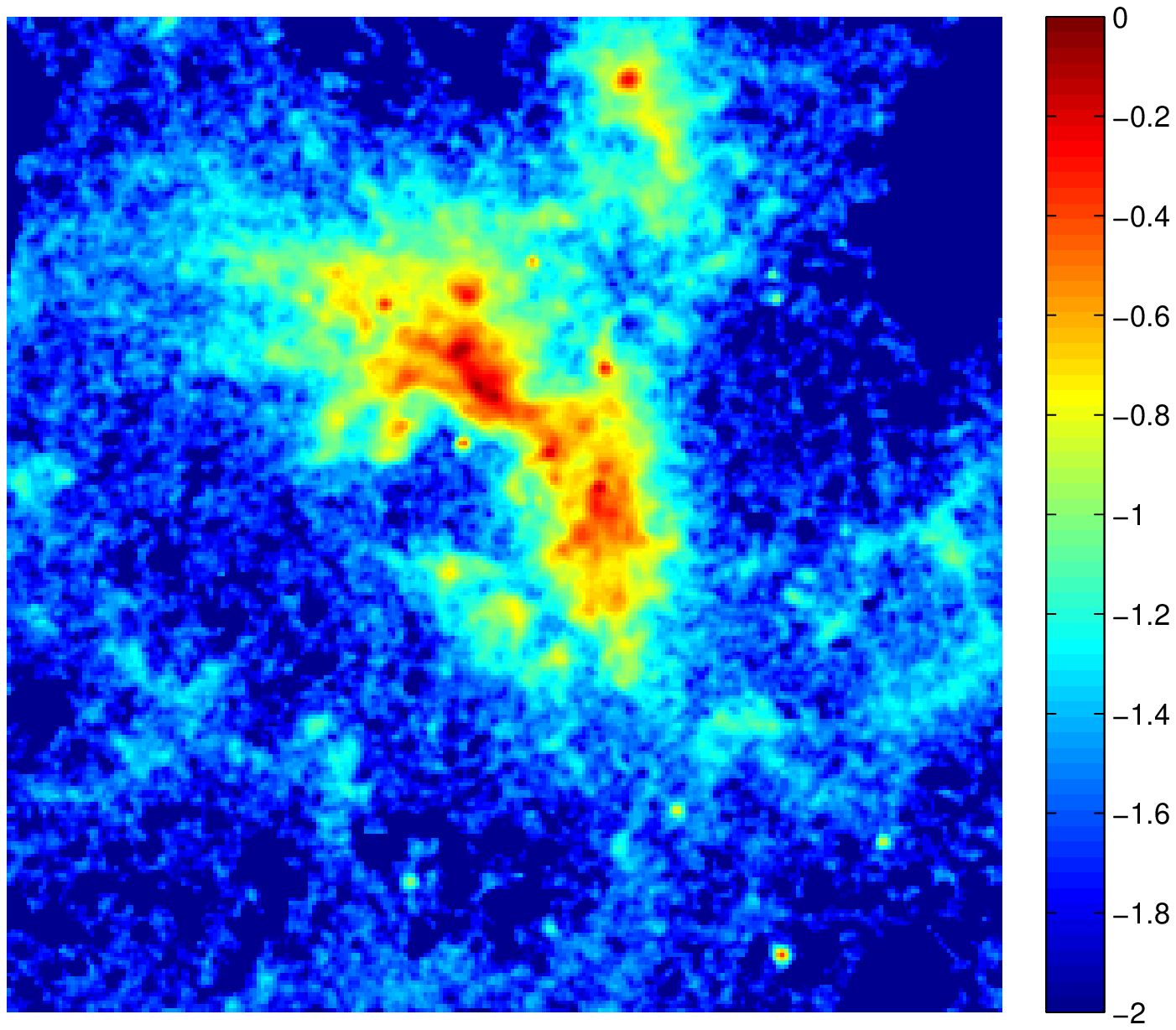}&
    \includegraphics[trim = 2cm 1.3cm 1cm 0.5cm, clip, keepaspectratio, width = 5.8cm]{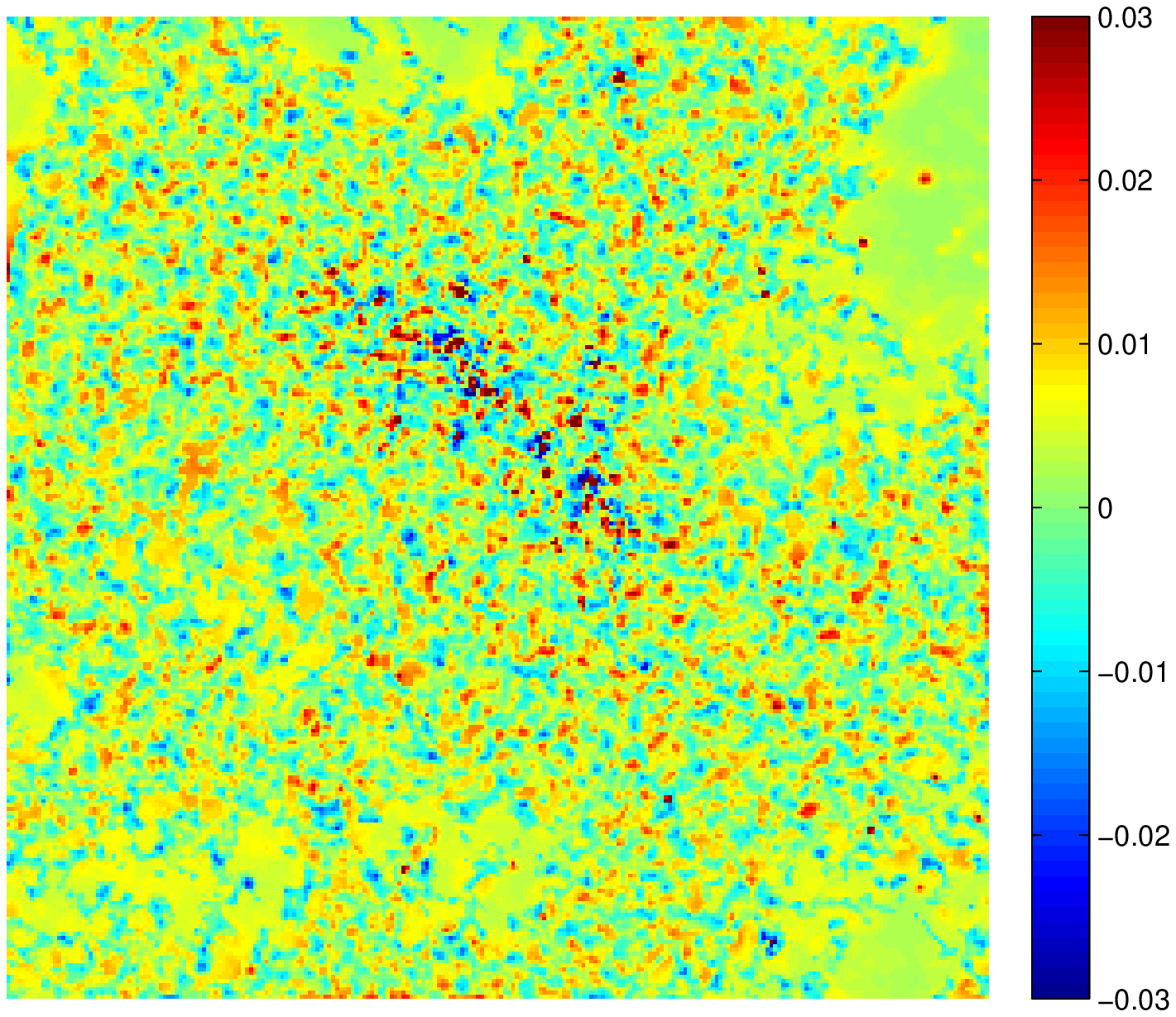}&
    \includegraphics[trim = 2cm 1.3cm 1cm 0.5cm, clip, keepaspectratio, width = 5.8cm]{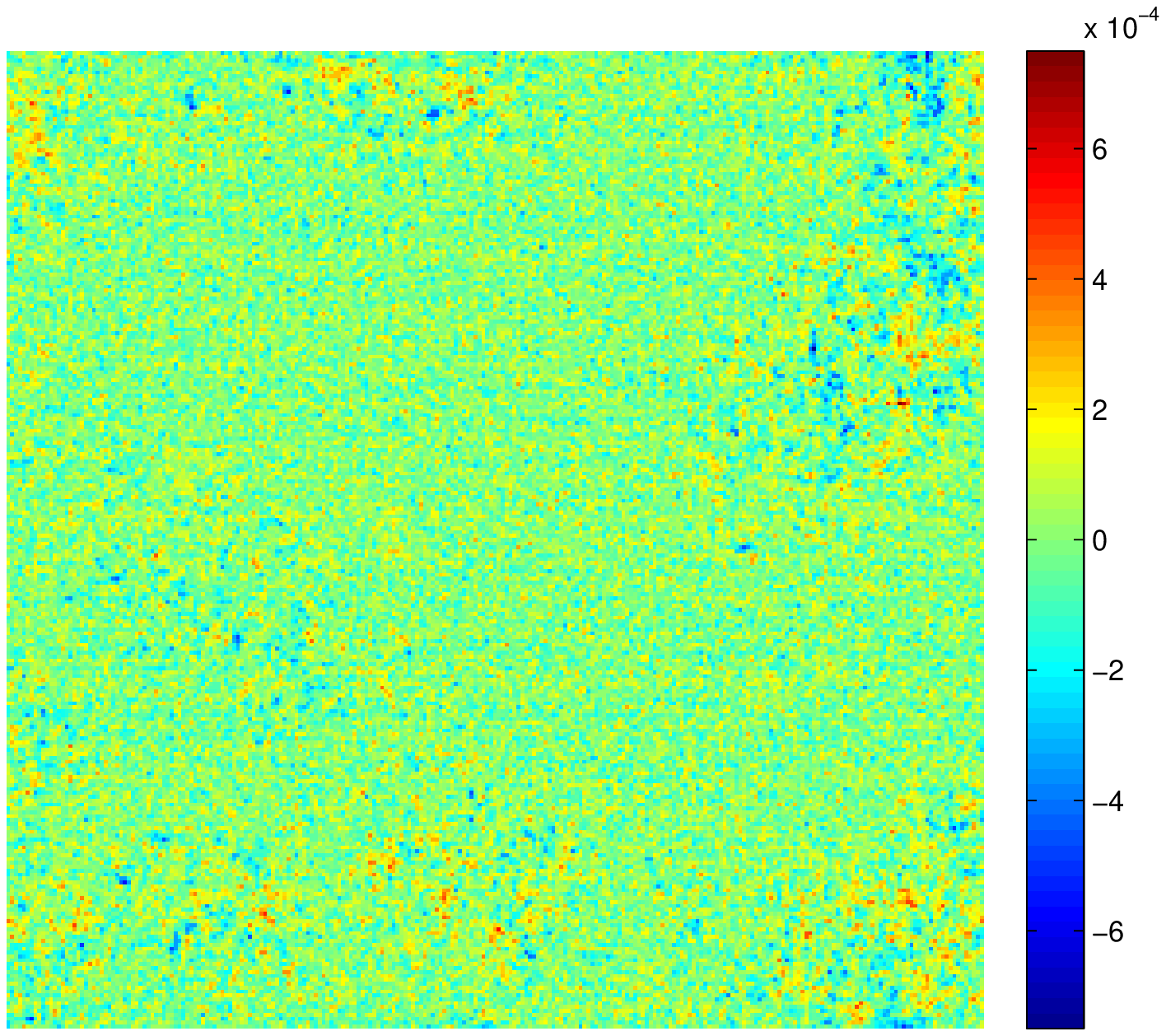}\\
    
    \includegraphics[trim = 2cm 1.3cm 1cm 0.5cm, clip, keepaspectratio, width = 5.8cm]{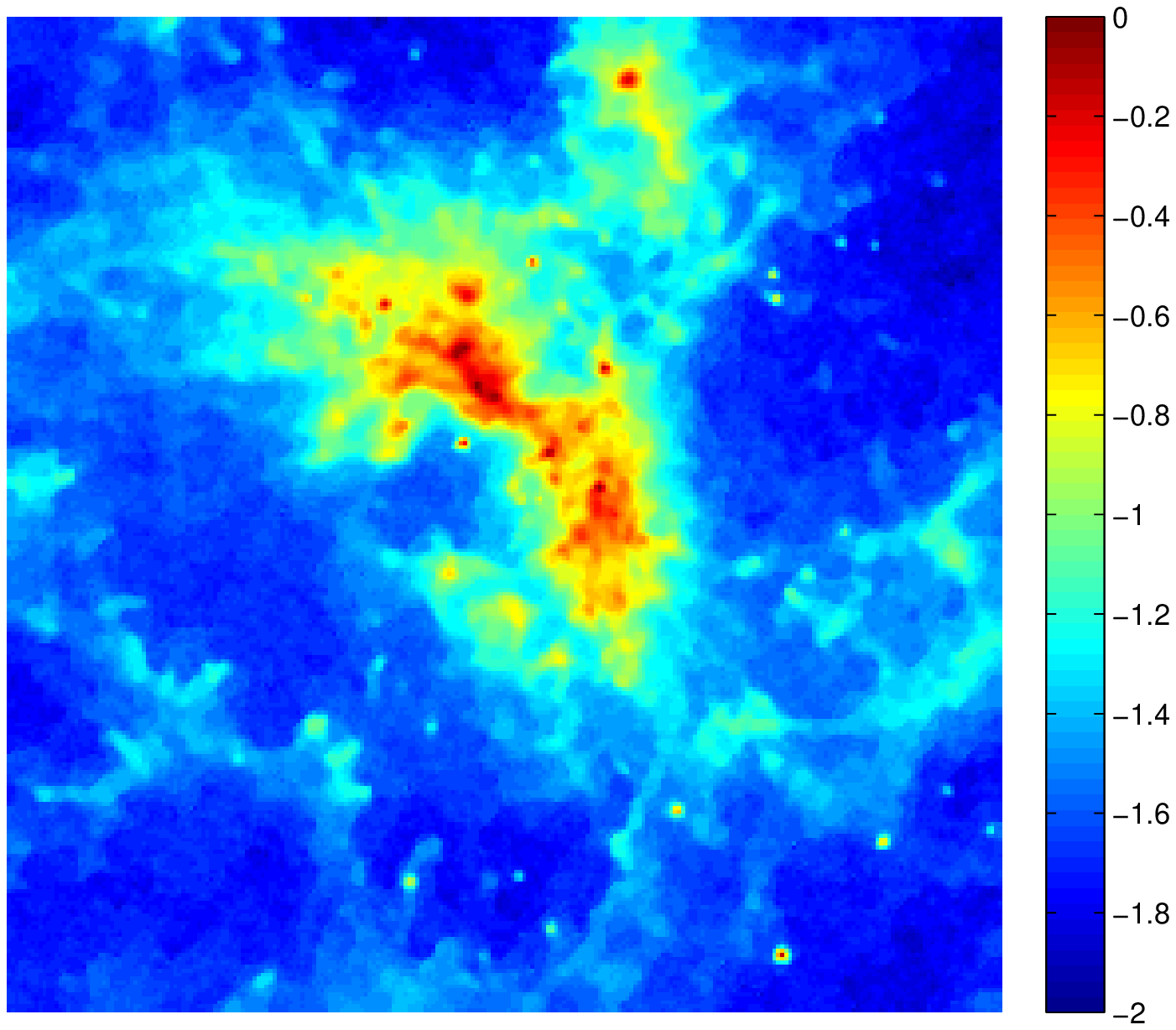}&
    \includegraphics[trim = 2cm 1.3cm 1cm 0.5cm, clip, keepaspectratio, width = 5.8cm]{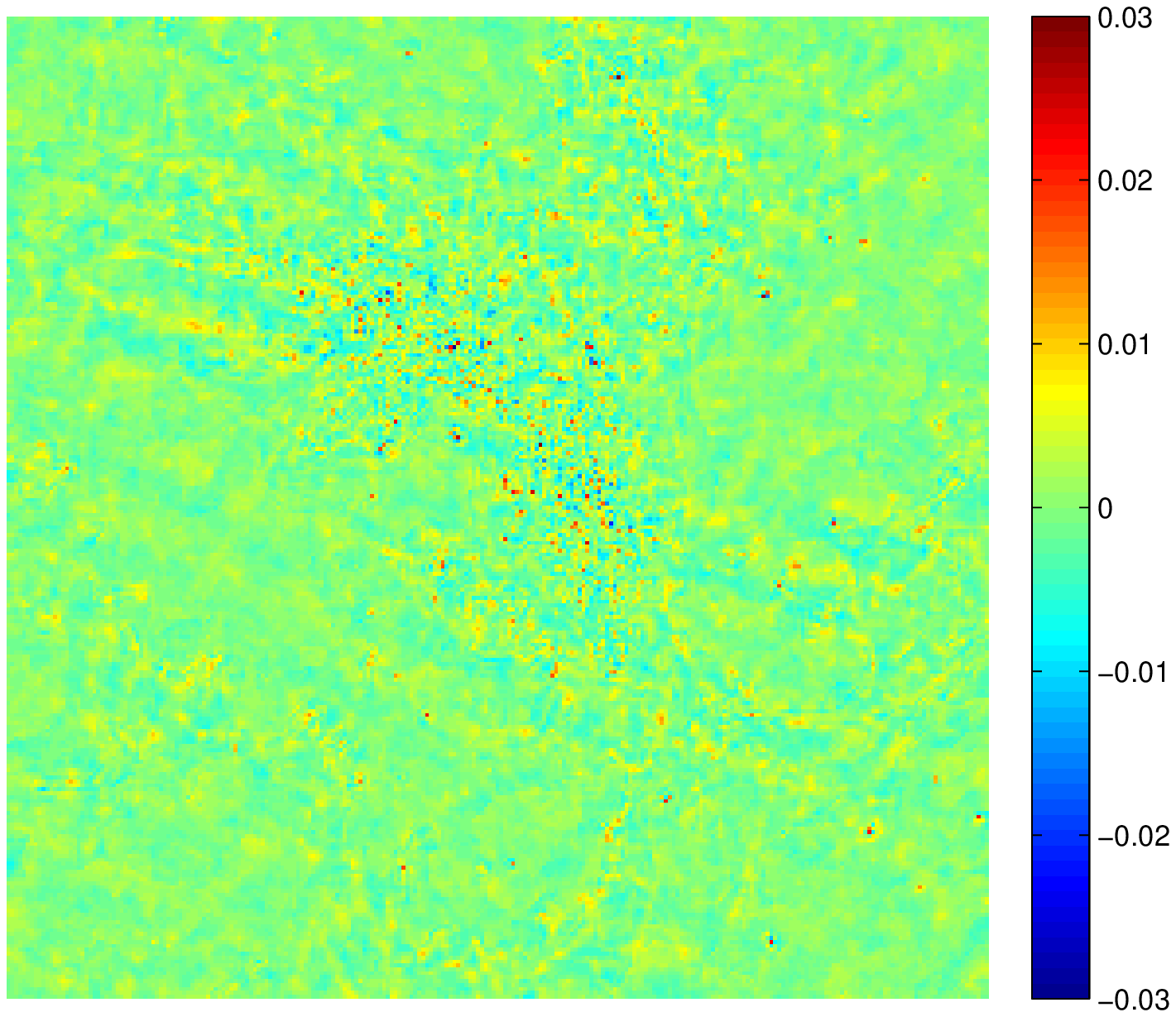}&
    \includegraphics[trim = 2cm 1.3cm 1cm 0.5cm, clip, keepaspectratio, width = 5.8cm]{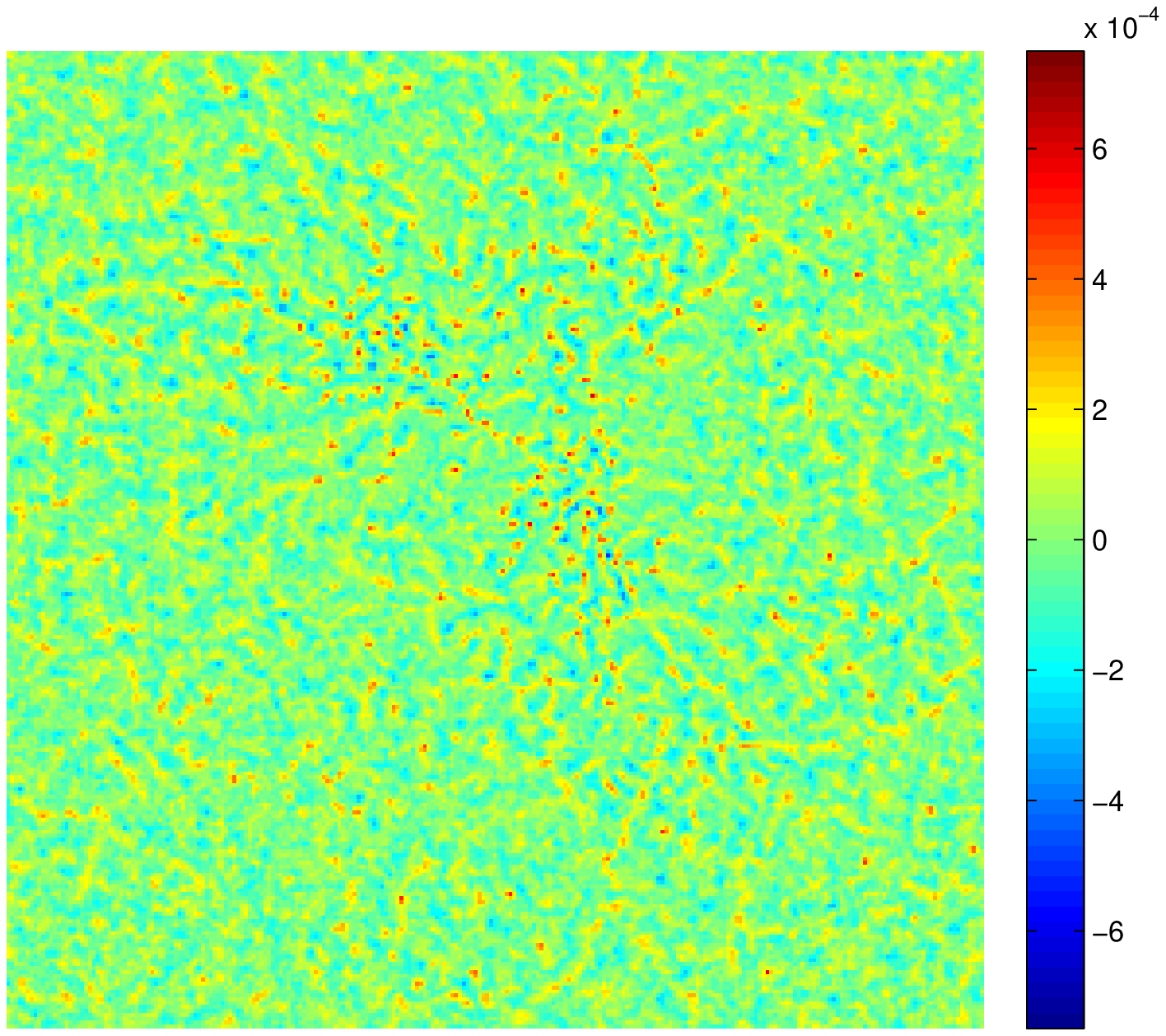}\\
    
    \includegraphics[trim = 2cm 1.3cm 1cm 0.5cm, clip, keepaspectratio, width = 5.8cm]{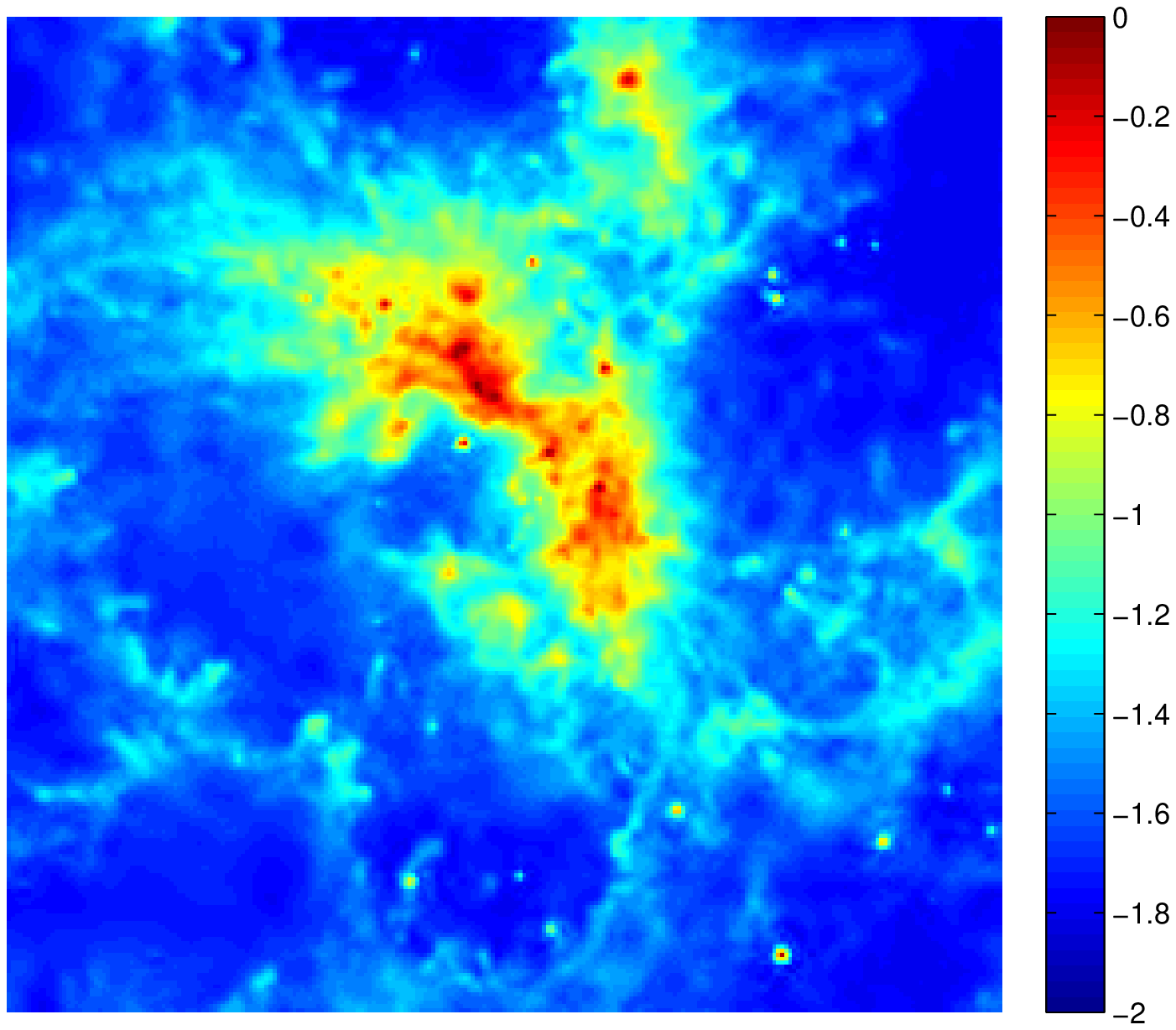}&
    \includegraphics[trim = 2cm 1.3cm 1cm 0.5cm, clip, keepaspectratio, width = 5.8cm]{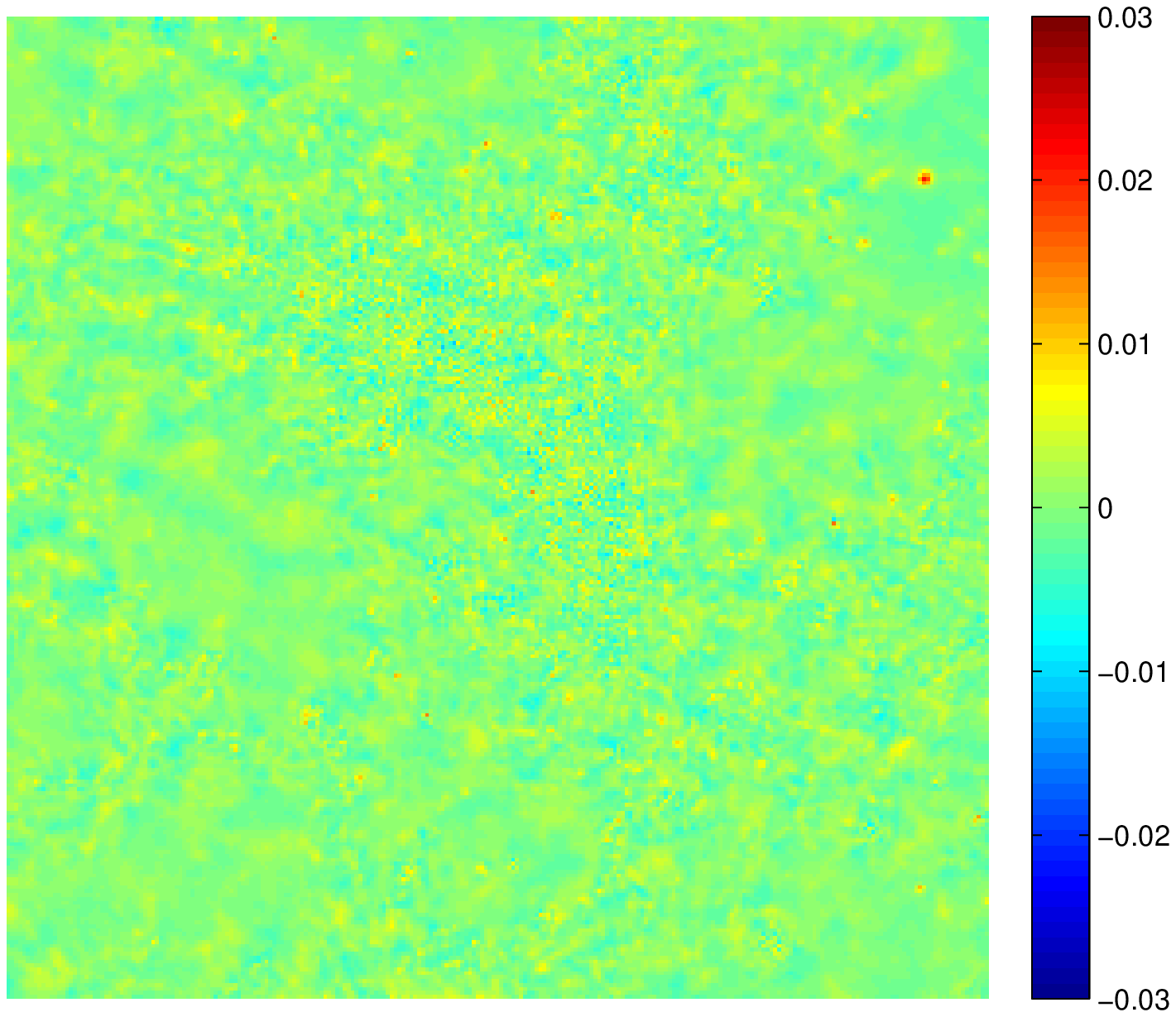}&
    \includegraphics[trim = 2cm 1.3cm 1cm 0.5cm, clip, keepaspectratio, width = 5.8cm]{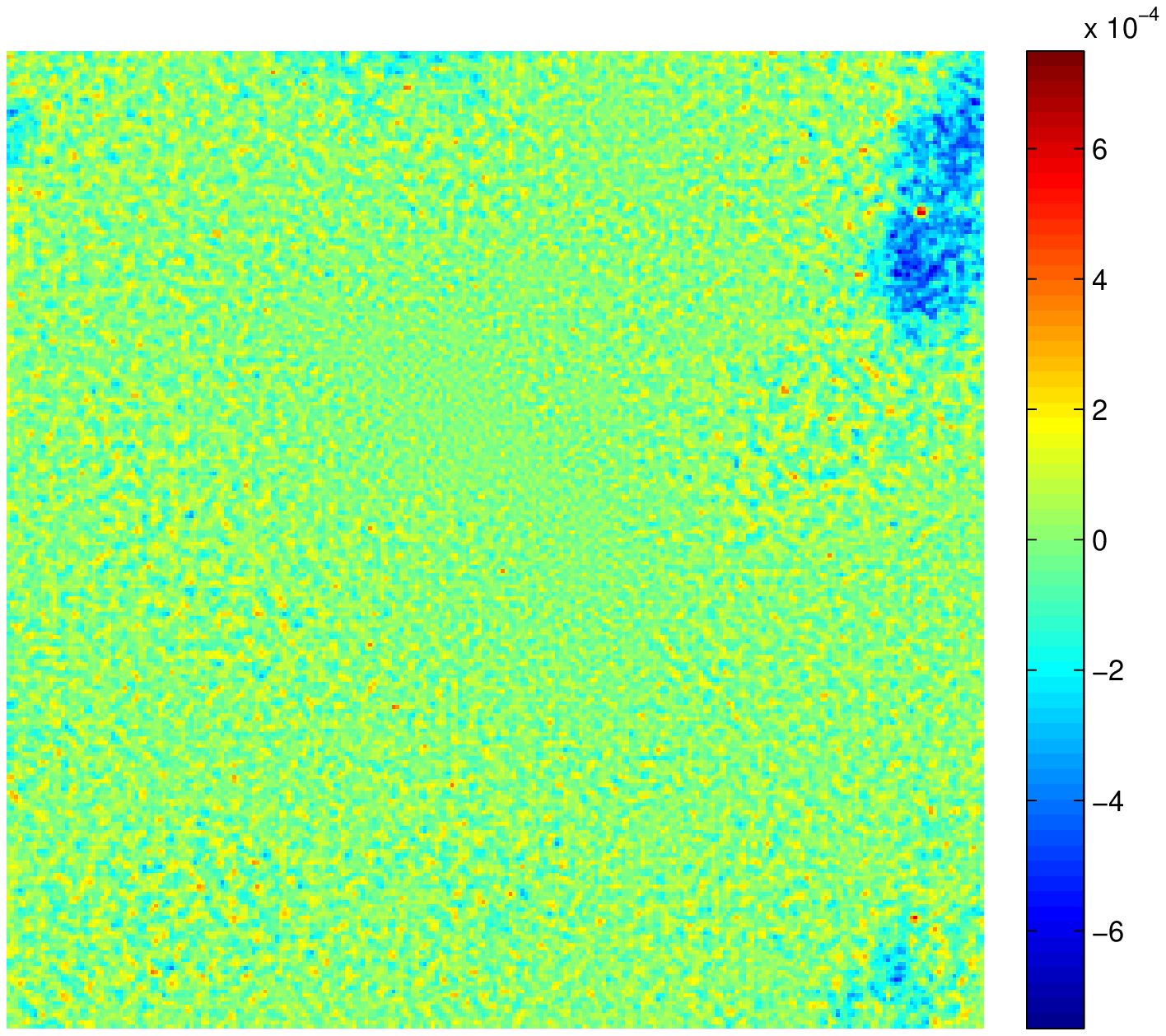}\\
   
    \end{tabular}

\caption{(color online). Reconstruction example of 30Dor. The results are shown from top to bottom for BP (SNR=16.67~dB), BPDb8 (SNR=24.53~dB), IUWT (SNR=17.87~dB), TV (SNR=26.47~dB) and SARA (SNR=29.08~dB) respectively. The first column shows the reconstructed images in a $\log_{10}$ scale, the second column shows the error images in linear scale, and the third column shows the residual images also in linear scale.}
\label{fig:6}
\end{figure*}

Next we present a visual assessment of the reconstruction quality of SARA compared to the benchmark algorithms. Figure~\ref{fig:5} and Figure~\ref{fig:6} show the results from M31 and 30Dor respectively for a coverage of 30\%. The results are shown from top to bottom for BP, BPDb8, IUWT, TV and SARA respectively. The first column shows the reconstructed images in a $\log_{10}$ scale, the second column shows the error images, defined as $\bm{x}-\hat{\bm{x}}$, in linear scale, and, the third column shows the real part of the residual images, defined as the difference between dirty images and dirty images constructed from recovered images, i.e. $\bm{r}=\mathsf{F}^{T}\mathsf{M}^{T}\bm{y}-\mathsf{F}^{T}\mathsf{M}^{T}\mathsf{\Phi}\hat{\bm{x}}$, also in linear scale. We use the residual images as a visual quality measure because it is commonly used in radio interferometry.  In a few words, beyond a significant SNR increase, SARA provides an impressive reduction of visual artifacts relative to the other methods. 

More specifically, for M31 we can see that BP and BPDb8 yield a good reconstruction of the inner structures but also give a lot of artifacts in the constant background, with BP having a lot of point-like errors as expected. TV also yields a good reconstruction quality, since the original image has well defined edges, even though TV suffers from bias problems and is not capable of estimating the correct background having a slight shift in the brightness value. The IUWT method yields a nearly flat residual map. However, this does not necessarily translate into a better reconstruction quality as can be observed in the error image. This highlights the fact that the common criterion of flatness of residual image is not an optimal measure of reconstruction fidelity. SARA yields a restored image with few artifacts in the background and small errors in the inner structures, showing the advantage of multiple basis representations.

For 30Dor we see that BP and IUWT do not yield good results, with reconstructions having a lot of spurious point-like structures. BPDb8 yields a good reconstruction but also yields a lot of visual artifacts. TV achieves a fair reconstruction of the original image. However, the resulting image has a piecewise structure (leading to a cartoon-like visual effect) due to the TV prior. SARA yields the best recovery of the original image, being able to recover point-like structures as well as continuous extended structures. Note that all methods yield noise-like residual maps for this example but the actual reconstruction quality differs for all methods.

\subsection{Spread spectrum illustration}
\begin{figure*}

\centering
    \begin{tabular}{cc}
   
    \includegraphics[trim = 2.5cm 3.9cm 1.5cm 3.4cm, clip, keepaspectratio, width = 7.0cm]{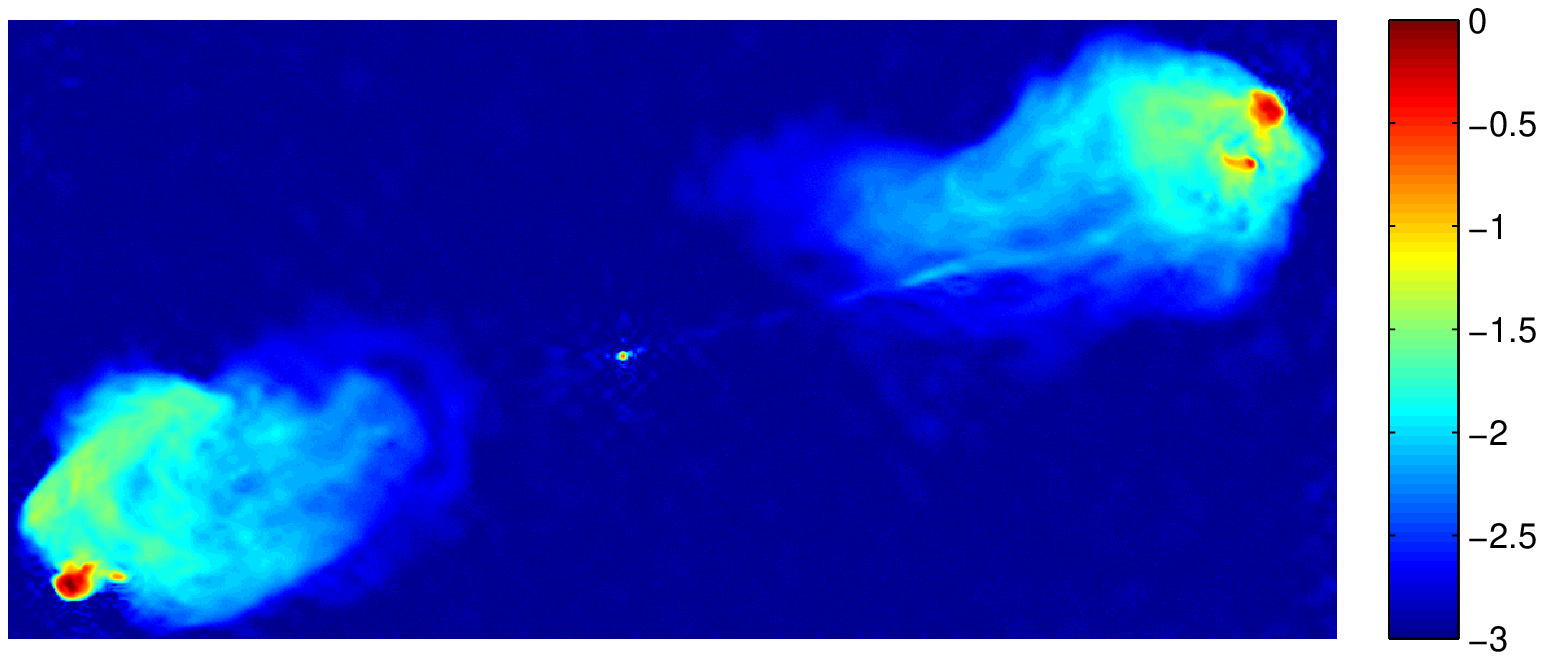}&
    \includegraphics[trim = 2.5cm 3.9cm 1.5cm 3.4cm, clip, keepaspectratio, width = 7.0cm]{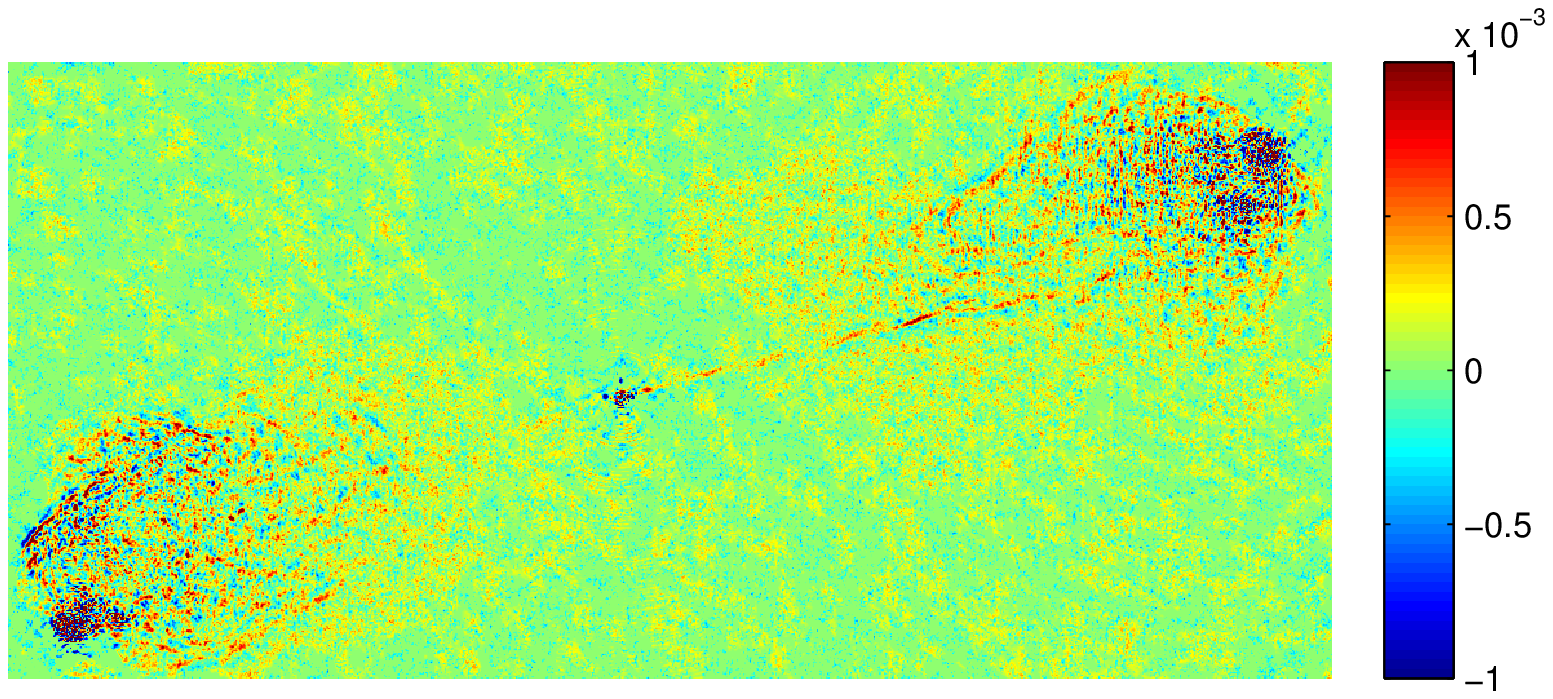}\\
    
\includegraphics[trim = 2.5cm 3.9cm 1.5cm 3.4cm, clip, keepaspectratio, width = 7.0cm]{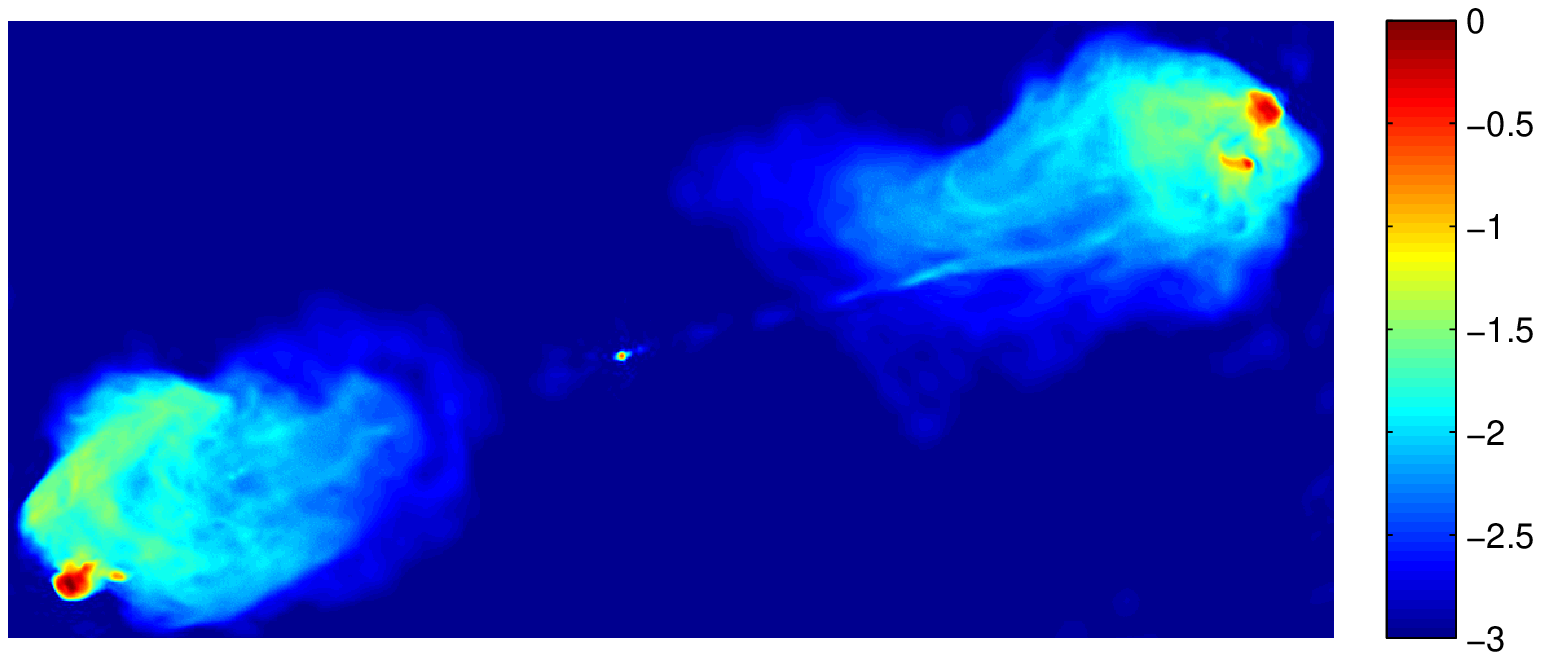}&
    \includegraphics[trim = 2.5cm 3.9cm 1.5cm 3.4cm, clip, keepaspectratio, width = 7.0cm]{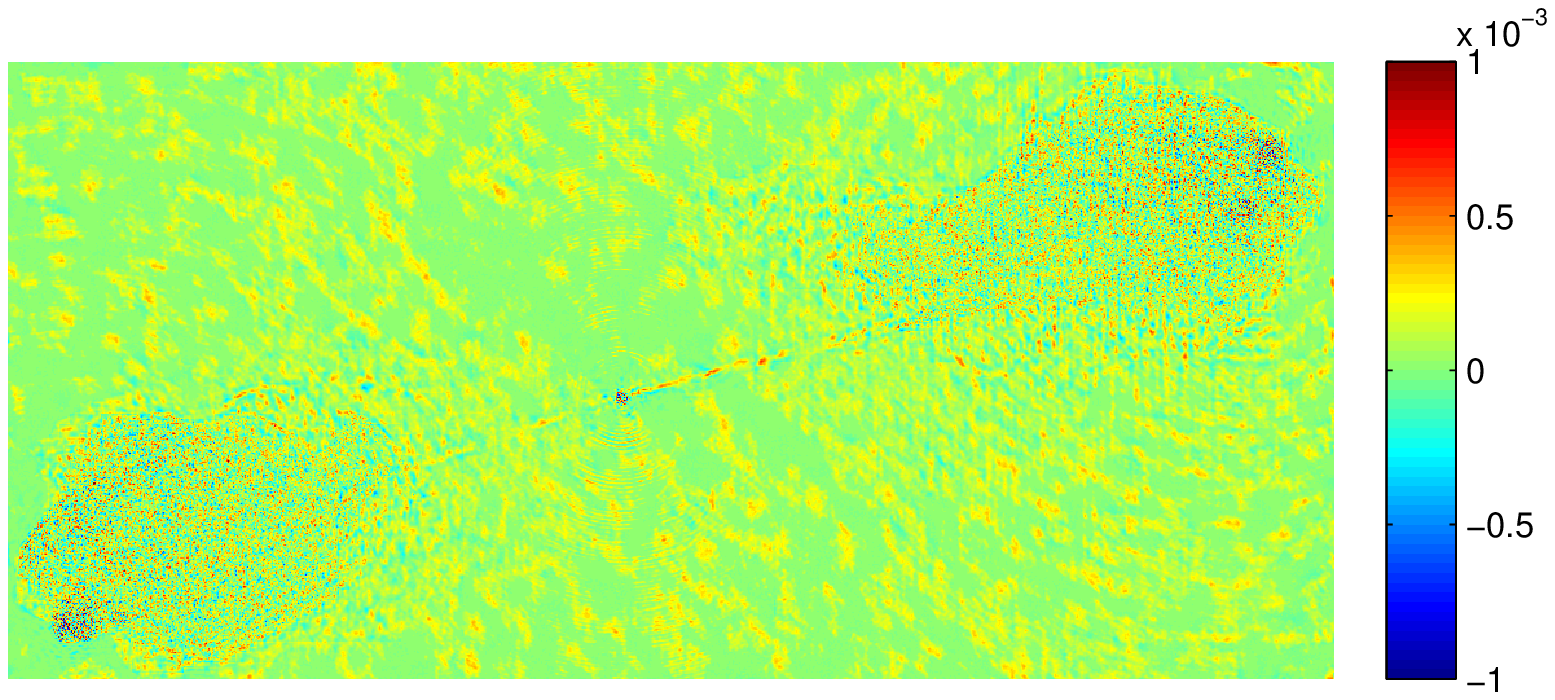}\\    
    
    \end{tabular}

\caption{(color online). Reconstruction example of Cygnus A with spread spectrum acquisition for 30\% coverage and 30 dB of input SNR. The results are shown from top to bottom for BPDb8 (SNR=36.0~dB) and SARA (SNR=40.2~dB). The first column shows the reconstructed images in a $\log_{10}$ scale and the second column shows the error images in linear scale.}
\label{fig:7}
\end{figure*}

In this subsection we present an illustrative example of the performance of SARA in the presence of the spread spectrum phenomenon. Recall that the spread spectrum phenomenon arises by partially relaxing the small FOV assumption and including a first order $w$ term. It was introduced by \citet{wiaux09b} as a potential optimization of the acquisition, leading to enhanced image reconstruction quality for sparsity bases that are not maximally incoherent with the measurement basis. Spread spectrum incorporates a modulating sequence in the measurement operator redefining it as $\mathsf{\Phi}\equiv\mathsf{MF}\mathsf{A}\mathsf{C}$, where $\mathsf{C}$ is a diagonal matrix with diagonal elements with unit norm. For the sake of simplicity we consider sequences with random phase instead of quadratic phase as considered by \citet{wiaux09b}. For our illustration we use Cygnus A as a test image~\citep{carilli96} for 30\% coverage and 30 dB of input SNR. We compare SARA against BP, BPDb8, BPIU, and TV. The SNR of the recovered image for each algorithm is as follows: BP (16.6 dB), BPDb8 (36.0 dB), BPIU (29.9 dB), TV (28.7 dB) and SARA (40.2 dB). The superior reconstruction quality of SARA is again clear. In Figure~\ref{fig:7} we show reconstructed images for SARA and BPDb8.

\section{Concluding Remarks}
\label{sec:Conclusion}
In this paper we have proposed a novel algorithm for image reconstruction in radio interferometry dubbed Sparsity Averaging Reweighted Analysis (SARA). The algorithm relies on the conjecture that astrophysical signals are simultaneously sparse in multiple bases, in particular the Dirac basis, wavelet bases, or in their gradient, so that promoting average signal sparsity over multiple wavelet bases represents an extremely powerful prior. Experimental results demonstrate that SARA outperforms state-of-the-art imaging methods in the field, all based on the assumption of signal sparsity in a single basis or signal gradient sparsity. 

In future work we plan to focus on extending the current algorithm to handle continuous visibilities. In this respect, a stable version of the algorithm must be implemented in a low level programming language. Also, the final evolution should take advantage of proximal splitting algorithms with parallel and distributed structures allowing implementation in multicore architectures or in computer clusters. Such approaches are crucial for the scalability of the proposed algorithm to very high dimensions when dealing with continuous visibilities.

\section*{Acknowledgments}
We thank Pierre Vandergheynst, Jean-Philippe Thiran and Dimitri Van De Ville for providing the infrastructure to support our research. REC is supported by the Swiss National Science Foun- dation (SNSF) under grant 200021-130359. JDM is supported by a Newton International Fellowship from the Royal Society and the British Academy and, during this work, was also supported by a Leverhulme Early Career Fellowship from the Leverhulme Trust. YW is supported in part by the Center for Biomedical Imaging (CIBM) of the Geneva and Lausanne Universities, EPFL and the Leenaards and Louis-Jeantet foundations, and in part by the SNSF under grant PP00P2-123438.

\label{lastpage}


\begin{thebibliography}{\protect\citeauthoryear{Baraniuk}{2007}}


\bibitem[\protect\citeauthoryear{Ables}{1974}]{ables74}Ables, J.G., 1974, A\&AS, 15, 383

\bibitem[\protect\citeauthoryear{Baraniuk}{2007}]{baraniuk07a}Baraniuk R., 2007, IEEE Signal Process. Mag., 24, 118

\bibitem[\protect\citeauthoryear{Bhatnagar \& Cornwell}{2004}]{bhatnagar04}Bhatnagar S., Cornwell T. J., 2004, A\&AS, 426, 747

\bibitem[\protect\citeauthoryear{Blythe}{1957}]{blythe57}Blythe J. H., 1957, MNRAS, 117, 644

\bibitem[\protect\citeauthoryear{Blumensath \& Davies}{2009}]{blumensath09}Blumensath T., Davies M., 2009, App. Comp. Harmonic Anal., 27, 265

\bibitem[\protect\citeauthoryear{Cand\`{e}s}{2006}]{candes06}Cand\`{e}s E. J., 2006, in Sanz-Sol\'e M., Soria J., Varona J. L., Verdera J., eds, Proc. Int. Congress Math. Vol. 3. Euro. Math. Soc. Z\"urich, p. 1433 

\bibitem[\protect\citeauthoryear{Cand\`{e}s}{2008}]{candes08}Cand\`{e}s E. J., 2008, C. R. Acad. Sci., 346, 589

\bibitem[\protect\citeauthoryear{Cand\`{e}s et al.}{2006}]{candes06a}Cand\`{e}s E. J., Romberg J., Tao T., 2006, IEEE Trans. Inf. Theory, 52, 489


\bibitem[\protect\citeauthoryear{Cand\`{e}s et al.}{2008}]{candes08a}Cand\`{e}s E. J., Wakin M., Boyd S., 2008, J. Fourier Anal. Appl., 14, 877

\bibitem[\protect\citeauthoryear{Cand\`{e}s et al.}{2010}]{candes10}Cand\`{e}s E. J., Eldar Y., Needell D., Randall P., 2010, App. Comp. Harmonic Anal., 31, 59

\bibitem[\protect\citeauthoryear{Carilli \& Barthel}{1996}]{carilli96}Carilli C., Barthel P., 1996, A\&A Reviews, 7, 1


\bibitem[\protect\citeauthoryear{Chen et al.}{2001}]{chen01}Chen S., Donoho D.L., Saunders M., 2001, SIAM Rev., 43, 129

\bibitem[\protect\citeauthoryear{Combettes \& Pesquet}{2007}]{combettes07}Combettes P. L., Pesquet J.-C., 2007, IEEE Sel. Top. Signal Process., 1, 564

\bibitem[\protect\citeauthoryear{Combettes \& Pesquet}{2011}]{combettes11}Combettes P. L., Pesquet J.-C., 2011,  Proximal splitting methods in signal processing, in Bauschke H. H., Burachik R.S., Combettes P.L., Elser V., Luke D.R., Wolkowicz H., eds, Fixed-Point Algorithms for Inverse Problems in Science and Engineering

\bibitem[\protect\citeauthoryear{Cornwell}{2008}]{cornwell08b}Cornwell T. J., 2008, IEEE Sel. Top. Signal Process., 2, 793

\bibitem[\protect\citeauthoryear{Cornwell \& Evans}{1985}]{cornwell85}Cornwell T. J., Evans K. F., 1985, A\&A, 143, 77

\bibitem[\protect\citeauthoryear{Daubechies}{1992}]{daubechies92}Daubechies I., 1992, Ten Lectures on Wavelets, CBMS-NSF Reg. Conf. Series in Applied Math., SIAM, Philadelphia

\bibitem[\protect\citeauthoryear{DeVore et al.}{1992}]{devore92}DeVore R. A., Jawerth B., Lucier B.J., 1992, IEEE Trans. Inf. Theory, 38, 719

\bibitem[\protect\citeauthoryear{Donoho}{2006}]{donoho06}Donoho D. L., 2006, IEEE Trans. Inf. Theory, 52, 1289

\bibitem[\protect\citeauthoryear{Donoho \& Tanner}{2009}]{donoho09}Donoho D. L., Tanner J., 2009, J. Amer. Math. Soc., 22, 1

\bibitem[\protect\citeauthoryear{Elad et al.}{2007}]{elad07}Elad M, Milanfar P., Rubinstein R., 2007, Inverse Probl., 23, 947

\bibitem[\protect\citeauthoryear{Fornasier \& Rauhut}{2011}]{fornasier11}Fornasier M., Rauhut H., 2011,  Compressed sensing, in Scherzer O., eds, Handbook of Mathematical Methods in Imaging

\bibitem[\protect\citeauthoryear{Gull \& Daniell}{1978}]{gull78}Gull S. F., Daniell G. J., 1978, Nat, 272, 686

\bibitem[\protect\citeauthoryear{H\"ogbom}{1974}]{hogbom74}H\"ogbom J. A., 1974, A\&AS, 15, 417

\bibitem[\protect\citeauthoryear{Li et al.}{2011}]{li11}Li F., Brown S., Cornwell T. J., De hoog F., 2011, A\&AS, 531, A126

\bibitem[\protect\citeauthoryear{Mallat \& Zhang}{1993}]{mallat93}Mallat S. G., Zhang Z., 1993, IEEE Trans. Signal Process., 41, 3397

\bibitem[\protect\citeauthoryear{Marsh \& Richardson}{1987}]{marsh87}Marsh K. A., Richardson J. M., 1987, A\&A, 182, 174

\bibitem[\protect\citeauthoryear{Mattingley \& Boyd}{2010}]{mattingley10}Mattingley J., Boyd S., 2010, IEEE Signal Process. Mag., 27, 50

\bibitem[\protect\citeauthoryear{McEwen \& Wiaux}{2011}]{mcewen11a}McEwen J. D., Wiaux Y., 2011, MNRAS, 413, 1318


\bibitem[\protect\citeauthoryear{Needell \& Tropp}{2008}]{needell08}Needell D., Tropp J., 2008, App. Comp. Harmonic Anal., 26, 301

\bibitem[\protect\citeauthoryear{Needell}{2009}]{needell09}Needell D., 2009, in Proc. ASILOMAR  Conf.,  p. 196

\bibitem[\protect\citeauthoryear{Nocedal \& Wright}{2006}]{nocedal06}Nocedal J., Wright S., 2006, Numerical Optimization. New York: Springer.

\bibitem[\protect\citeauthoryear{Puy et al.}{2011}]{puy11}Puy G., Vandergheynst P., Wiaux Y., 2011, IEEE Signal Proc. Letters, 18, 595

\bibitem[\protect\citeauthoryear{Rauhut et al.}{2008}]{rauhut08}Rauhut H., Schnass K., Vandergheynst P., 2008, IEEE Trans. Inf. Theory, 54, 2210

\bibitem[\protect\citeauthoryear{Rudin et al.}{1992}]{rudin92}Rudin L. I., Osher S., Fatemi E., 1992, Physica D, 60, 259

\bibitem[\protect\citeauthoryear{Ryle \& Vonberg}{1946}]{ryle46}Ryle M., Vonberg D. D., 1946, Nat, 158, 339

\bibitem[\protect\citeauthoryear{Ryle et al.}{1959}]{ryle59}Ryle M., Hewish A., Shakeshaft J. R., 1959, IRE Trans. Antennas Propag., 7, 120

\bibitem[\protect\citeauthoryear{Ryle \& Hewish}{1960}]{ryle60}Ryle M., Hewish A., 1960, MNRAS, 120, 220

\bibitem[\protect\citeauthoryear{Suksmono}{2009}]{suksmono09}Suksmono A., 2009, in ICEEI, volume 1, 110–116

\bibitem[\protect\citeauthoryear{Schwarz}{1978}]{schwarz78}Schwarz U. J., 1978, A\&A, 65, 345

\bibitem[\protect\citeauthoryear{Thompson et al.}{2004}]{thompson04}Thompson A. R., Moran J. M., Swenson G. W. Jr, 2004, Interferometry and Synthesis in Radio Astronomy. WILEY-VCH Verlag GmbH \& Co. KGaA, Weinheim

\bibitem[\protect\citeauthoryear{Tropp \& Gilbert}{2007}]{tropp07}Tropp J.A.,  Gilbert A.C., 2007, IEEE Trans. Inf. Theory, 53, 4655

\bibitem[\protect\citeauthoryear{Wenger et al.}{2010}]{wenger10}Wenger S., Darabi S., Sen P., Glassmeier K.H., Magnor M., 2010, in Proc. IEEE Int. Conf. on Image Process.. IEEE Signal Process. Soc., p. 1381

\bibitem[\protect\citeauthoryear{Wiaux et al.}{2009a}]{wiaux09}Wiaux Y., Jacques L., Puy G., Scaife A. M. M., Vandergheynst P., 2009a, MNRAS, 395, 1733

\bibitem[\protect\citeauthoryear{Wiaux et al.}{2009b}]{wiaux09b}Wiaux Y., Puy G., Boursier Y., Vandergheynst P., 2009b, MNRAS, 400, 1029

\bibitem[\protect\citeauthoryear{Wiaux et al.}{2010}]{wiaux10a}Wiaux Y., Puy G., Vandergheynst P., 2010, MNRAS, 402, 2626

\end{thebibliography}
\end{document}